\documentclass[11pt,a4paper]{article}
\usepackage{amsmath,geometry}
\allowdisplaybreaks
\usepackage{amssymb}
\usepackage{color}
\usepackage{graphicx}
\usepackage[T1]{fontenc}
\usepackage[applemac]{inputenc}
\usepackage{textcomp}
\usepackage{braket}
\usepackage{epsdice}
\usepackage{float}
\usepackage{enumerate,mathtools}
\usepackage{cancel, comment}
\usepackage[scr]{rsfso}
\usepackage{hyperref}
\usepackage{booktabs}
\usepackage[sc,small]{caption}
\usepackage{subcaption}

\usepackage{titlesec}

\usepackage{tikz}
\usepackage{hyperref}
\usepackage{fontawesome5}

\newcommand{\be}{\begin{equation}}
	\newcommand{\ee}{\end{equation}}
\newcommand{\bea}{\begin{eqnarray}}
	\newcommand{\eea}{\end{eqnarray}}
\newcommand{\tft}{{\cal T}_{F^3}}	

\newcommand{\ra}{\rightarrow}
\newcommand{\lra}{\leftrightarrow}
\newcommand{\FTa}{\xrightarrow[\text{}]{\text{f.t}}}
\newcommand{\SKa}{\xrightarrow[\text{}]{\text{skinny}}}

\newcommand{\RE}{\mathrm{Re}}
\newcommand{\IM}{\mathrm{Im}}
\newcommand{\dd}{\text{d}}
\newcommand{\JD}{\mathcal J}
\newcommand{\ID}{\mathcal I}
\newcommand{\bX}{\bar{X} }
\newcommand{\tC}{\tilde{C} }
\newcommand{\tP}{\tilde{P} }

\newcommand{\half}{\frac{1}{2}}
\newcommand{\tho}{\vartheta_1}

\newcommand{\CG}{\mathcal G}
\newcommand{\FTM}{\mathscr{M}}
\newcommand{\euv}{\epsilon}
\newcommand{\efn}{\varepsilon}
\newcommand{\esm}{\epsilon}
\newcommand{\pT}{\mathsf{p}}
	
\newcommand{\co}{\ , \ \ \ \ \ \ }
\newcommand{\efrak}{\mathfrak e}
\newcommand{\ffrak}{\mathfrak f}
\newcommand{\te}{\textrm}
\newcommand{\OrbSum}{\Xi}
\newcommand{\hs}{\hat{s}}
\newcommand{\hatt}{\hat{t}}
\newcommand{\Lgsh}{\log\!\left(\frac{\hs}{\hatt}\right)}
\newcommand{\FO}{F_{1/2}^{(2)}}
\newcommand{\FOT}{\tilde{F}_{1/2}^{(2)}}

\titleformat{\paragraph}
{\normalfont\normalsize\bfseries}{\theparagraph}{1em}{}
\titlespacing*{\paragraph}
{0pt}{3.25ex plus 1ex minus .2ex}{1.5ex plus .2ex}

\numberwithin{equation}{section}

\begin{document}
	\hfill 
	
	\vspace{2cm}
	\begin{center}
		{\Large {\bf One-loop Amplitudes: \\[2mm] String Methods, Infrared Regularization, and Automation}}
		
		\vspace{2cm}
		Marcus Berg${}^1$, Michael Haack${}^2$, Yonatan Zimmerman${}^2$\\ \ \\
		{\it ${}^1$ Karlstad University, Universitetsgatan 1, 651 88  Karlstad, Sweden\\ [8pt]
			${}^2$ Arnold Sommerfeld Center for Theoretical Physics,\\ 
			Ludwig--Maximilians--Universit\"at M\"unchen, Theresienstr. 37,
			80333 Munich, Germany }

		\thispagestyle{empty}
			\setcounter{page}{1}
		\vspace{2cm}
		{\bf Abstract}\\
		
	\end{center}
		
		We calculate field theory loop amplitudes by string methods, applied to half-maximal 4-point one-loop graviton amplitudes.
		Infrared divergences are regulated similarly to
		soft-collinear effective field theory (SCET): new mass scales are introduced, here by higher-point kinematics.
		We use an analytically continued single-valued polylogarithm as generating function. The Feynman integrals for the new tensor structures are infrared finite. We provide code as a step towards automation.

	\newpage
	
	\tableofcontents
	
	
	\section{Introduction}
	The main theme of this paper is how to handle infrared (soft and collinear) divergences in string and field theory. 
	Let us first very briefly outline the big picture. Weinberg's 1965 {\it soft theorem} is clearly explained in Schwartz's textbook \cite{Schwartz:2014sze}, Ch.~9.
	Consider an $n$-point amplitude. If we add one photon of momentum $k$ and polarization $e$ to the $i$th leg with momentum $p_i$, this produces the  factor
	\begin{equation}
		\frac{p_i \cdot e}{p_i \cdot k} \; . 
	\end{equation}
	Weinberg emphasized that we should expect the infrared limit $p_i \cdot k\rightarrow 0$ where the denominator vanishes to be cancelled in the total $S$-matrix element, by gauge invariance.
	For example, in \cite{Schwartz:2014sze}, Ch.~20 there is an explicit computation of the 4-point function of leptons
	at one-loop, and it is shown how the infrared divergences of this one-loop amplitude are cancelled by a 5-point function at tree-level, where the extra external state is a photon.
	
	It is not so convenient for perturbation theory to have cancellations across different orders (tree-level vs.\ one-loop). There are now many methods that improve on
	this classic cancellation from QED.
	One modern example that has some resemblance to the spirit (if not detail) of this paper is the {\it hard S-matrix} constructed by 
	Hannesdottir and Schwartz \cite{Hannesdottir:2019opa}. They work at fixed order of perturbation theory, say one-loop, but systematically use cuts
	 to reproduce the above cancellation. The philosophy is that of soft-collinear effective field theory (SCET). There, one
	gives up full manifest Lorentz invariance, which provides additional kinematic variables. In practice, these variables appear as low-energy mass scales
	in Feynman integrals, that regulate those integrals, sometimes completely. This is a simple but crucial calculational point. Here, 
	 we keep full manifest Lorentz invariance but harvest additional invariants from higher-point functions.
	
	In more detail: we introduce a method to compute the field theory (worldline) limit of string theory amplitudes at fixed order of perturbation theory, using as generating functions certain Feynman integrals with {\it external masses}. The use of generating functions in this context is of course not new, in fact we essentially recycle the Bern-Dixon-Kosower (BDK) differentiation method from 1993 \cite{Bern:1993kr}. 
	Using external masses as regulators is also hardly new. But the combination of computing {\it massless} string amplitudes in terms of amplitudes with external {\it masses} is if not completely new, then at least unusual. (To avert confusion: we are {\it not} computing amplitudes with massive external string states,
	the masses are only regulators.)
	
	Allowing for external masses might seem like introducing complications in an already complicated calculation. But we claim the situation is the opposite: 
	a key feature of our approach is that Feynman diagrams with all masses generic, ``all-mass diagrams'', sometimes
	enjoy greatly increased symmetry compared to when some masses are taken to zero, and postponing the taking of the soft/collinear limit to the very end of the calculation increases
	 {\it simplicity}, not complexity. In particular, as is well known, the scalar $D$-point function in exactly 
	$D$ dimensions (so the 4-point function in $D=4$, but not in $D=4-2\epsilon$) has 
	{\it dual conformal invariance}, as explained for example in \cite{Schnetz:2010pd,Bourjaily:2019exo}. To be clear, this is a generic property
	of the nonsupersymmetric Feynman integral. In fact, it is an important feature of the formalism in this paper that it makes no use of supersymmetry.
	(We do apply it to an example with a specific amount of supersymmetry, but we believe that it could equally well be applied to any piece in
	the supersymmetry decomposition of a nonsupersymmetric loop amplitude.)
	
	In string theory, there are other methods that bear some resemblance to ours, for example  momentum from a pinch \cite{Stieberger:2022lss,Stieberger:2023nol},
	and chiral splitting \cite{Balli:2024wje}. We do not make explicit connections to these or other string methods in this paper, so this is an interesting question for the future.
	
	There is of course a vast literature on graviton loop amplitude more broadly. However, most of it is in a helicity basis,
	that for reasons that will become clear, we do not employ. (One obvious reason is there is no dimension-independent helicity basis.) Early papers calculating the maximally supersymmetric one-loop 4-point graviton amplitude are \cite{BGS,Green:1999pv}. Another early paper 
	that computed the one-loop pure gravity 4-point function in helicity basis is Dunbar-Norridge in 1994 \cite{DunbarNorridge}.
	We use some details from \cite{Tourkine:2012vx} that used string theory to recreate and extend the Dunbar-Norridge helicity result. 
	A few other papers that make progress on and summarize previous results are
	\cite{Edison:2020uzf,Edison:2021ebi,Porkert:2022efy}. We make particular use of \cite{Bern:2017tuc}, which gives useful comments on the basis of curvature tensors.
	The alternative point of view in \cite{Chowdhury:2019kaq}, using representation theory to constrain the nonsupersymmetric S-matrix, 
	is very relevant to our calculations, but we do not attempt a complete match 
	to their formalism in this paper. We do use explicit analytic continuations in field theory, in particular \cite{Chavez:2012kn,Bourjaily:2020wvq,Corcoran:2020akn,Corcoran:2020epz}.
	 
    Finally, we rely heavily on \cite{paper1,paper2}: in a nutshell, the main technical result of this paper is to perform the Feynman integrals whose integrands were constructed in those two papers.
    
\begin{figure}[htbp]
		\begin{center}
			\includegraphics[width=\textwidth]{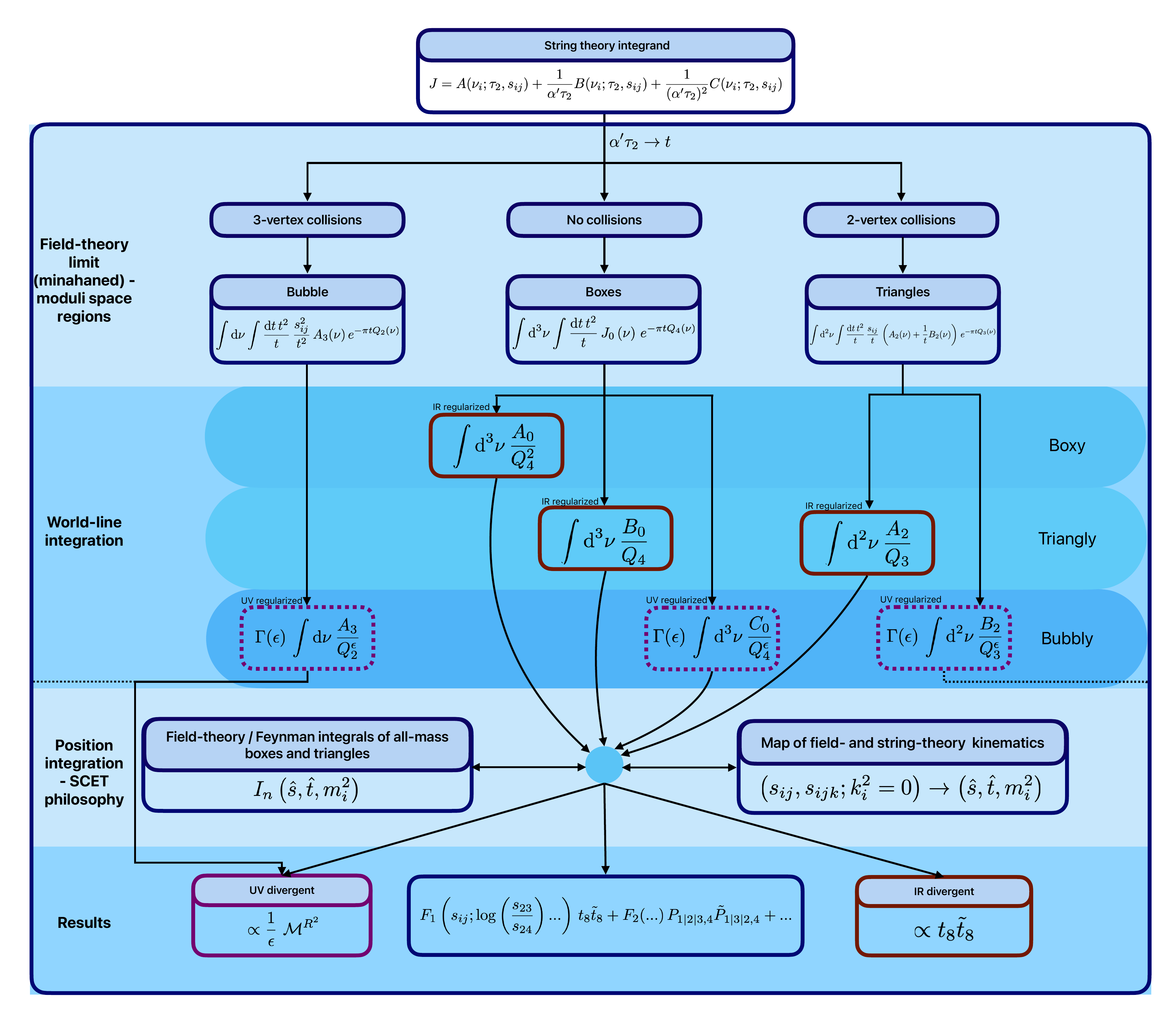}
			\caption{Flow chart that summarizes the method developed in this work to calculate the field-theory limit integrals of amplitude \eqref{eq:our_amplitude}. The horizontal sections correspond to sections \ref{sec:FT_limit}, \ref{sec:WL_int}, \ref{sec:position_int} and \ref{sec:results}, respectively. The circle in the center of the position integration section contains the machinery by which the integrals are calculated. For its expansion, see fig.\ref{fig:flow_chart_dot}.}
			\label{fig:flow_chart}
		\end{center}
	\end{figure}

\begin{figure}[htbp]
		\begin{center}
			\includegraphics[width=0.9\textwidth]{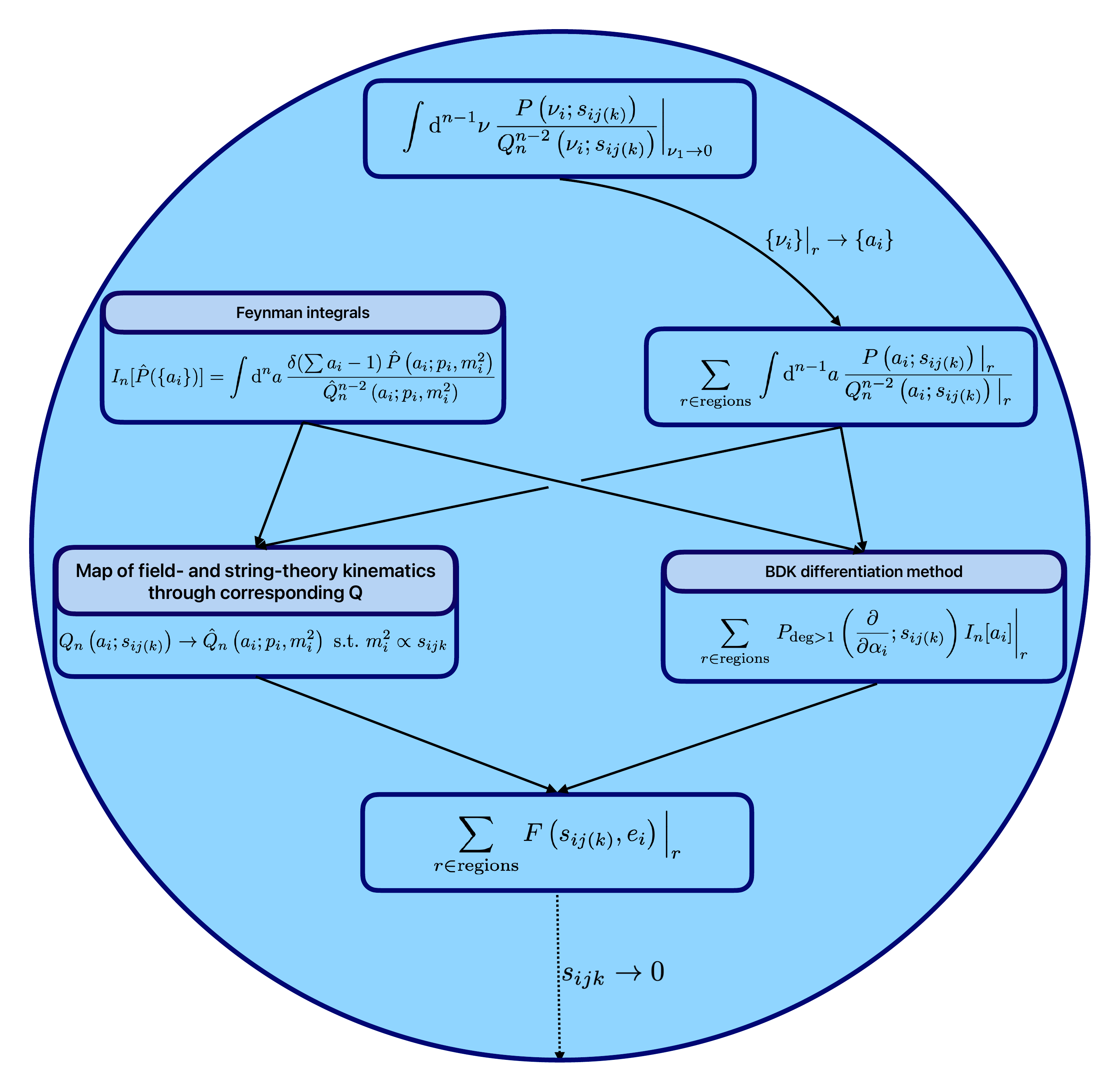}
			\caption{Flow chart corresponding to the circle in the center of the position integration section of fig.\ref{fig:flow_chart}. Here summarizing the method of calculating the field-theory embedded integrals, explained in detail in section \ref{SCET}, specifying to the most complicated box integrals. }
			\label{fig:flow_chart_dot}
		\end{center}
	\end{figure}

    We provide code for most of our calculations as a step towards automation (the link to the repository can be found at the beginning of sec.\ \ref{sec:review_string_comp}). Moreover, in figs.\ \ref{fig:flow_chart} and \ref{fig:flow_chart_dot}, we offer a flow chart that the reader can revisit when getting lost in the details. Table \ref{sec:results} is included for a similar purpose, serving as a reference point to help orient the reader with respect to the various contributions to our final results.
	
		\begin{table}[ht]
		\centering
		\begin{tabular}{lcccccc}
			\toprule
			& \multicolumn{3}{c}{Boxes} & \multicolumn{2}{c}{Triangles} & Bubble \\
			\cmidrule(lr){2-4} \cmidrule(lr){5-6}
			Contributing terms & Boxy & Triangly& Bubbly & Triangly & Bubbly & Bubbly \\
			\midrule
			Finite        & $\checkmark$  & $\checkmark$  & $\times$ & $\times$ & $\times$ & $\times$ \\
			IR divergent  & $\checkmark$  & $\times$ & $\times$ & $\times$ & $\times$ & $\times$ \\
			UV divergent  & $\times$ & $\times$  & $\checkmark$ & $\times$ & $\checkmark$ & $\times$ \\
			\bottomrule
		\end{tabular}
		\caption{Summary of contributing terms to the full amplitude, presented in section \ref{sec:results}.}
	\end{table}
	
	\section{Review of string computation} \label{sec:review_string_comp}
	
	In this section we give a relatively brief review of how correlation functions in worldsheet conformal field theory are assembled to an {\it integrand} of the  four-point one-loop scattering amplitudes of the open- and closed-string at half-maximal supersymmetry. The rest of this paper, after this review, is about {\it integrating} this integrand. 	
	The computation will trace the work in sections 4 and 6 of \cite{paper1}. The code used in these calculations is available from 
	\begin{center}
	the repository \href{https://github.com/YonatanZimmerman/Field-theory-limit-of-half-maximal-string-loop-amplitudes}{\faGithub}.
	\end{center}
	In the main body of this work, however, the main part of the algorithm is presented in a programming-language-agnostic procedure.

	\subsection{Spin summations} \label{sec:spinSums}
	
	\subsubsection{Open-string half-maximal supersymmetry spin sums}
	
	In this review section, we will focus on the half-maximal $n$-point case and start with the open string.
	
	The open-string integrand has a correlation function of the form
	\begin{equation} \label{corr}
		\braket{V_1^{(0)}(z_1)\dots V_n^{(0)}(z_n)}_{\nu} \ ,
	\end{equation}
	where $\nu$ labels the spin structure and gauge bosons are represented by vertex operators
	with polarization $e^m$ and momenta $k^m$ \footnote{\label{fn:Convention_iX}Here we absorbed a factor of $i$ in $ X^m$. Compare
		the  standard form $e^{ik\cdot X}$ in a vertex operator.} 
		
	\be
	V^{(0)}(e,k) \equiv e_m (\partial X^m + (k\cdot \psi) \psi^m) e^{k\cdot X} ,
	\label{red21}
	\ee
	and the $(0)$ superscript denotes zero superghost picture. (For non-string theorists: this is a discrete gauge choice for the vertex operator
	representing the external gauge boson, among an infinity of  choices. The worldsheet quantum fields $X$ and $\psi$ are bosons and fermions, respectively.)
	
	For the purposes of this review, let us focus on contributions to \eqref{corr} with only worldsheet fermions $\psi$. They are of the form
	\begin{equation} \label{SpinSums_1}
		\braket{[(e_i \cdot \psi_i)(k_i \cdot \psi_i)] \dots [(e_j \cdot \psi_j)(k_j \cdot \psi_j)] }_{\nu} \ ,
	\end{equation}
	where $\psi_i \equiv \psi(z_i)$. Relabel the indices in \eqref{SpinSums_1} and concentrate exclusively on the fermions themselves, factoring
	out the constant factors $e_i$ and $k_i$:
	\begin{equation} \label{SpinSums_2}
		\braket{(\psi \psi)_1 \, (\psi \psi)_2 \dots (\psi \psi)_m }_{\nu} \ .
	\end{equation}
	Pairwise contractions produce closed cycles of the two-point function (Szeg\"o kernel) $S_{\nu}$,
	\begin{equation} \label{SpinSums_3}
		S_{\nu}(x_1) \, S_{\nu}(x_2) \dots S_{\nu}(x_m) \ , \qquad \sum_{i=1}^m x_i = 0 \ .
	\end{equation} 
	The Szeg\"o kernel has a representation in terms of Jacobi theta functions:
	\be \label{Skernel}
	S_{\nu}(z) \equiv \frac{\tho'(0,\tau) \, \vartheta_{\nu}(z,\tau)}{\vartheta_{\nu}(0,\tau) \, \tho(z,\tau)} \ .
	\ee
	
	In \cite{paper1}, the summation over $\nu$ was performed for the reduced supersymmetric sector, namely the half- and quarter-maximal even sectors.
	This is done by rewriting the orbifold-twisted partition functions in terms of fermion Green's functions with the twist vector as the argument:
	\begin{equation} \label{SpinSumHalf_1}
		\ID_{n,1/2}^e (\vec v_k) = \frac{1}{\Pi_n} \, \sum_{\nu=2}^{4} (-1)^{\nu-1} \Bigg [ \frac{\vartheta_{\nu}(0,\tau)}{\tho'(0,\tau)} \Bigg ]^4 \, S_{\nu}(kv) S_{\nu}(-kv)\, \braket{V_1^{(0)}(z_1)\dots V_m^{(0)}(z_m)}_{\nu} \ ,
	\end{equation}
	where the Koba-Nielsen factor $\Pi_n$ is given in eq.~\eqref{KNdef}.
	The two $S_{\nu}$ combine with the explicit Jacobi theta functions to form the K3 $\times\, T^2$ orbifold partition function: these two  $S_{\nu}$ can be thought of as background field representation of the nontrivial background.
	The  fermion Wick contractions in \eqref{SpinSumHalf_1} will yield the same cycles as \eqref{SpinSums_2}:
	\begin{equation}\label{SpinSumsHalf_2}
		\sum_{\nu=2}^{4} (-1)^{\nu-1} \Bigg [ \frac{\vartheta_{\nu}(0,\tau)}{\tho'(0,\tau)} \Bigg ]^4 \, S_{\nu}(kv) S_{\nu}(-kv)\, S_{\nu}(x_1) \, S_{\nu}(x_2) \dots S_{\nu}(x_m) \equiv  \CG_{m+2} (\gamma, -\gamma, x_1,\dots , x_m) \ ,
	\end{equation}
	where ${\mathcal G}$ is just notation for the left-hand side, and the orbifold twist vector is:
	\begin{equation} \label{orbifoldgamma}
		\gamma \equiv kv \ , \quad \gamma_i \equiv kv_i \ ,
	\end{equation}
	and we have $\gamma_1 = -\gamma_2$ for K3 $\times\, T^2$ orbifold compactifications. Note that we are relating the half-maximally supersymmetric spin sum for $m$ two-point functions to that of maximal supersymmetry with $m+2$ two-point functions,
	hence the notation $\CG_{m+2}(\gamma,-\gamma,x_1,\dots,x_m)$. 
	
	To write this out explicitly, we follow the notation of \cite{Broedel:2014vla} where a doubly-periodic function  $f^{(n)}$ for each non-negative integer $n$ is defined by 
	using a non-holomorphic Kronecker-Eisenstein series as generating function (closely related to the Szeg\"o kernel $S_{\nu}$)
	\be
	\Omega(z,\alpha,\tau) \equiv \exp\Big(2\pi i \alpha \frac{  {\rm Im\,} z}{ {\rm Im\,} \tau} \Big) \frac{ \vartheta_1'(0,\tau) \vartheta_1(z+\alpha,\tau) }{\vartheta_1(z,\tau)\vartheta_1(\alpha,\tau)} 
	\equiv
	\sum_{n=0}^{\infty} \alpha^{n-1} f^{(n)}(z,\tau) \ ,
	\label{red11}
	\ee
	and the functions $f_i$ so generated begin with:
	\begin{align}
		f^{(0)}(z,\tau) &\equiv 1 \ , \quad f^{(1)}(z,\tau) =   \partial \ln \vartheta_1(z,\tau) +  2\pi i \, \frac{  {\rm Im\,} z }{ {\rm Im\,}\tau}\ , 
		\label{red12}
		\\
		f^{(2)}(z,\tau) & \equiv \frac{1}{2} \Big\{  \Big( \partial \ln \vartheta_1(z,\tau) +  2\pi i \, \frac{ {\rm Im\,} z }{ {\rm Im\,} \tau} \Big)^2 + \partial^2 \ln \vartheta_1(z,\tau) - \frac{\vartheta_1'''(0,\tau)}{ 3\vartheta_1'(0,\tau)} \Big\} \ ,
		\label{red13}
	\end{align}
	which are actually all the $f^{(i)}$ we need for the purposes of this paper.
	Here, $f^{(1)}$ is the only singular term of (\ref{red11}) with a simple pole at lattice points $z=n+m\tau$ with $m,n\in {\mathbb Z}$. We will suppress the dependence on the modular parameter $\tau$ in the following. (For the reader who is not comfortable with this type of function, we only ask for patience: in the next sections, taking the field theory limit will collapse them to polynomials.)
	
	By using results from the maximally supersymmetric case (cf.\ (3.20)-(3.22) in \cite{paper1}), we find:
	\begin{equation} \label{CG_22}
		\mathcal G_{2+2}(\gamma,-\gamma,x_1, x_2)  = 1 \ , 
	\end{equation}  
	and 
	\begin{equation} \label{CG_32}
		\mathcal G_{3+2}(\gamma,-\gamma,x_1, x_2,x_3)  = \sum_{i=1}^3 f_i^{(1)} \ , 
	\end{equation}
	since $f^{(1)}(\gamma) =-f^{(1)}(-\gamma) $, and
	\begin{equation} \label{CG_42}
		\mathcal G_{4+2}(\gamma,-\gamma,x_1, x_2,x_3,x_4)  = F^{(2)}_{1/2}(\gamma) + \sum_{i=1}^4 f_i^{(2)} + \sum_{1\leq j  < k}^4 f_j^{(1)} \, f_k^{(1)}  \ . 
	\end{equation}
	For the interested reader, this can be seen as follows:
	\begin{equation} \begin{aligned}
			\mathcal G_{4+2}(\gamma,-\gamma,x_1, x_2,x_3,x_4)  & = f^{(2)}(\gamma) + f^{(2)}(-\gamma) + \sum_{i=1}^4 f_i^{(2)} +f^{(1)}(\gamma) f^{(1)}(-\gamma) \\
			& + f^{(1)}(\gamma) \sum_{i=1}^4 f_i^{(1)} + f^{(1)}(-\gamma) \sum_{i=1}^4 f_i^{(1)} + \sum_{1\leq j  < k}^4 f_j^{(1)}  f_k^{(1)} \\
			& = 2f^{(2)}(\gamma) - (f^{(1)}(\gamma))^2 + \sum_{i=1}^4 f_i^{(2)} + \sum_{1\leq j  < k}^4 f_j^{(1)} \, f_k^{(1)}  \ ,
		\end{aligned} 
	\end{equation} 
	where we used even- and oddness of $f^{(2)}$ and $f^{(1)}$, respectively, and
	\begin{equation} \label{F2_def}
		F^{(2)}_{1/2}(\gamma) \equiv  2f^{(2)}(\gamma) - (f^{(1)}(\gamma))^2 \ .
	\end{equation}
	We see that the orbifold twist
	$\gamma$ only appears in spin sums with at least four fermion insertions, $\CG_{4+2}$.

	\subsection{Automating the worldsheet Wick contractions}
	This algorithm of Wick's theorem application is based on \cite{Vieira}. We put the operators into a list (with the exponentials being placed last), that we call {\it list}, and apply the following recursive steps:
	\begin{enumerate}
		\item Extract the pair-wise contraction of the first and $k$\textsuperscript{th} elements of $list$ with the appropriate sign, according to the number of anti-commuting operators that had to be switched. Start with $k=2$.
		\item Alter the $list$ according to the following criteria: if the two-point function of the contracted elements does not contain an operator, drop the two elements from the $list$. The other case manifests only in the bosonic sector, where contractions of $\partial X$ with the exponentials retain the exponentials. In this case drop only the non-exponential element from the $list$.
		\item Apply the procedure to the altered $list$.
		\item Run until $k$ equals the length of $list$.
		\item Break if $list$ is empty or if it contains only the exponentials. In the latter case, the correlation function is the Koba-Nielsen factor 
\begin{equation}
\Pi_n \equiv \biggl\langle \prod_{r=1}^{n} e^{k_r \cdot X(z_r)} \biggr\rangle
= \prod_{1 \le i < j}^{n} e^{s_{ij} G_{ij}} \, . \label{KNdef}
\end{equation}
	\end{enumerate}
	
				\subsection{Closed-string four-point scattering amplitudes} \label{sec:cs_calc}
				
				The integration measure of $n$-point string scattering amplitudes on the torus in $D=4$ or $6$ uncompactified dimensions is given by
				\be \label{cs_measure}
				\int d\rho^{D}{_{12...n}}  = \frac{(\alpha')^n V{_D}}{8N} \int_{\mathcal F} \frac{\dd^2\tau}{(4\pi^2 \alpha ' \tau_2)^{D/2}} \int_{\mathcal T (\tau)^n}  
				\dd^2z_1 \dd^2z_2 ...\dd^2z_n \,\delta^2(z_1,\bar{z}_1) \Pi_n \ ,
				\ee
				as in eq.(3.61) in \cite{paper1}\footnote{There is a factor of $(\alpha')^n$ missing in that formula.}. The torus modulus $\tau$ is integrated over the fundamental domain,
				\be
				\mathcal F : \left | \RE(\tau) \right | \leq \half  \text{ and } \left | \tau \right | \geq 1 \ ,
				\ee
				and the insertions are integrated over the torus $\mathcal T (\tau)$, parametrised by $\tau$.
				
				The integrand for a half-maximal supersymmetric orbifold is the sum of the maximally supersymmetric term and the orbifold twisted terms, eqs.~(3.64)-(3.67) in \cite{paper1}, so that the amplitude becomes
				\be \label{cs_amplitude}
				\mathcal M_{1/2} (1,2,\dots,n) = \int \dd \rho_{12\dots n}^{D}\ \Bigg \{ \Gamma \, \JD_{n, \text{max}} +  \sum_{\substack{k,k'=0 \\ (k,k') \neq (0,0)}}^{N-1} \! \hat \chi_{k,k'} \, \JD_{n,1/2}(\vec v_{k,k'}) \Bigg \} \ ,
				\ee
				where $\Gamma$ is the sum over winding and KK momentum states, $\vec v_{k,k'} \equiv (k+k'\tau)(v,-v)$ and $\hat \chi_{k,k'}$ denotes the degeneracies of twisted states, see discussion in appendix A in \cite{paper1}.
				
				To calculate the closed-string \emph{four-point} one-loop amplitudes in half-maximal supersymmetry, we divide and conquer: divide each sector of the amplitude into the contributions from the zero-mode contractions of left-and right moving bosonic operators, and contributions from the left and right correlation functions.
			
				\subsubsection{Closed-string separated correlators}
				Due to the exponentials remaining in the correlation function after a contraction with a bosonic operator, we can think of them as appearing in both left- and right-moving bosonic correlation functions. We only need to remember that when combining the two contributions there is only one Koba-Nielsen factor $\Pi_4$, not two. 
				Hence, the complete contribution from the non-zero-modes (non-Z.M.) to the half-maximal closed-string correlator is 
				\begin{equation} \begin{aligned} \label{Jhalf_nonZM} 
						\JD_{4,1/2}(\vec v_{k,k'}) \Big |_{\text{non-Z.M.}} &=  \Big \{\JD_{4,1/2}^{e,\tilde e}(\vec v_{k,k'}) + \JD_{4,1/2}^{e,\tilde o}(\vec v_{k,k'})+ \JD_{4,1/2}^{o,\tilde e}(\vec v_{k,k'})+ \JD_{4,D=6}^{o, \tilde o} \Big \} \Bigg |_{\text{non-Z.M.}} \\
						& = \ID_{4,1/2}^{e} \, \tilde \ID_{4,1/2}^{e} + \ID_{4,1/2}^{e} \, \tilde \ID_{4,D=6}^{o}  + \ID_{4,D=6}^{o} \, \tilde \ID_{4,1/2}^{e}  + \ID_{4,D=6}^{o} \, \tilde \ID_{4,D=6}^{o} \\
						& = \ID_{4,1/2} \, \tilde \ID_{4,1/2} \ ,
				\end{aligned} \end{equation} 
				where $ \ID_{4,1/2}^{e}$ and $ \ID_{4,D=6}^{o}$ refer to the contribution of even and odd spin structures,and are given explicitly in eqs.~(4.22) and (4.48) in \cite{paper1}, respectively. The $D=6$ delineation in the odd terms points to the fact that they only occur in this case and drop out for $D=4$, and where $\tilde \ID_{4,1/2}$ is understood to map $e_i^m \ra \tilde e_i^m$ and $f_{ij}^{(1)} \ra \bar f_{ij}^{(1)}$.

				\subsubsection{Closed-string zero-mode bosonic contractions}
				
				In the four-point case we only need to consider up to two zero-mode contractions since otherwise there would not be enough world-sheet fermions left in the correlator  in order to lead to a non-vanishing result after spin structure summation. In the case of left-right interactions sector by sector, one gets:
				\begin{equation} \label{cs_ZM_ee}
					\JD_{4,1/2}^{e,\tilde e} \Big |_{\text{Z.M.}}= \zeta \ \ID_{4,1/2}^{m,\,e} \, \tilde \ID_{4,1/2}^{m,\, e} + \half \zeta^2 \ \ID_{4,1/2}^{mn,\,e} \, \tilde \ID_{4,1/2}^{mn,\, e} \ ,
				\end{equation}
				\begin{equation} \label{Jhalf_eo_corroboration}
					\JD_{4,1/2}^{e,\tilde o} \Big |_{\text{Z.M.}}= \zeta \ \ID_{4,1/2}^{m,\,e} \, \tilde \ID_{4,D=6}^{m,\, o} + \half \zeta^2 \ \ID_{4,1/2}^{mn,\,e} \, \tilde \ID_{4,D=6}^{mn,\, o} \ ,
				\end{equation}
				and
				\begin{equation} \label{cs_ZM_oo}
					\JD_{4,D=6}^{o,\tilde o} \Big |_{\text{Z.M.}}= \zeta \ \ID_{4,D=6}^{m,\,o} \, \tilde \ID_{4,D=6}^{m,\, o} +\half  \zeta^2 \ \ID_{4,D=6}^{mn,\,o} \, \tilde \ID_{4,D=6}^{mn,\, o} \ ,
				\end{equation}
				where $\JD_{4,1/2}^{o,\tilde e}$ is trivially implied from \eqref{Jhalf_eo_corroboration} upon switching left- and right-movers and $\zeta \equiv \frac{\pi}{\IM(\tau)}$ is the factor emerging from the zero-mode contractions themselves (setting $\alpha' = 2$ for the closed-string). The even-sector vectorial and tensorial integrands are given in eqs.~(6.12) and (6.13) in \cite{paper1}, respectively, and the odd-sector corresponding integrands are in eqs.~(6.15) and (6.16) in the reference.
				
				The above conspire together to
				\begin{equation}
					\ID_{4,1/2}^m = \ID_{4,1/2}^{m,\, e} + \ID_{4,D=6}^{m,\, o} \ , \qquad \ID_{4,1/2}^{mn} = \ID_{4,1/2}^{mn,\, e} + \ID_{4,D=6}^{mn,\, o} \ ,
				\end{equation}
				corresponding to eq.~(6.7) in \cite{paper1}.

				\subsubsection{Full amplitude}
				
				Accumulating the above results, we can summarise the closed-string four-point one-loop half-maximal amplitude:
				\begin{equation} \label{JHalf_calculated} 
					\JD_{4,1/2} = \ID_{4,1/2} \, \tilde \ID_{4,1/2} +\Big ( \frac{ \pi }{\IM(\tau)} \Big ) \, \ID_{4,1/2}^m \, \tilde \ID_{4,1/2}^m +\half \Big ( \frac{ \pi }{\IM(\tau)} \Big )^2\, \ID_{4,1/2}^{mn} \, \tilde \ID_{4,1/2}^{mn} \ ,
				\end{equation}
				which agrees with eq.(6.29) in \cite{paper1}. The accompanying code \href{https://github.com/YonatanZimmerman/Field-theory-limit-of-half-maximal-string-loop-amplitudes/blob/main/Half-Maximal%20Four-Point%20One-Loop%20Amplitudes%20.nb}{\faGithub} corroborates these findings.
				The end result of this review section is the closed-string integrand
				\begin{equation} \label{cs_integrand}
					\begin{aligned}
						\JD_{4,1/2} & = \Big | X_{23,4}C_{1|234} + X_{24,3}C_{1|243} + [s_{12}(f_{12}^{(2)} + f_{34}^{(2)})P_{1|2|3,4} + (2 \leftrightarrow 3,4 )] -2 F_{1/2}^{(2)}(\gamma) t_8 (1,2,3,4)  \Big |^2  \\
						&+  \Big ( \frac{2 \pi }{\alpha' \,\IM(\tau)} \Big ) (X_{23}C_{1|23,4}^m + X_{24}C_{1|24,3}^m + X_{34}C_{1|34,2}^m ) (\bX_{23} \tC_{1|23,4} ^m  + \bX_{24}\tC_{1|24,3}^m + \bX_{34}\tC_{1|34,2}^m) \\
						&+\Big ( \frac{2 \pi }{\alpha' \,\IM(\tau)} \Big )^2 (\frac{1}{2} C_{1|2,3,4}^{mn} \tC_{1|2,3,4}^{mn} - P_{1|2|3,4}\tP_{1|2|3,4} - P_{1|3|2,4}\tP_{1|3|2,4} - P_{1|4|2,3}\tP_{1|4|2,3}) \ ,
					\end{aligned}
				\end{equation}
				which is eq.(6.32) in \cite{paper1}, except that we restored factors of $\alpha'$, that in \cite{paper1} were set to 2 in closed-string calculations. This makes it clear that the terms in all three lines of \eqref{cs_integrand}  have the same mass dimension 8, see table \ref{tab:alpha}.
				In the equation above, $\gamma$ is the orbifold twist from eq.\ \eqref{orbifoldgamma} and the contracted~$t_8$ is given in eq.~\eqref{t8}. Additionally, we use the definitions of \cite{paper1}:
				\be
				X_{ij} = s_{ij}f^{(1)}_{ij}\; , \quad X_{ij,k} =s_{ij}f_{ij}^{(1)}(s_{ik}f_{ik}^{(1)}+s_{jk}f_{jk}^{(1)}) \ .
				\ee
				In general we define the Mandelstam variables:
				\begin{equation} \label{Mand.def}
					s_{i_1 ... i_p} \equiv \half k^2_{i_1 ... i_p} \ \ ; \ \ k_{i_1 ... i_p} \equiv k_{i_1} + ... + k_{i_p} \ .
				\end{equation}
				The kinematic factors $C_{1|234}$, $C^m_{1|234}$, $C^{mn}_{1|234}$, and $P_{1|234}$ are Berends-Giele currents, defined recursively in terms
				of polarizations and momenta.
				We provide them in auxiliary Mathematica files, and  appendix \ref{sectappBG} has a short introduction to how they are constructed.  
					\begin{table}[t]
					\centering
					\begin{tabular}{lcccccccc}
						\toprule
						Quantity & $k^m$ & $s_{ij}$ & $C$ & $C^m$ & $C^{mn}$ & $P$ & $X_{ij}$ & $X_{ij,k}$ \\
						\midrule
						Mass dimension & $1$ & $2$ & $0$ & $1$ & $2$ & $2$ & $2$ & $4$ \\
						\bottomrule
					\end{tabular}
					\caption{\label{tab:alpha}Mass dimensions of the expressions in eq.~\eqref{cs_integrand}. See appendix~C of \cite{paper2} for relations between them.}
				\end{table}
				The maximally supersymmetric integrand in \eqref{cs_amplitude}, specified to the four-point amplitude is
				\begin{equation} \label{J4-max}
					\mathcal J _{4,\text{max}} = 4 \, t_8(1,2,3,4)\, \tilde{t}_8(1,2,3,4) \ .
				\end{equation}
				
				\section{Infrared regularization: from minahaning to virtuality} \label{sec:IR_virtualities}
				\begin{figure}[htbp]
					\begin{center}
						\includegraphics[width=0.3\textwidth]{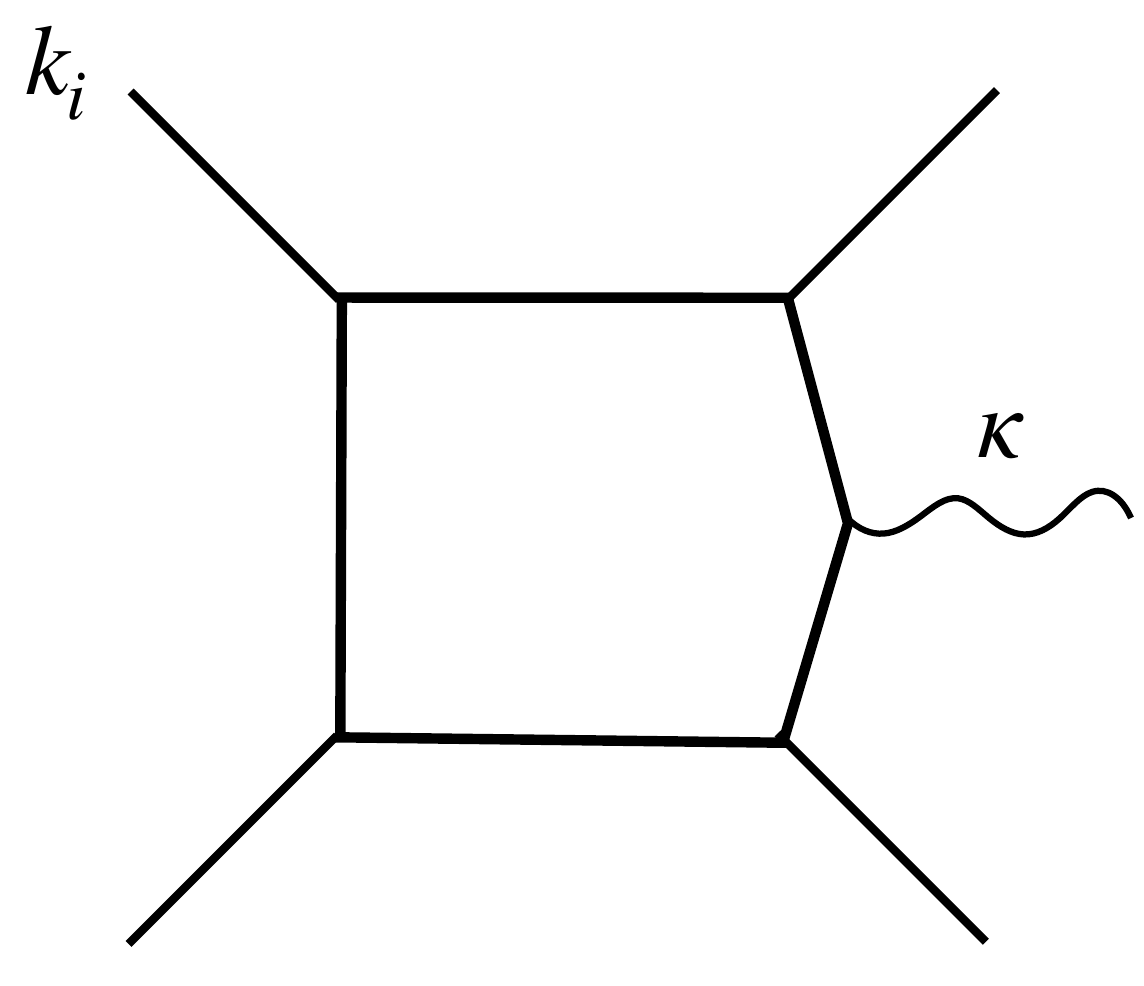}
						\caption{Minahaning a 4-point function means including a 5th massless momentum $\kappa$ and taking $\kappa\rightarrow 0$.  Having $\kappa \neq 0 $ allows 3-index Mandelstam variables $s_{ijk}$ mimicking 5-point kinematics.}
						\label{pentagon}
					\end{center}
				\end{figure}
				
				As in e.g.\ \cite{paper1,paper2}, minahaning the 4-point function is illustrated in figure \ref{pentagon}.  A minahaned $n$-point function satisfies
				\begin{equation} \label{min}
					\sum_{i=1}^n k_i = -\kappa\ ,
				\end{equation}
				where $\kappa$ is the additional lightlike momentum. 
				Minahaning does {\it not} mean to 
				insert additional external states: $\kappa$ is only a deformation of overall momentum conservation. So the amplitude we compute is not an actual pentagon amplitude,
				and figure \ref{pentagon} should only be taken as an aid for memory concerning the role of the 5th momentum $\kappa$.
				(As we shall see in sec.\ \ref{sec:position_int}, it may be related to an {\it octagon} amplitude, but we will not pursue that relation in this work.)
				The point of introducing the additional massless momentum $\kappa$ is to resolve 0/0 indeterminacies. 
				(The original paper \cite{Minahan} had $p=-\kappa$. It will become evident below why this notational change is convenient here.)
				
				\subsection{\texorpdfstring{4-point function: $k_1+k_2+k_3+k_4=-\kappa$}%
					{4-point function: k1+k2+k3+k4 = -kappa}}
				\label{sec:minahaning_4point}
				We distinguish between four-point external momenta, $k_i$ of massless string states $k_i^2=0$, on the one hand, 
				and field theory four-point momenta, $p_i$, on the other hand. The field theory momenta can have virtuality, parametrized as masses $m_i^2=p_i^2 \neq 0$ (that can be complex, in general). 
				
				For the 4-point function, we introduce three-index Mandelstam variables for the external string states, which according to eq.(\ref{Mand.def}) are
				\begin{equation} \label{eq:sijk_sij_sik_sjk}
					s_{ijk} = k_i \cdot k_j + k_i \cdot k_k + k_j \cdot k_k + \half \cancelto{0}{ \left( k_i^2+k_j^2+k_k^2 \right) }= s_{ij}+s_{ik}+s_{jk} \ ,
				\end{equation}
				in terms of the two-index Mandelstam variables
				\begin{equation}
					\label{stringsij}
					s_{ij}=k_i \cdot k_j \ .
				\end{equation}
				As emphasized in figure \ref{pentagon}, the three-index Mandelstam variables $s_{ijk}$ really belong to 5-point ki\-ne\-ma\-tics, but  here we view them as regulators
				of the 4-point function. In particular, it may seem unfamiliar that $s_{ij}+s_{ik}+s_{jk} \neq 0 $ for a massless 4-point function. We now attempt to clarify this.
				
				Square the lightlike 4-point regulator, $\kappa=-k_{1234}$, to obtain
				\begin{equation}
					\cancelto{0}{k_1^2 + k_2^2+k_3^2+k_4^2} + 2k_1 \cdot k_2 +2k_1 \cdot k_3 + 2k_1 \cdot k_4 +2k_2 \cdot k_3 + 2k_2 \cdot k_4 +2k_3 \cdot k_4 = 0 \ ,
				\end{equation}
				which gives the following relation between all possible  three-index Mandelstam variables:
				\begin{equation}  \label{minahaningcond}
					s_{123}+s_{124}+s_{134}+s_{234} = 0 \ .
				\end{equation} 
				This illustrates that minahaning is not an arbitrary regularization: the condition \eqref{minahaningcond} is less restrictive than $s_{ijk}\equiv 0$
				(which would have followed from the traditional $s_{ij}+s_{ik}+s_{jk} = 0 $), but 
				more restrictive than allowing arbitrary regulators $s_{ijk}$. It turns out that \eqref{minahaningcond} is the condition for
				the regularization to respect modular invariance \cite{Minahan}. 
				
				We also introduce \textit{field theory} Mandelstam variables:
				\begin{equation}
					\hat{s}_{ij} \equiv (p_i + p_j)^2 \ .
				\end{equation} 
				They are denoted with a hat to distinguish them from the string theory Mandelstam variables \eqref{stringsij}.

				\section{Field theory limit} \label{sec:FT_limit}
				
				The field theory limit of string amplitudes is computed by examining the S-matrix in the low-energy limit, i.e., $\alpha' \ra 0$, which sends the massive states to infinity, while simultaneously degenerating the genus-one surface to a worldline by sending $\tau_2 \equiv \IM (\tau) \ra \infty$. The limits are taken such that $\alpha' \tau_2$ is kept finite and reduces to the worldline length, $\alpha' \tau_2 \FTa t$, in the Schwinger parametrization of the corresponding field theory diagrams.\footnotemark{} Defining $\nu_i$ by $\IM(z_i)\equiv \tau_2 \, \nu_i$ identifies them with the proper times in the worldline reduction of the original string worldsheet.
				\footnotetext{In order to avoid the proliferation of meanings of the arrow sign, $\FTa$ will henceforth denote field theory limits only.} 
				
				\subsection{Limit ordering and singularities} \label{sec:Limit_ordering}
				
				Following the above prescription, our aim is to explore the behaviour of the integrals in the string amplitudes under the field theory limit. This entails understanding whether we may interchange the limit and the integral; for example, whether
				\begin{equation}	\label{Integral_limit_exchange}
					\lim\limits_{\text{f.t.}} \int \dd \rho \, \JD =  \int \lim\limits_{\text{f.t.}} \, \dd \rho \, \JD = \int \lim\limits_{\text{f.t.}} \dd \rho \, \lim\limits_{\text{f.t.}} \JD
				\end{equation}  
				holds, where $\dd \rho$ and $\JD$ represent the measure in \eqref{cs_measure} and the integrand in \eqref{cs_integrand}, respectively. By the uniform convergence theorem, \eqref{Integral_limit_exchange} holds provided no singularities are present. Therefore, determining the singularities and the regions of moduli space in which they occur, and then resolving them, is the path to performing the desired limit.
				
				As shown in section \ref{sec:skinny_limit}, the only poles within the worldsheet functions present in the string amplitude materialize exclusively for $z \in \Lambda$, and for the representative $z_{ij}=0$, these functions take the forms\footnote{As previously, in order to distinguish between different limits, $\SKa$ indicates a limit of $z\ra \Lambda$, or in some cases the limit to the representative $z \ra 0 \in \Lambda$.}:
				\begin{equation} \label{SK_fone}
					f_{ij}^{(1)}  \SKa \frac{1}{z_{ij}} \ ,
				\end{equation}
				\begin{equation}
					f_{ij}^{(2)}  \SKa 0 \ ,
				\end{equation}
				and
				\begin{equation} \label{SK_KN_closed}
					\Pi_n^{\text{closed}}  \SKa |z_{ij}|^{\alpha' s_{ij}} \, \prod^n_{\substack{1 \leq k < l \\ {kl} \neq {ij}}} e^{s_{kl} \, G_{kl}} \Big |_{z_i = z_j} \ .
				\end{equation}
				Here we used the torus Green's function\footnote{Notice that here we have an opposite overall sign in the Green's function compared to Polchinski \cite{Polchinski1}, eq.(7.2.3). This is due to the fact that we absorbed a factor of $i$ into $X$, cf. footnote \ref{fn:Convention_iX}.}
				\be \label{G_closed}
				G_{ij} \equiv G(z_{ij},\tau)  = \frac{\alpha'}{2} \Big \{ \ln \Big | \frac{\vartheta_1 (z_{ij},\tau) }{\vartheta_1' (0,\tau) } \Big |^2 - \frac{2 \pi}{\IM \, \tau} \IM ^2 (z_{ij}) \Big \} \ ,
				\ee
				and the definition
				\be
				z_{ij} \equiv z_i - z_j \ .
				\ee
				
				With the above arsenal, we can assess once more the amplitude integrand \eqref{cs_integrand}. It is evident that the singularities, and therefore the obstacles to interchanging the limit and the integral, arise only from $X_{ij}$ and $X_{ij,k}$, which contain $f^{(1)}$ and $f^{(1)}f^{(1)}$, respectively, in the region of moduli space where $z_i \lra z_j$, $z_i \lra z_k$ or $z_j \lra z_k$. We shall refer to these limits as collisions since they arise from inserting external legs, i.e., vertex operators, at the same point on the worldsheet.
				The moduli space is consequently split into three classes of regions:
				\begin{enumerate}[(i)]
					\item No collisions -- we can perform the field theory limit on the measure and integrand.
					\item Collisions of two vertices $z_i \lra z_j$ -- possible singularities due to $\frac{1}{|z_{ij}|^2}$ terms. We require further examining of the integral before performing the field theory limit.
					\item Collision of three vertices $z_i \lra z_j \lra z_k$ -- possible singularities due to $\frac{1}{|z_{ij}|^2 |z_{jk}|^2}$ terms, similarly requiring further examining of the integral before performing the field theory limit.
				\end{enumerate}
				
				Let us note that, \emph{a priori}, one can argue that the above singularities appear not only for $z_{ij} = 0$, i.e., $z_i \lra z_j$, but also for $z_{ij}$ on the periodicity lattice of $\tho$, $\Lambda$, i.e., singularities appear for $z_i = z_j + n + m\tau$. However, in order for both $z_i$ and $z_j$ to lie in the fundamental parallelogram of the torus, $\mathcal T(\tau)$, we are constrained to $z_{ij} \rightarrow 0$.
			
				\subsubsection{Collision of two vertices} \label{sec_2col}
				
				In this case we will explore more concretely the relevant behaviour of the four-point one-loop closed-string amplitude integrals -- without loss of generality -- in the region of moduli space $z_2 \lra z_3$. The other single-collision regions, i.e., $z_3 \lra z_4$ and $z_2 \lra z_4$, follow immediately by permutation. (Note 
				that in the integrand \eqref{cs_integrand} there are no collisions between vertex operator number 1 and operators 2,3,4.)\footnote{We
				are not aware of a general reason in string theory that this is always possible: in  \eqref{cs_integrand} it was implemented
				by judicious integration by parts. We believe there is a reason, we just do not know what it is.}
				
				Expanding on the work of \cite{paper2}, we approach the collision $z_2 \lra z_3$ in the closed string. There are two types of terms in the integrand \eqref{cs_integrand} to address: ``diagonal'' combinations $X_{23}\bX_{23}$ and ``non-diagonal'' combinations $X_{23}\bX_{kl}$ with $\{k,l\} \neq \{2,3\}$. The former, along with the Koba-Nielsen factor, according to \eqref{SK_fone} and \eqref{SK_KN_closed}, reduce to
				\begin{equation}
					X_{23}\bX_{23} \, \Pi_4 \SKa \frac{s_{23}^2}{|z_{23}|^{2-\alpha' s_{23}}}\  \tilde \Pi_4^{23} \ ,
				\end{equation}
				where we defined
				\begin{equation}
					\tilde \Pi_4^{ij} \equiv \prod^4_{\substack{1 \leq k < l \\ kl \neq ij}} e^{s_{kl} \, G_{kl}} \Big |_{z_i = z_j} \ .
				\end{equation}
				Consequently, using identity \eqref{col_id3}, the potentially kinematically singular terms in the amplitude become
				\begin{equation} \label{cs_2col_SK}
					\int \dd^2 z_2 X_{23}\bX_{23} \, \Pi_4 \SKa \frac{4\pi}{\alpha'} \, s_{23} \ \tilde \Pi_4^{23} \Big |_{z_2 = z_3} \ ,
				\end{equation}
				in the $\alpha'$ expansion, i.e.\ there is actually no obstacle to take the field-theory limit after employing \eqref{cs_2col_SK}. The latter, non-diagonal combinations reduce, along with the Koba-Nielsen factor, to $|z_{23}|^{\alpha' s_{23}} z_{23}^{-1}$ which do not contribute to the limit we are tackling due to angular integration. 
				
				\subsubsection{Collision of three vertices} \label{sec_3col}
				
				Collisions of three insertions on the torus manifest exclusively in the region of moduli space $z_2\lra z_3\lra z_4$. Again there are two types of terms in \eqref{cs_integrand}: ``diagonal" combinations $X_{23,4}\bX_{23,4}$ and ``non-diagonal" combinations $X_{23,4}\bX_{24,3}$. Compared to the collision of two vertices, there is a subtlety to take into account, based on section 3.3 in \cite{OchirovTourkine}: the occurrence of simultaneous limits points to possible issues with limit ordering once again. The solution is to split the moduli space further into successive-limit regions, i.e., regions in which two vertices collide and then the remaining vertex collides with the former cluster. For example, a $z_3 \ra z_2$ collision succeeded by $z_4 \ra z_2=z_3$, which we denote by $|z_{23}| \ll |z_{24}| \sim |z_{34}| \ll 1$.  
				
				In contrast to the treatment of single collisions, the effect of one of the ordered successive limits discussed above does not transfer trivially to the remaining cases and requires nuanced modifications. For bookkeeping, let us note the three cases: 
				\begin{enumerate}[(i)]
					\item $|z_{23}| \ll |z_{24}| \sim |z_{34}| \ll 1$, 
					\item $|z_{24}| \ll |z_{23}| \sim |z_{43}| \ll 1$, and 
					\item $|z_{34}| \ll |z_{32}| \sim |z_{42}| \ll 1$.
				\end{enumerate}
				Notice that regions (ii) and (iii) are obtained from (i) by permuting $(3 \lra 4)$ and cyclically $(2,3,4)$, respectively. 
				
\begin{figure}[htbp]
					\centering
					\begin{subfigure}[b]{0.25\textwidth}
						\centering
						\includegraphics[width=\textwidth]{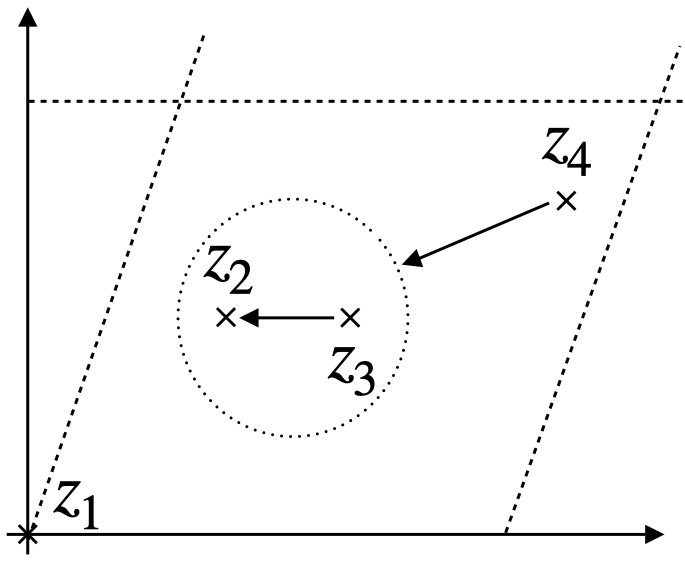}
						\caption{}
						\label{subfig:3VC1}
					\end{subfigure}
					\hfill
					\begin{subfigure}[b]{0.5\textwidth}
						\centering
						\includegraphics[width=\textwidth]{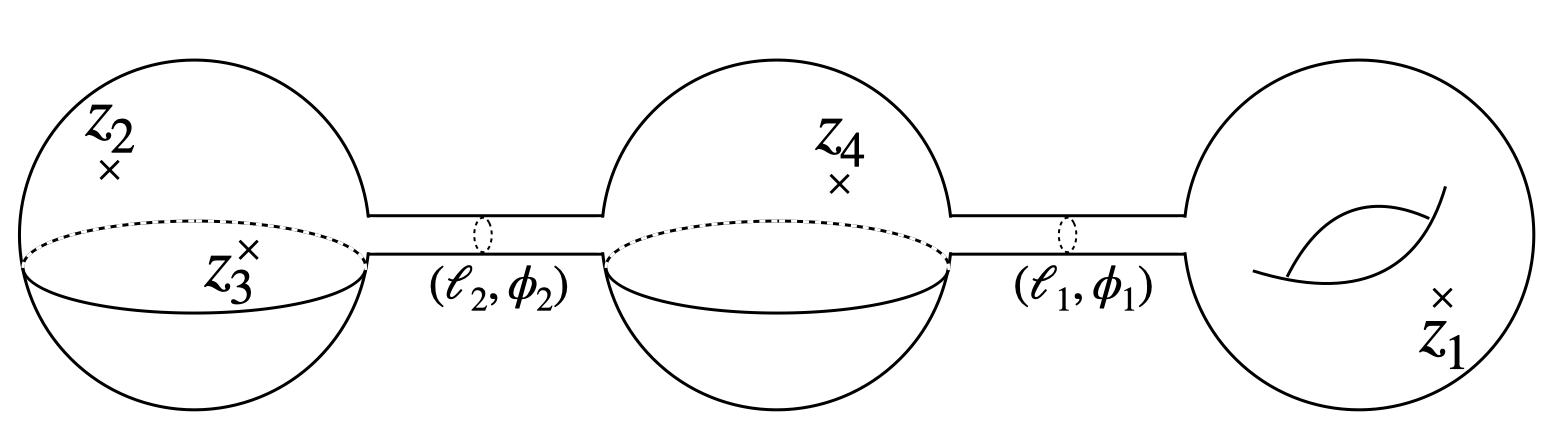}
						\caption{}
						\label{subfig:i3VC2}
					\end{subfigure}
					\hfill
					\begin{subfigure}[b]{0.3\textwidth}
						\centering
						\includegraphics[width=\textwidth]{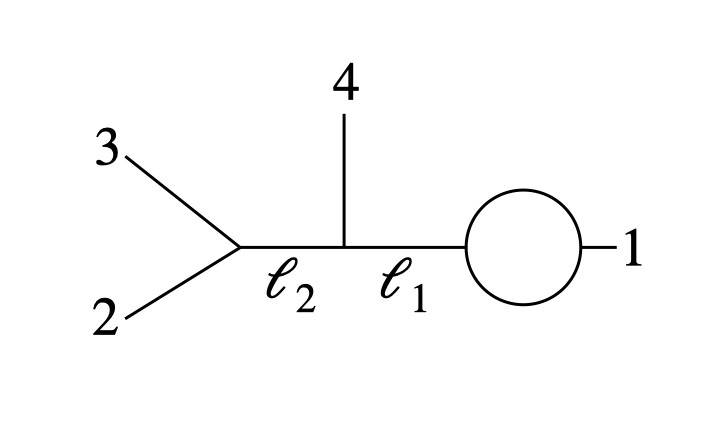}
						\caption{}
						\label{subfig:3VC3}
					\end{subfigure}
					\caption{Three-vertex collision in the four-point one-loop low-energy limit, adapted from figure 3 in \cite{OchirovTourkine}. From left to right: (A) the complex plane, (B) the worldsheet and (C) the worldline graph.}
					\label{fig:three_vertex_col}
				\end{figure}

				In these regions (following \cite{OchirovTourkine}), the original closed-string toroidal worldsheet reduces to spheres connected by thin, long tubes; see figure \ref{fig:three_vertex_col} (\subref{subfig:i3VC2}) for a depiction of case (i). One therefore has to integrate out the angular coordinates along the tubes to obtain graph lines; see figure \ref{fig:three_vertex_col} (\subref{subfig:3VC3}). 
				
				We parametrise the tube connecting $z_4$ to the fixed vertex operator $z_1$ by the phase $\phi_1$ of the cylindrical coordinate and the length parameter $\ell_1$. The larger $\ell_1$ gets (greater than some UV cutoff of order $\alpha'$), the shorter the distance of $z_4$ to the $z_2 \lra z_3$ cluster relative to $1$.\footnote{This is a tropical parametrization. For a broader scope see \cite{TropicalTourkine}.} Hence,
				\begin{equation} \label{3Col_cs_parameterisation1}
					z_{24} \sim z_{34} \SKa e^{i\phi_1} \, e^{-\ell_1/\alpha'} \ .
				\end{equation}
				Similarly, the tube connecting $z_4$ to the cluster of $z_2 \lra z_3$ is parametrised by the phase $\phi_2$ and length $\ell_2$, and the distance $z_{23}$ decreases with larger $\ell_2$ (with the same caveat) relative to \eqref{3Col_cs_parameterisation1}:
				\begin{equation} \label{3Col_cs_parameterisation2}
					z_{23} \SKa z_{24} \, e^{i\phi_2} \, e^{-\ell_2/\alpha'} = e^{i(\phi_1 + \phi_2)} \, e^{-(\ell_1+\ell_2)/\alpha'} \ .
				\end{equation}
				The influence on the measure is then computed by means of the Jacobian:
				\begin{equation}
					\dd^2 z_{24} \SKa i \begin{vmatrix} -\frac{1}{\alpha'} e^{i\phi_1}  e^{-\ell_1/\alpha'}  & i e^{i\phi_1} e^{-\ell_1/\alpha'}  \\ -\frac{1}{\alpha'} e^{-i\phi_1}  e^{-\ell_1/\alpha'} & -i e^{-i\phi_1} e^{-\ell_1/\alpha'}  \end{vmatrix} \dd \ell_1 \dd \phi_1 = -\frac{2}{\alpha'}e^{-2\ell_1/\alpha'} \,  \dd \ell_1 \dd \phi_1 \ ,
				\end{equation}
				and by the same procedure
				\begin{equation}
					\dd^2 z_{23} \SKa -\frac{2}{\alpha'}e^{-2(\ell_1+\ell_2)/\alpha'} \,  \dd \ell_2 \dd \phi_2 \ .
				\end{equation}
				Next, we examine the influence of this limit on the Koba-Nielsen factor \eqref{KNdef}, taking into account that for small $z$, the Green's function behaves as in \eqref{G_SK}:
				\begin{equation} \begin{aligned} \label{3Col_cs_KN}
						\Pi_4 & \SKa \exp \Big \{ \alpha' s_{23}\, \ln |z_{23}| + \alpha' s_{24}\, \ln |z_{24}| + \alpha' s_{34}\, \ln |z_{34}| \Big \} \, e^{(\sum_{j=2}^4 s_{1j}G_{1j})} \\[5pt]
						& = \exp \Big \{ -(\ell_1+\ell_2) s_{23} - \ell_1 s_{24} - \ell_1 s_{34} \Big \} \, e^{(\sum_{j=2}^4 s_{1j}G_{1j})} \\[5pt]
						& = e^{-\ell_1 s_{234} - \ell_2 s_{23}}\, e^{(\sum_{j=2}^4 s_{1j}G_{1j})} \ .
				\end{aligned} \end{equation} 
				
				Combining the above, the potential singularities of the closed-string amplitude in a three-vertex collision have the form:
				\begin{equation} \label{OchT_integral}
					\frac{4 s_{ij}s_{kl}s_{mn}s_{pq}}{ \alpha'^2} \int \dd \ell_1 \dd \ell_2 \dd \phi_1 \dd \phi_2 \frac{e^{-2\ell_1/\alpha'}  \, e^{-2(\ell_1+\ell_2)/\alpha'}}{z_{ij}z_{kl}\bar z_{mn} \bar z_{pq}} \, e^{-\ell_1 s_{234} - \ell_2 s_{23}} \ ,
				\end{equation}
				similar to eq. (3.22) in \cite{OchirovTourkine}, where a factor of $\exp \{ \sum_{j=2}^4 s_{1j}G_{1j}\}$ was omitted as it does not affect the current calculations. A quartet of $\frac{s_{ij}}{z_{ij}}$ appears due to the quartet of $X_{ij}$ (from $X_{ij,k} \bar X_{ij,k}$ or $X_{ij,k} \bar X_{ik,j}$) in the integrand \eqref{cs_integrand}, where two collisions materialize. 
				
				It is clear that for \eqref{OchT_integral} not to integrate to zero, the phases $\phi_1$ and $\phi_2$ should cancel in the de\-no\-mi\-na\-tor quartet. Moreover, the denominator is required to cancel the numerator that stemmed from the Jacobian; otherwise the integration over $\ell_1$ or $\ell_2$ would yield, at least, a factor of $\alpha'$, thereby removing this contribution from the leading order of the $\alpha'$ expansion. Consequently, the only contributing quartet combinations are:
				\begin{equation} \label{3col_contribution1}
					|X_{23}|^2 \ \text{combined with} \ \ |X_{24}|^2 \, , \, |X_{34}|^2 \, , \, X_{24}\bX_{34} \, \text{or} \, X_{34}\bX_{24} \ .
				\end{equation}
				
				Plugging into \eqref{OchT_integral} any of these combinations produces\footnote{\label{3Col_lim_note} There is a slight subtlety in this equation. We neglect here a factor of $\exp \{ \sum_{j=2}^4 s_{1j}G_{1j}\}$. Consequently, this equality should be understood as producing the first order of the $\alpha'$ expansion, and the omitted factor becomes $\exp \{ \sum_{j=2}^4 s_{1j}G_{1j}\} \Big |_{z_2=z_3=z_4}$ in the skinny limit.}
				\begin{equation} \begin{aligned} \label{3Col_lim}  
						\frac{4 s_{23}^2 s_{k4}s_{l4}}{ \alpha'^2} \int \dd \ell_1 \dd \ell_2 \dd \phi_1 \dd \phi_2  \, e^{-\ell_1 s_{234} - \ell_2 s_{23}} &= \frac{4 s_{23}^2 s_{k4}s_{l4}}{ \alpha'^2} \frac{4\pi^2}{s_{23}s_{234}} + \ldots\\
						& = \frac{16 \pi^2}{\alpha'^2} \frac{s_{23} s_{k4}s_{l4}}{s_{234}} + \ldots\ ,
				\end{aligned} \end{equation} 
				for $k,l \in \{2,3\}$. A potential $s_{234}^{-1}$ pole manifests in the integration of one quartet element for each of the three regions of moduli-space splitting (cases (i)--(iii)). Our saving grace is that we ought to sum the skinny limit in eq. \eqref{3Col_lim} for all three regions and for the combinations actually appearing in integrand \eqref{cs_integrand}. 
				
				The skinny limit of the $X\bX$ combinations affected by the collision of three vertices in the closed string is:
				\begin{equation} \begin{aligned} \label{cs_3col_SK}
						\int \dd^2 z_2 \,  \dd^2 z_3 \,  \dd^2 z_4\ X_{23,4}\bX_{23,4} \, \Pi_4 & \SKa \frac{16 \pi^2}{\alpha'^2} \, s_{23}(s_{24}+s_{34}) \, \int \dd^2 z_4 \, \Pi_4 \Big |_{z_2 = z_3=z_4}  \ , \\ &~ \\
						\int \dd^2 z_2 \,  \dd^2 z_3 \,  \dd^2 z_4 \ X_{24,3}\bX_{24,3} \, \Pi_4 & \SKa \frac{16 \pi^2}{\alpha'^2} \, s_{24}(s_{23}+s_{34}) \,  \int \dd^2 z_4 \, \Pi_4 \Big |_{z_2 = z_3=z_4}  \ , \\ \text{and} \\
						\int \dd^2 z_2 \,  \dd^2 z_3 \,  \dd^2 z_4 \ X_{23,4}\bX_{24,3} \, \Pi_4 & \SKa \frac{16 \pi^2}{\alpha'^2} \, s_{23}s_{24} \,  \int \dd^2 z_4 \, \Pi_4 \Big |_{z_2 = z_3=z_4}  \ , 
				\end{aligned} \end{equation} 
				where $\Pi_4 \Big |_{z_2 = z_3=z_4} =\exp \{ \sum_{j=2}^4 s_{1j}G_{1j}\}$ and the detaild calculations are in appendix \ref{sec:skinny_limit}.
				
				This shows that, under the above treatment, no kinematic singularities of order $s_{234}^{-1}$ (or worse), stemming from the pole of $f^{(1)}$, prevent us from proceeding to take the field theory limit of the closed string. 
				
				We are now fully equipped to tackle the field theory limit without fearing kinematic poles. 
				
				\subsection{Prescription for field theory limit application} \label{sec:Limit_components} 
				
				The strategy we deploy in determining the field theory limit is dictated by the above discussion. First, we determine the limit of the components of the measure and integrand, e.g., $G_{ij}$, $f^{(1)}$, etc. Second, we split the integration region of the moduli space of the theory into regions of no vertex collisions, two-vertex collisions and three-vertex collisions. For each of these regions, the limit of the integration measure is then calculated.\footnote{Notice that in the collision regions, the integrand is also affected since a combination of $X_{ij}$ elements is integrated out; see sections \ref{sec_2col} and \ref{sec_3col}. Nevertheless, the integrand is simply modified by dropping the involved $X_{ij}$ elements.} These steps, once implemented, provide us with the desired limit in Schwinger representation. 
				
				\subsubsection{Limit of integrand components} \label{sec:FT_limit_components}
				
				Following the detailed calculations in appendix \ref{sec:WL_functions_limit}, the field theory limits of the worldsheet functions are
				\begin{equation} \begin{aligned} \label{WS_FTlimit}
						f^{(1)}(z, \tau) &\FTa 2\pi i \Big [ \nu - \frac{1}{2} \, \text{sgn}(\nu) \Big ]\ , \\
						f^{(2)}(z, \tau) &\FTa (2\pi i)^2 \Big [ \half \nu^2 - \half \nu \, \text{sgn}(\nu)  + \frac{1}{12} \Big ] \ ,\\
						F_{1/2}^{(2)} (kv,\tau)  &\FTa \pi^2 \Big [ \frac{1}{3} - \ \frac{1}{\sin^2(\pi k v)} \Big ] \ , \\
						G(z,\tau) &\FTa - \pi t (\nu^2 - |\nu|)\ ,
				\end{aligned} \end{equation} 
				in agreement with eqs. (6.15) and (6.16) in \cite{paper2}. Subsequently, the Koba-Nielsen factor, defined in \eqref{KNdef}, reduces to
				\begin{equation} \label{KN_FT_lim}
					\Pi_n \FTa e^{-  \pi t Q_n(k_1, k_2, \dots, k_n)} \ , 
				\end{equation}
				where, generally,
				\begin{equation} \label{Qm_FT_lim}
					Q_m(k_{A_1}, k_{A_2}, \dots, k_{A_m})\equiv \sum_{1 \leq i < j}^m (k_{A_i} \cdot k_{A_j}) \Big (\nu_{ij}^2 - |\nu_{ij}| \Big ) \ ,
				\end{equation}
				and $A_i$ is a multi-index as in \eqref{Mand.def}. 
				
				Additionally, the terms in \eqref{cs_integrand} that stem from the interaction of zero-mode left- and right-movers have a straightforward limit:
				\begin{equation} \label{ZM_FT_limit}
					\frac{2\, \pi}{\alpha' \, \IM(\tau)} \FTa \frac{2\,\pi}{t} \ .
				\end{equation}
				
				Finallly the lattice sum $\Gamma$ over winding modes and compactified momenta becomes trivial in the field-theory limit and we set it to $1$ in the limit. 

				\subsubsection{Limit of integration measure} 
				
				Starting with the integration measures of the closed-string theory, eq. \eqref{cs_measure}, we shall calculate their field theory limits.
				
				\begin{itemize}
					\item No vertex collisions:
					
					In this case no poles are present, and we determine the limit of the original integration measure:\footnote{\label{double_int_conventions} For the factors of $2$ in the second line, note that we use the conventions $\int \dd^2z = i\int \dd z \dd \bar{z} =2 \int \dd(\IM z) \, \dd(\RE z)$, as in  \cite{Polchinski1, BLT}.}
					
					\begin{equation} \begin{aligned} \label{cs_measure_FT_lim_s1}
							\int \dd \rho^{D}{_{12...n}} & = \frac{V{_D}}{8N(4\pi^2 )^{D/2}} \int_{\mathcal F} \frac{\alpha' \, \dd^2 \tau}{(\alpha ' \tau_2)^{D/2}} \, (\alpha' \tau_2)^{n-1} \, \int_{\mathcal T (\tau)^n}  \prod_{i=2}^n \frac{\dd^2 z_i}{\tau_2}  \Pi_n \Big |_{z_1=\bar z_1 = 0} \\
							& = \frac{ 2^{n} \, V{_D}}{8N(4\pi^2 )^{D/2}} \int_0^{\infty} \frac{\alpha' \, \dd \tau_2}{( \alpha ' \tau_2)^{D/2 - n + 1}} \int_0^{\tau_2} 
							\prod_{j=2}^n \frac{\dd \, \IM(z_j)}{\tau_2}  \Pi_n \Big |_{z_1=\bar z_1 = 0} \ ,
					\end{aligned} \end{equation} 
					where in the last step the integrations over $\RE(\tau)$ and $\RE(z_j)$ were performed in the field-theory limit. These integrations are trivial as the integrand does not depend on the real parts at all in this limit, cf.\ sec.\ \ref{sec:FT_limit_components}. Therefore, in the field-theory limit the measure becomes
					\begin{equation} \begin{aligned} \label{cs_measure_FT_lim}
							\int \dd \rho^{D}{_{12...n}} \FTa  \frac{2^{n+D/2} \,V{_D}}{8N(8\pi^2 )^{D/2}}  \int_{0}^{\infty} \frac{\dd t}{t} \, t^{n-D/2}\int_0^1 
							\dd\nu_2 \dots \dd\nu_n \;  e^{-\pi t Q_n[k_1,\dots, k_n]} \Big |_{\nu_1 = 0} \ .
					\end{aligned} \end{equation} 
					
					\item Collision of two vertices
					
					Here we need to perform the skinny limit on the components of the integrands that admit poles in this one-collision region of moduli space and then evaluate the field theory limit of the modified measure. As a representative, we take the region of moduli space in which $z_2 \lra z_3$ collide:
					\begin{equation} \begin{aligned} \label{cs_measureLimit_2col}
							\int & \dd \rho^{D}{_{12...n}} \ X_{23} \bX_{23} \\
							& \SKa \frac{4 \pi V_{D}}{8N(4\pi^2 )^{D/2}} \, s_{23} \, \int_{\mathcal F} \frac{\alpha' \, \dd^2 \tau}{(\alpha ' \tau_2)^{D/2}} \, (\alpha' \tau_2)^{n-2} \, \int_{\mathcal T (\tau)^n}  \prod_{i=3}^n \frac{\dd^2 z_i}{\tau_2}  \tilde \Pi_n^{23} \Big |_{z_1=\bar z_1 = 0} \\
							& \FTa  \frac{2^{n+1+D/2} \, \pi \,V_{D}}{8N(8\pi^2 )^{D/2}} \, s_{23} \,  \int_{0}^{\infty} \frac{\dd t}{t} \, t^{n-1-D/2}\int_0^1 
							\dd\nu_3 \dots \dd\nu_n \;  e^{-\pi t Q_n[k_1,k_{23},\dots, k_n]} \Big |_{\substack{\nu_1=0 \\ \nu_2=\nu_3}} \ .
					\end{aligned} \end{equation} 
					Notice that in the equation above, the integration over the worldline parameter, $t$, is reduced by one in the degree of the polynomial of said parameter compared to the case when no collisions occur -  $t^{n-1-D/2}$ compared to $t^{n-D/2}$ in eq.\eqref{cs_measure_FT_lim}. This impacts the structure of the worldline integral as seen in eq.\eqref{eq:WL_integral_Ddim}.
					
					\item Collision of three vertices
					
					We apply the same procedure as in the two-vertex collision case to the sole three-vertex collision that appears in integrand \eqref{cs_integrand}, namely $z_2 \lra z_3 \lra z_4$:
				\begin{equation} \begin{aligned} \label{cs_measureLimit_3col}
						\int & \dd \rho^{D}{_{12...n}} \ X_{23,4}\bX_{23,4} \\
						& \SKa \frac{(4 \pi)^2 V_{D}}{8N(4\pi^2 )^{D/2}} \, s_{23}(s_{24}+s_{34}) \,\int_{\mathcal F} \frac{\alpha' \, \dd^2 \tau}{(\alpha ' \tau_2)^{D/2}} \, (\alpha' \tau_2)^{n-3} \, \int_{\mathcal T (\tau)^n}  \prod_{i=4}^n \frac{\dd^2 z_i}{\tau_2}  \tilde \Pi_n^{234} \Big |_{z_1=\bar z_1 = 0} \\
						& \FTa  \frac{2^{n+2+D/2} \, \pi^2\,V_{D}}{8N(8\pi^2 )^{D/2}} \, s_{23}(s_{24}+s_{34}) \ \int_{0}^{\infty} \frac{\dd t}{t} \, t^{n-2-D/2} \ \times \\ 
						&\qquad \qquad \qquad \qquad \qquad \qquad \int_0^1 \dd\nu_4 \dots \dd\nu_n \;  e^{-\pi t Q_n[k_1,k_{234},\dots, k_n]} \Big |_{\substack{\nu_1=0 \\ \nu_2=\nu_3 = \nu_4}} \ .
				\end{aligned} \end{equation} 
					As in the collision of two vertices, in this case the polynomial degree of the wordline parameter, $t$, is reduced again, in this case by order of two with respect to the limit of the integration measure in regions where no collisions appear, eq.\eqref{cs_measure_FT_lim}.
				\end{itemize}

				\subsection{Field theory limit of the closed string} \label{sec:FT_limit_integrand}
				
				In this section we start taking the field-theory limit of the closed-string four-point one-loop amplitude at half-maximal supersymmetry,
				\begin{equation} \label{eq:our_amplitude}
					\mathcal M _{1/2}(1,2,3,4) = \int \dd \rho^{D}_{1234} \  \Bigg \{ \Gamma \mathcal J _{4,\text{max}} + \sum_{k,k'=1}^{N-1} \hat{\chi}_{k,k'} \mathcal J_{4,1/2} (v_{k,k'}) \Bigg \}
				\end{equation}
				specifying \eqref{cs_amplitude} to $n=4$. The integrands are given in eqs.~\eqref{J4-max} and \eqref{cs_integrand}.
				
				The strategy is to apply the limit prescription of section \ref{sec:Limit_components} to the three types of moduli-space regions discussed in section \ref{sec:Limit_ordering}. In order to organize the following formulas, we split the integrand, eq. \eqref{cs_integrand}, according to
				\begin{equation}
					 \mathcal J_{4,1/2} = A(\nu_i;\tau_2,s_{ij})+ \frac{2\pi}{\alpha' \tau_2} B(\nu_i;\tau_2,s_{ij})+\left(\frac{2\pi}{\alpha' \tau_2}\right )^2 C(s_{ij}) \ ,
				\end{equation}
				where
				\begin{equation}
					\begin{aligned} \label{cs_integrand_lines}
						A &=  \Big | X_{23,4}C_{1|234} + X_{24,3}C_{1|243} + [s_{12}(f_{12}^{(2)} + f_{34}^{(2)})P_{1|2|3,4} + (2 \leftrightarrow 3,4 )] -2 F_{1/2}^{(2)}(\gamma) t_8 (1,2,3,4)  \Big |^2  \ , \\
						B &= (X_{23}C_{1|23,4}^m + X_{24}C_{1|24,3}^m + X_{34}C_{1|34,2}^m ) (\bX_{23} \tC_{1|23,4} ^m  + \bX_{24}\tC_{1|24,3}^m + \bX_{34}\tC_{1|34,2}^m)  \\
						&\text{and} \\
						C &= (\frac{1}{2} C_{1|2,3,4}^{mn} \tC_{1|2,3,4}^{mn} - P_{1|2|3,4}\tP_{1|2|3,4} - P_{1|3|2,4}\tP_{1|3|2,4} - P_{1|4|2,3}\tP_{1|4|2,3}) \ .
					\end{aligned}
				\end{equation}

				\subsubsection{ No vertex collision -- box}
				
				The contribution to the amplitude from integration regions that do not contain vertex collisions are box-diagram-like:
				\begin{equation} \begin{aligned} \label{eq:box_integral_full}
						\mathcal M _{1/2} & (1,2,3,4  ) \Big |_{\text{no coll.} } \ra \FTM_0 = 2^{4+D/2} \int_{0}^{\infty} \frac{\dd t}{t} \, t^{4-D/2}\int_0^1 
						\dd\nu_2 \dots \dd\nu_4 \;  e^{-\pi t Q_4[k_1,\dots, k_4]}  \times \\
						& \qquad \Bigg \{ 4\,t_8 \tilde t_8 + \sum_{k,k'=1}^{N-1} \hat{\chi}_{k,k'} \, \JD_0 \Bigg \} \Bigg |_{\nu_1 = 0} \\
						& = 2^{4+D/2}  \, \int_{0}^{\infty} \frac{\dd t}{t} \, t^{4-D/2} \int_0^1 \dd^3 \nu \;  e^{-\pi t Q_4[k_1,\dots, k_4]}  \times \\
						& \qquad \Bigg \{ 4\,t_8 \tilde t_8 + \sum_{k,k'=1}^{N-1} \hat{\chi}_{k,k'} \,  [ A_0(s,\nu) +  \frac{2\pi}{t} \, B_0 (s,\nu) +  \frac{(2\pi)^2}{t^2} \, C_0(s)  ] \Bigg \}  \Bigg |_{\nu_1 = 0}  \ ,
				\end{aligned} \end{equation} 
				where $A_0, B_0$ and $C_0$ delineate the classifications introduced in eq. \eqref{cs_integrand_lines} after performing the field theory limit of their components; see eq. \eqref{WS_FTlimit}. We will spare the reader from the full expressions here; they are available in the accompanying code \href{https://github.com/YonatanZimmerman/Field-theory-limit-of-half-maximal-string-loop-amplitudes/blob/main/Assembly%20and%20Results.nb}{\faGithub}. Nevertheless, note that $C$ does not depend on any of the worldsheet functions and therefore is not affected by the field-theory limit, i.e.\ $C_0=C$. Factors of $\frac{V_{D}}{8N(8\pi)^{D/2}}$ are not carried through the limits in \eqref{eq:box_integral_full} (and in the following) since they are an overall prefix for every amplitude.
				
				\subsubsection{ Two-vertex collision -- triangles}
				
				Let us look at one representative collision -- $z_2 \lra z_3$ -- for which the triangle integral is:
				\begin{equation} \begin{aligned} \label{eq:triangle_23_integral_full}
						\mathcal M _{1/2}(1,2,3,4 & ) \Big |_{z_2 \lra z_3 } \ra \FTM_{2|23} = 2^{5+D/2} \, \pi \, s_{23} \,\sum_{k,k'=1}^{N-1} \hat{\chi}_{k,k'}  \, \int_0^{\infty} \frac{\dd t}{t} \, t^{3-D/2}  \int_0^{1} \ \dd \nu_3 \, \dd \nu_4 \ \times \\
						&  \Bigg \{ A_{2|23} + \frac{2\pi}{t} \, B_{2|23} \Bigg \} \, e^{- \pi t Q_3[k_1,k_{23},k_4]} \Bigg |_{\nu_2=\nu_3} \ ,
				\end{aligned} \end{equation} 
				where, as previously, $A_{2|ij}$ and $B_{2|ij}$ represent the tiered classifications -- now with collision of vertices $i$ and $j$. As these terms are not very long in this case, we can spell them out:
				\begin{equation}
					B_{2|23} = \Big | C_{1|23,4}^m \Big |^2 \ ,
				\end{equation}
				and
				\begin{equation} \label{eq:A2|23}
					\begin{aligned}
						A_{2|23} &=(2 \pi)^2 \Big | 
						{\textstyle
							\left[ s_{24} \left( \nu_{24}- \frac{1}{2}\,\text{sgn}(\nu_{24})\right)
							+s_{34} \left( \nu_{34}- \frac{1}{2}\,\text{sgn}(\nu_{34}) \right) \right]
							C_{1|234}
							+s_{24} \left( \nu_{24}- \frac{1}{2}\,\text{sgn}(\nu_{24})\right)
							C_{1|243}
						}
						\Big |^2 
						\\
						&=  (2 \pi)^2  \, s^2_{234}
						\Big | {\textstyle \left(  \nu_{24}- \frac{1}{2}\,\text{sgn}(\nu_{24}) \right)
							C_{1|243} }  \Big |^2 \ ,
					\end{aligned}
				\end{equation}
				where in the last step we used the identities in appendix \ref{sec:BCJ_ids} to simplify the expression. Interestingly, we notice that if the integral of $A_{2|23}$ diverges slower than $s_{ijk}^{-2}$, reinstating momentum conservation, i.e., $s_{ijk} \ra 0$, will mean that there is no contribution to the amplitude from this part. Similarly, there will be no contribution from triangle $A_2$'s from the other two vertex collisions:
				\begin{equation}\label{eq:A2|24}
					A_{2|24} = (2 \pi)^2  \frac{s_{23}}{s_{34}}\, s^2_{234} \Big | {\textstyle \left( \nu_{23}- \frac{1}{2} \, \text{sgn}(\nu_{23} )\right)C_{1|243}  } \Big |^2 \ ,
				\end{equation}
				and, more drastically,
				\begin{equation}\label{eq:A2|34}
					A_{2|34} = 0 \ ,
				\end{equation}
				is identically zero. Obviously, the permutation symmetry is not manifest in this result, even though it is present in the full amplitude. This is due to the fact that permutation invariance in $A$ of \eqref{cs_integrand_lines} is implemented on just two terms, cf.\ the discussion in appendix \ref{app:bg}, below eq.\ \eqref{BG21var}. Consequently, $A_{2|23}$ and $A_{2|24}$ also vanish on setting the regulator to zero, i.e.\ $s_{234} \ra 0$.
				
				\subsubsection{ Three-vertex collision -- bubble}
				
				Only one collision, $z_2 \lra z_3 \lra z_4$, contributes to the bubble integrals:
				\begin{equation} \label{eq:bubble_integral_full}
					\mathcal M _{1/2} \Big |_{z_2 \lra z_3 \lra z_4} \ra \FTM_3=2^{6+D/2} \, \pi^2 \, \sum_{k,k'=1}^{N-1} \hat{\chi}_{k,k'}  \, \int_0^{\infty} \frac{\dd t}{t} \, t^{2-D/2}  \int_0^{1} \ \dd \nu_4 \ A_3 \, e^{- \pi t Q_2[k_1,k_{234}]} \Bigg |_{\nu_2=\nu_3=\nu_4}\ .
				\end{equation} 
				with
				\begin{equation} \label{eq:bubble_numerator}
					\begin{aligned}
						A_{3} &= s_{23}(s_{24}+s_{34})\, |C_{1|234}|^2 + s_{24}(s_{23}+s_{34})\, |C_{1|243}|^2 + s_{23}s_{24} \Big [ C_{1|234} \tilde C_{1|243} + C_{1|243} \tilde C_{1|234} \Big ] \\
						&=  \frac{s_{23}(s_{23}+s_{24})}{s_{24}}  \, s_{234} \, |C_{1|234}|^2 \ ,
				\end{aligned} \end{equation}
				where in the last step we again used the identities in appendix \ref{sec:BCJ_ids} to simplify the expression. 
				
				At this point we have the field theory limit of our amplitude in integral form. \emph{All} that is left to do is perform those integrals. 
	
				\section{Worldline integration} \label{sec:WL_int}
				There are of course many books on Feynman integrals, but we found \cite{Weinzierl:2022eaz} particularly useful for some of the following analysis.
				First one comment when comparing to standard works: a commonly used regularization is to allow
				arbitrary exponents in the denominator. Here we let those exponents keep their original values,
				to keep separate the various types of diagrams. 
				
				We start the integral computation by performing the worldline integration, in the spirit of \cite{Tourkine:2012vx}. As can be seen in the previous section, there are three cases of $t$ integrals we need to consider:
				
				\begin{equation} \label{eq:WL_integral_Ddim}
					\begin{aligned}
						\int_{0}^{\infty} & \frac{\dd t}{t}  \, t^{4-D/2} \,  \Big [ \Omega_1 + \frac{\Omega_2}{t} + \frac{\Omega_3}{t^2} \Big ]  \;  e^{-\pi t Q_n} \\
						& =  \pi^{\frac{D}{2}-4} \, \Gamma\left(4-\frac{D}{2}\right) \, Q_n^{\frac{D}{2}-4} \, \Omega_1 + \pi^{\frac{D}{2}-3} \, \Gamma\left(3-\frac{D}{2}\right) \, Q_n^{\frac{D}{2}-3} \, \Omega_2 + \pi^{\frac{D}{2}-2} \, \Gamma\left(2-\frac{D}{2}\right) \, Q_n^{\frac{D}{2}-2} \, \Omega_3 \ ,
					\end{aligned}
				\end{equation}
				where $n=2,3,4$.
				
				From this point onwards, we specify our calculations to $D \ra 4-2\, \euv$ dimensions, omitting the dimensional regulator when terms are UV finite. The above equation is reduced to
				\begin{equation} \label{eq:WL_integral}
						\int_{0}^{\infty}  \frac{\dd t}{t} \, t^{2+\euv} \,  \Big [ \Omega_1 + \frac{\Omega_2}{t} + \frac{\Omega_3}{t^2} \Big ]  \;  e^{-\pi t Q_n}
						\ra \pi^{-2} \, Q_n^{-2} \, \Omega_1 +  \pi^{-1} \, Q_n^{-1} \, \Omega_2 + 
						\Gamma(\euv) \, \pi^{-\euv} \, Q_n^{-\euv} \, \Omega_3 \ .
				\end{equation}
				
				Let us comment on the validity of equation \eqref{eq:WL_integral_Ddim}. If we had demanded
				$s_{ij} + s_{ik}  + s_{jk} = 0$  there
				would have been no choice of real kinematic variables $s_{ij}$ for which Koba-Nielsen factor in the integral \eqref{eq:WL_integral_Ddim} damps the integral for each ordering of the $\nu_i$. This problem is discussed for example in \cite{DHoker:1993hvl}  for the
				one-loop 4-graviton amplitude. Here, the infrared regularization we use is to move $s_{ij} + s_{ik}  + s_{jk}$ away from zero
				using the $s_{ijk}$. Then,
				the integral can be made to converge for sufficiently large $|s_{ijk}|$, as in \eqref{eq:sijk_sij_sik_sjk}.
				As usual in analytic continuation,
				we will then in fact expand it in the opposite limit $|s_{ijk}| \ll |s_{ij}|$, i.e.\ where $|s_{ijk}|$ is not large. So it would take some additional work
				to study convergence issues in that regime. In this work,  we do not offer a complete discussion of convergence issues, we merely calculate
				the Feynman integral contributions using our regularization method, and find that
				the integrals for the tensor structure that are new compared to maximal supersymmetry appear to be infrared finite. 
				
				There is a proposal for a solution of this problem in the underlying string theory, as discussed in \cite{Witten:2013pra}, that we now very briefly review.
				 In field theory, the Euclidean propagator can be expressed in terms of the integral
				\be
				\frac{1}{p^2+m^2} = \int_0^\infty d \mathfrak{t} \exp{\Big( - \mathfrak{t} (p^2 + m^2) \Big)}\ ,
				\ee
				which converges only for $p^2+m^2 > 0$. On the other hand, the Lorentzian propagator is given via the integral
				\be \label{lorentzprop}
				\frac{-i}{p^2+m^2-i \efn} = \int_0^\infty d \mathfrak{t} \exp{\Big( - i \mathfrak{t} (p^2 + m^2 - i \efn) \Big)}\ ,
				\ee
				which converges for arbitrary values of $p^2+m^2$ due to the convergence factor $e^{-\mathfrak{t} \efn}$. In analogy to this, one can make the integrals \eqref{eq:WL_integral_Ddim} convergent independently of the ordering of the $\nu_i$ by considering the deformed integration contour
				\be 
				KN = \exp \Big( - t \big( \ldots \big) \Big) \longrightarrow \exp \Big( - i t \big( \ldots - i \efn \big) \Big)\ .
				\ee
				This is consistent with the prescription in \cite{Witten:2013pra} given that we are considering the field-theory limit in which $\tau_2$ is large. It also automatically incorporates the $i \efn$ on the right hand side of \eqref{eq:WL_integral_Ddim} that we need in order to make contact with the field theory integrals in sec.\ \ref{sec:position_int}.  However, it would sprinkle some relative factors  of $-i$ from \eqref{lorentzprop} into the integrals above. A more careful treatment would be to take our final results, 
				in particular the infrared-finite ones, and backtrack to the string theory variables to see how our regularization connects
				to the contour deformation of  \cite{Witten:2013pra}. We will not study this further here.
				
				Applying eq.\eqref{eq:WL_integral} to the integrals in section \ref{sec:FT_limit_integrand} leaves the following position, $\nu$, integrals\footnote{We omit a global $2^6$ factor from these formulas and reinstate it in the results.}:
				
				\subsection{Boxes}
				
				As is evident from eq.\eqref{eq:box_integral_full}, we can further classify the box integrals:
				
				\begin{itemize}
					\item ``boxy box'' \label{sec:boxy_box_integral}
					
					\begin{equation} \label{eq:boxy_box_integral}
						\FTM_0^A =  \pi^{-2} \int_0^1 \dd^3 \nu \; \frac{4\,t_8 \tilde t_8 + \sum_{k,k'=1}^{N-1} \hat{\chi}_{k,k'} \,  A_0(s,\nu)}{Q_4^{2}(s,\nu)} \ . 
					\end{equation}
					
					\item ``triangly boxes'' \label{sec:triangly_box_integral}
					
					\begin{equation}  \label{eq:triangly_box_integral}
						\FTM_0^B =2\, \sum_{k,k'=1}^{N-1} \hat{\chi}_{k,k'} \,  \int_0^1 \dd^3 \nu \; \frac{ B_0(s,\nu)}{Q_4 (s,\nu)} \ ,
					\end{equation}
					which gets its delineation from being a box integral, i.e. integrating over three positions and four external states -- $Q_4$ -- yet the power of $Q_4^{-1}$ is reminiscent of triangle integrals.
					
					\item ``bubbly boxes'' \label{sec:bubbly_box_integral}
					
					Similarly, the following case contains $Q_4^{-\euv}$, which occurs in bubble integrals:
					\begin{equation} \label{eq:bub_box_integral_full}
						\FTM_0^C= (2\pi)^2\, \Gamma(\euv)\, \sum_{k,k'=1}^{N-1} \hat{\chi}_{k,k'} \,  \int_0^1 \dd^3 \nu \; \frac{ C_0(s)}{Q_4^{\euv}(s,\nu)} \ .
					\end{equation}
				\end{itemize}
				
				\subsection{Triangles}
				
				In the same manner as above, the triangle integrals in eq.\eqref{eq:triangle_23_integral_full} can be further classified:
				
				\begin{itemize}
					\item ``triangly triangles''
					
					\begin{equation} \label{eq:triangly_triangles}
						\begin{aligned}
							\FTM_2^A= 2\,  &\sum_{k,k'=1}^{N-1} \hat{\chi}_{k,k'} \, \Biggl\{ s_{23} \, \int_0^1 \dd \nu_3 \dd \nu_4 \; \frac{ A_{2|23}(s,\nu)}{Q_3(k_1,k_{23},k_4)} +\\
							&s_{24} \, \int_0^1 \dd \nu_3 \dd \nu_4 \; \frac{A_{2|24}(s,\nu)}{Q_3(k_1,k_{24},k_3)} +s_{34} \, \int_0^1 \dd \nu_2 \dd \nu_4 \; \frac{ A_{2|34}(s,\nu)}{Q_3(k_1,k_{2},k_{34})}\Biggr\}  \ ,
					\end{aligned}\end{equation}
					and
					\item ``bubbly triangles''
				\end{itemize}
				\begin{equation} \label{eq:bubbly_triangle_integral}
					\begin{aligned}
						\FTM_2^B = (2\pi)^2\, \Gamma(\euv)\, &\sum_{k,k'=1}^{N-1} \hat{\chi}_{k,k'} \, \Biggl\{ s_{23} \, \int_0^1 \dd \nu_2 \dd \nu_4 \; \frac{ B_{2|23}(s)}{Q_3^{\euv}(k_1,k_{23},k_4)} +\\
						&s_{24} \, \int_0^1 \dd \nu_2 \dd \nu_3 \; \frac{B_{2|24}(s)}{Q_3^{\euv}(k_1,k_{24},k_3)} +s_{34} \, \int_0^1 \dd \nu_2 \dd \nu_3 \; \frac{ B_{2|34}(s)}{Q_3^{\euv}(k_1,k_{2},k_{34})}\Biggr\}  \ .
				\end{aligned}\end{equation}
				
				\subsection{Bubble}
				
				The sole ``bubbly bubble'' gives
				\begin{equation} \label{eq:bubble_integral}
					\FTM_3 = \FTM_3^A = (2\pi)^2\, \pi^{-\euv}\, \Gamma(\euv)\, \sum_{k,k'=1}^{N-1} \hat{\chi}_{k,k'} \,  \int_0^1 \dd \nu_2 \; \frac{ A_3(s)}{Q_2^{\euv}(k_1,k_{234})} \ .
				\end{equation}
	
				\section{Position / Coordinates integration} \label{sec:position_int}
				
				In computing the coordinate, $\nu_i$, integrals we used different methods for the box, triangle and bubble diagrams. Some integrals were straightforward enough for Mathematica to be able to calculate them directly; see section \ref{sec:pos_int_bubbles_direct}. Other integrals were less straightforward, so  we developed methods based on the differentiation method of 't Hooft-Veltman \cite{tHooft:1978jhc} and its elaboration in BDK \cite{Bern:1993kr}. In appendix \ref{app:BDK}, we give the shortest of reviews of the relevant parts to our calculations and demonstrate a proof of concept by applying this method to already established results in \cite{Tourkine:2012vx}.
				
				\subsection{Boxes - SCET philosophy} \label{SCET}
				
				After computing the field theory limit and the worldline integral, we are left with boxy box diagram integrals, eq.\eqref{eq:boxy_box_integral}, in the form:
				\begin{equation} \label{eq:ST_general_boxybox}
					\FTM_0^A \sim \int \dd^3 \nu \  \frac{P(\nu_i ; s_{ij}, s_{ijk})}{Q_4^{2}(\nu_i ; s_{ij}, s_{ijk})} \Bigg |_{\nu_1\ra 0} \ ,
				\end{equation}
				where $P$ is a polynomial in positions $\nu_i$ and also depends polynomially on the momenta $s_{ij}$ (we emphasised the dependence on the momentum regulators $s_{ijk}$). In our case, $P$ contains terms up to degree $4$ in the $\nu_i$. Importantly, we should remember that our amplitude is IR regulated; see section \ref{sec:minahaning_4point}. This means that these box integrals cannot be related to massless four-point results in the literature, as the latter use strict momentum conservation. Inspired by the methods of soft/collinear effective theory (SCET), see \cite{Becher:2014oda} for a review, we attach to each leg of the diagram -- with hard string momentum $k_i$ ($k_i^2=0$) -- a portion $\kappa_i$ of the soft momentum regulator such that $\sum \kappa_i=\kappa $; see the l.h.s.\ of fig \ref{fig:box_SCET}. This formally leads to field theory amplitudes for massive particles, with virtualities $p_i^2$, which obey strict momentum conservation:
				\begin{figure}[htbp]
					\centering
					\includegraphics[width=0.6\textwidth]{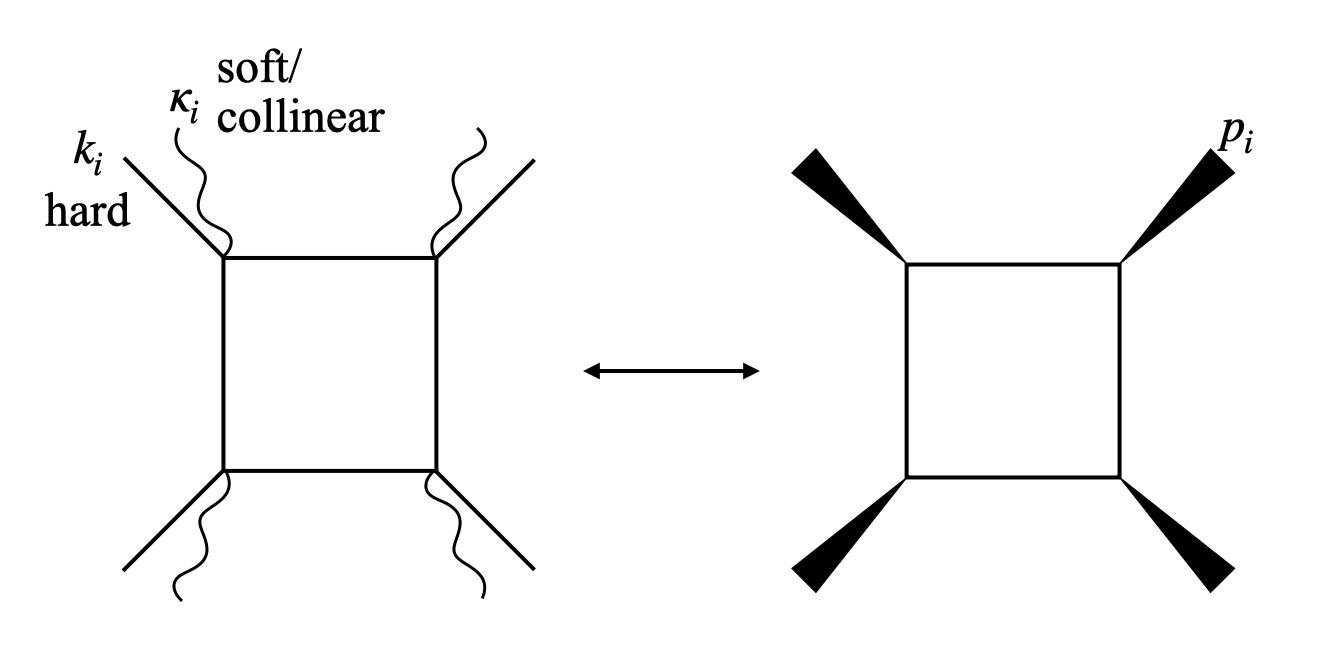}
					\caption{SCET-inspired field-theory box diagrams. On the left-hand side: string theory kinematics with massless gravitons and minahaning; on the right-hand side: four-mass field-theory diagram.}
					\label{fig:box_SCET}
				\end{figure}
				\begin{equation}
					m_i^2=p_i^2 = (k_i+\kappa_i)^2 = \cancelto{0}{ k_i^2 } + 2 k_i \cdot \kappa_i + \kappa_i^2 \ ,
				\end{equation}
				and
				\begin{equation}
					\sum p_i = \sum (k_i+\kappa_i) = \sum k_i + \kappa = 0 \ .
				\end{equation}
				We are left with: 
				\begin{itemize}
					\item 
					calculating the four-mass box diagram, which in the language of \cite{Bern:1993kr} eq.(2.5), also \eqref{eq:D4_nGon_BDK} in the review appendix, specifying $n\ra 4$, is\footnote{\label{fn:neglectUV}We neglect the UV dimensional regulator, $\euv$, in the integrals that are UV finite.}
					\begin{equation} \label{eq:FT_general_box}
						I_4[P\left(\left\{a_i\right\}\right)] =  \int_0^1 \dd^4 a \, \delta {\textstyle \left(1-\sum_i a_i\right) }\, \frac{P\left(\left\{a_i\right\}\right)}{\left[\sum_{i,j}^{4} S_{ij}a_ia_j - i \efn \right]^2} \ ,
					\end{equation}
					where the denominator,
					\begin{equation}
						\sum_{i,j=1}^4 S_{ij} a_i a_j =  -\hat{s}a_1a_3 -\hat{t}a_2a_4 -m_1^2a_1a_2-m_2^2a_2a_3-m_3^2a_3a_4-m_4^2a_4a_1 \ ,
					\end{equation}
					is given in eq.~(2.6) in the reference, and $P$ here is a general polynomial in the Feynman parameters, $a_i$. We distinguish the field theory Mandelstams -- $\hat{s}\equiv (p_1+p_2)^2$ and $\hat{t}\equiv (p_2+p_3)^2$ -- in order to avoid confusion with string theory kinematics.
					
					\item 
					finding a map that relates the field theory calculations, eq.\eqref{eq:FT_general_box}, to our actual integrals derived from string theory, eq.\eqref{eq:ST_general_boxybox}.
				\end{itemize}
				
				\subsubsection{String theory and field theory correspondence} \label{sec:ST_FT_correspondence}
				
				We bring \eqref{eq:ST_general_boxybox} to the form of \eqref{eq:FT_general_box}. The latter, rewritten as
				\begin{equation} \label{eq:FT_general_box2}
					I_4[\hat{P}\left(\left\{a_i\right\}\right)] =  \int_0^1 \dd^4 a  \, \delta {\textstyle \left(1-\sum_i a_i\right) }\, \frac{\hat{P}\left(\left\{a_i\right\}\right)}{\hat{Q}_4^2(a_i; p_i)} \ ,
				\end{equation}
				where
				\begin{equation}
					\hat{P} \left(\left\{a_i\right\}\right) \equiv P\left(\left\{a_i\right\}\right)\Big |_{a_4 \ra 1-(a_1+a_2+a_3)} \ , \qquad \hat{Q}_4(a_i; p_i) \equiv \sum_{i,j=1}^4 S_{ij} a_i a_j \Big |_{a_4 \ra 1-(a_1+a_2+a_3)} \ .
				\end{equation}
				
				\paragraph*{Hypercube variables}
				In eq.\eqref{eq:ST_general_boxybox} the positions, $\nu_i$, are integrated over a three-dimensional box. Yet, $Q_4$ depends on the relative ordering of the $\nu_i$, as can be seen in eq.\eqref{Qm_FT_lim}. Therefore,  the integrand is different in the six simplices of the box integration regions, e.g. $0\leq\nu_2\leq \nu_3\leq\nu_4\leq 1$ (remember that $\nu_1 \ra 0$). In order to retrieve the form of \eqref{eq:FT_general_box2}, with its hypercube integration, we change variables according to the prescription of Montag-Weisberger \cite{Montag:1991hd}:
				\begin{equation} \label{eq:MW_variables_box}
					\nu_i = a_1\ , \qquad \nu_j = a_1+a_2 \ , \qquad \nu_k= a_1+a_2+a_3 \ ,
				\end{equation}
				for the region/simplex represented by $0\leq\nu_i\leq \nu_j\leq\nu_k\leq 1$. For further details regarding the splicing of integration region, see discussion around eq.~(8) of \cite{Montag:1991hd}. This transformation has a trivial Jacobian. The basic idea is that when you \textit{add} variables that are all positive, you get a \textit{larger} value. So this gives the ordering. Then you use a fourth variable $a_4$ to "soak up" the danger of going above 1. We have thus brought \eqref{eq:ST_general_boxybox} to the desired form:
				\begin{equation} \label{eq:boxybox_in_MW}
					\FTM_0^A \sim \sum_{r\in \text{regions}}\int_0^1 \dd^4 a  \,  \delta {\textstyle \left(1-\sum_i a_i\right) }\, \frac{P\left(a_i; s_{ij(k)}\right)}{Q_4^2(a_i; s_{ij(k)})} \bigg|_r \ .
				\end{equation}
				\paragraph*{The map} \label{sec:box_STFT_map}
				Hence, by comparing denominators term by term and region by region, we obtain a map between the string theory and field theory kinematics:
				\begin{equation}
					Q_4(a_i; s_{ij(k)})\big|_r \ra \hat{Q}_4(a_i; \hat{s},\hat{t}, m_i^2) \ ,
				\end{equation}
				such that the embedding is given by
				\begin{equation} \label{embeddingstmsq}
					\hat{s} = \hat{s}(s_{ij},s_{ijk}) \ , \qquad  \hat{t} = \hat{t}(s_{ij},s_{ijk}) \ , \qquad m_i^2 = m_i^2(s_{ijk}) \ ,
				\end{equation}
				where the masses depend only on three-index Mandelstams. The last point could be dictated since we had freedom of choice in the above correspondence (the maps are not bijective). This freedom is related to the freedom in splitting $\kappa$ into $\kappa_i$, cf.\ fig.\ \ref{fig:box_SCET}. The embedding \eqref{embeddingstmsq} is very convenient, as the masses act as the new IR regulators: when we take $s_{ijk}\ra 0$, all $m_i\ra 0$. For example, the embedding of the region $0\leq\nu_2\leq \nu_3\leq\nu_4\leq 1$ can be chosen to be:
				\begin{equation} \label{eq:map_reg1}
					\begin{aligned}
						\hat{s} = -2s_{23}+s_{123} + s_{234}\ , \quad \hat{t}&= 2s_{23}+2s_{24}-s_{123}-s_{124}-2s_{234}\ , \\
						m_1^2=s_{123}+s_{124}+s_{234}\ , \quad m_2^2 &= -s_{124}\ , \quad m_3^2=-s_{123}\ , \quad m_4^2=-s_{234} \ .
					\end{aligned}
				\end{equation}
				Note that the four $m_i^2$ sum to zero, leaving five independent kinematic variables in field theory, matching the number of independent kinematic variables in string theory. 
				
				In the original Minahan paper \cite{Minahan}, for the 3-point function,
				 it was important that the $s+t+u=0$ condition due to modular invariance prohibited the ``obvious'' choice
				of symmetric kinematic regulators $s=t=u$. Here, the analogous statement is that the ``obvious'' symmetric distribution of regulator momenta $\kappa_i$ 
				among the four external states, $\kappa_i=\kappa/4$, is excluded by \eqref{embeddingstmsq}. By contrast, \eqref{eq:map_reg1}
				is an allowed choice.
				
				\paragraph*{Assembly} \label{sec:box_assembly}
				
				Once we have calculated the field-theory box building blocks, for which the polynomial $\hat{P}$ in \eqref{eq:FT_general_box2} is actually a monomial, see section \ref{sec:tensor_boxes}, we can implement the above string- and field-theory correspondence. In a particular region of integration, a string inherited polynomial $P$, e.g. in \eqref{eq:boxybox_in_MW}, is fixed. The resulting integral is then assembled via a linear combination corresponding to $P$ of the building blocks.
				
				For example, take $P= s_{23}\, a_1(s_{234}-s_{24}a_2)^2$, then the integral is 
				\begin{equation} \label{eq:example_assembly}
					I_4\left[P\right]_r = s_{23}\left(s_{234}^2\, I_4[a_1] + s_{24}^2\, I_4[a_1 a_2^2]-2s_{24}\,s_{234}\, I_4[a_1 a_2]\right) \big|_r \ ,
				\end{equation}
				in region $r$. Notice that this intermediate result depends also on field-theory kinematics through the building blocks.
				
				To calculate the full result of the integral, one then sums the results of each region over all the regions, mapping the field-theory kinematics back to the string-theory ones.
				
				\subsubsection{Differentiation method} \label{sec:BDK}
				The method in \cite{tHooft:1978jhc,Bern:1993kr} that we used to calculate the above integrals requires a generating function - for our case the vector boxes, $I_4[a_i]$, that are then differentiated with respect to projective parameters $\alpha_i$ which correspond to the Feynman integration variables, $a_i$, to obtain higher-order boxes. See appendix \ref{app:BDK} for a brief review.
				
				From Duplancic-Nizic \cite{Duplancic:2002dh} eq.(27):
				\begin{equation} \label{eq:DupNic_box}
					I_4[1] = I_4^{\text{scal}}(\hat{s},\hat{t},m_1^2,m_2^2,m_3^2,m_4^2) =  I_3^{\text{scal}}(\hat{s}\hat{t},m_1^2m_3^2,m_2^2m_4^2) , 
				\end{equation}
				one obtains a relation between the all-mass box and the triangle\footnote{Notice that this equation holds only under certain conditions which we are going to discuss at the beginning of the following section, i.e.\ \ref{sec:string_p}.}. 
				
				We used the eqs.\ (4.54) and (6.45) of \cite{Bourjaily:2020wvq} for the three-mass scalar triangle, based on the integral
				\begin{equation} \label{eq:scal_triangle}
					I_3^{\text{scal}}(z,\bar{z})= -\frac{1}{\pT_1^2}\frac{\Phi_1(z,\bar{z})}{z-\bar{z}} \ , 
				\end{equation}
				where \textsf{sans-serif} $\pT_i $s are used to denote triangle momenta, and with
				\begin{equation}
					\Phi_1(z,\bar{z}) = 2\text{Li}_2(z)-2\text{Li}_2(\bar{z})-\ln(z\bar{z})\left[ \text{Li}_1(z)-\text{Li}_1(\bar{z}) \right] \ , \label{Phieq}
				\end{equation}
				along with the relations in eq.(6.32) in \cite{Bourjaily:2020wvq}:
				\begin{equation} \label{eq:triangle_uv_rule1}
					u =\frac{\pT_2^2}{\pT_1^2}= z\bar{z} \ , \qquad v=\frac{\pT_3^2}{\pT_1^2}=(1-z)(1-\bar{z}) \ .
				\end{equation}
				
				The function $\Phi_1(z,\bar{z})$ in \eqref{Phieq} is closely related to the Bloch-Wigner dilogarithm, a single-valued
				polylogarithm. Its symmetries are discussed for example in appendix D of \cite{Bourjaily:2020wvq}. In appendix \ref{app:Mellin}, we review its contour representation.
				
				Therefore, the scalar four-mass box is obtained by the chain mapping:
				\begin{equation}
					I_4[1] = 	I_3^{\text{scal}}\left(z\left(u\left(\pT_i^2\right),v\left(\pT_i^2\right)\right),\bar{z}\left(u\left(\pT_i^2\right),v\left(\pT_i^2\right)\right)\right) \ ,
				\end{equation}
				and 
				\begin{equation}
					\pT_1^2 = \hat{s} \, \hat{t} \ , \qquad \pT_2^2 = m_1^2m_3^2 \ , \qquad \pT_3^2 = m_2^2m_4^2 \ ,
				\end{equation}
				where we reintroduce the \emph{causal} $i \efn$ in functions which contain branch cuts through
				\begin{equation}
					\pT_i^2 \mapsto \pT_i^2 + i \efn \ . 
				\end{equation}
				
					In order to bring \eqref{eq:FT_general_box2} to the form of eq.~\eqref{eq:BDK_general_red_int}, we can use the map between the field-theory kinematics and the differentiating parameters $\alpha_i$
				\begin{equation} \label{eq:alpha_choice}
					\hat{s}=-\frac{1}{\alpha_1\, \alpha_3}\, , \quad 	\hat{t}=-\frac{1}{\alpha_2\, \alpha_4}\, , \quad m_1^2=-\frac{u}{\alpha_1\, \alpha_2}\, , \quad m_2^2=-\frac{v}{\alpha_2\, \alpha_3}\, , \quad m_3^2=-\frac{1}{\alpha_3\, \alpha_4}\, , \quad m_4^2=-\frac{1}{\alpha_1\, \alpha_4}\, .
				\end{equation}
				
				A priori, having the scalar box is enough to start calculating higher-order tensor boxes. However, upon examination of eq.~\eqref{eq:BDK_general_red_int}, we notice that without dimensional regularisation, i.e. $\euv \ra 0$, the reduced generating scalar box does not depend on the $\alpha_i$'s. Since our box integrals are IR and not UV regulated, we cannot simply differentiate the scalar box. Hence, we set out to calculate the vector boxes, $I_4[a_1]$, $I_4[a_2]$ and $I_4[a_3]$, which in turn would serve as generating functions to higher tensor box calculations. This was done by applying the algebraic equations method of \cite{Bern:1993kr}, section 3, which is detailed for our case in section \ref{app:BDK_vec}.
				
				\subsubsection{Regions in field theory} \label{sec:string_p}
				
				As mentioned above, eq.\eqref{eq:DupNic_box} holds under certain conditions:
				\begin{itemize}
					\item Eq.(28) of \cite{Duplancic:2002dh}
					\begin{equation} \label{eq28DN}
						\left(u+i\efn \right)^{s''} \, \left(v+i\efn \right)^{s''} = \left(uv+i\efn\right)^{s''} \ ,
					\end{equation}
					is satisfied if $u$ and $v$ are not both negative. Here $u$ and $v$ are not to be confused with the variables in \eqref{eq:triangle_uv_rule1}. Rather, $(u,v)$ is any element of the set of pairs $\{ (\hat s, \hat t), (m_1^2, m_3^2), (m_2^2, m_4^2)\}$. When this condition does not hold, eq.\eqref{eq:DupNic_box} is modified by adding a correction term $K$, which is given in eq.(36) in \cite{Duplancic:2002dh}.\footnote{See also \cite{Corcoran:2020akn,Corcoran:2020epz} for the scalar box amplitude in all kinematic regions.}
					\item More positive than negative kinematics ($\hs,\hatt, m_i^2$). This is easily circumnavigated since, according to eq.(40) in \cite{Duplancic:2002dh}, the scalar box is invariant to flipping all signs of the kinematics and the causal~$i \efn$.
				\end{itemize}
				
				Having this in mind, we set out to find a point in \emph{string-theory} kinematics that would satisfy \eqref{eq28DN} for the six different embeddings corresponding to the six simplices of the box integration region. As a proof of principle, we settled on the following "string-point":
				\begin{equation} \label{eq:string_point}
					s_{23} = 1 \, , \quad s_{24} = -2 \quad \text{and} \qquad \{s_{123},s_{124},s_{234}\} = 10^{-3}\left\{\pi,-3\sqrt{2},-\frac{e}{2} \right\} \ .
				\end{equation}
				Notice that the three-index-Mandelstams are orders of magnitude smaller than the box-Mandelstams, so that \eqref{embeddingstmsq} is fulfilled and they can serve as IR regulators\footnote{We chose distinct irrational values for these regulators to avoid accidental numerical cancellations.}.
				This point, through the mapping to the first region, eq.\ \eqref{eq:map_reg1}, corresponds to mostly-negative field-theory kinematics ($\hs,\hatt, m_i^2$). It is easy to see that in this region the first condition above is satisfied, whereas we need to flip the sign of these kinematics to satisfy the second condition. This is also trivial as the \emph{triangle} kinematics, $\pT_i^2$, are just products of two box kinematics each. Hence, we only need to change $\efn \mapsto -\efn$.
				
				Indeed, point~\eqref{eq:string_point} maps in \emph{all} regions to field-theory kinematics that satisfy condition \eqref{eq28DN}, and, thus, the correction term $K$ is never needed in our calculations. See table~\ref{tab:string_point_regions} for a summary of the complete correspondence of the six different box integration regions and field-theory kinematics. In the following one can have this concrete kinematics in mind to be explicit.
				
			\begin{table}[htbp]
					\centering
					\setlength{\tabcolsep}{24pt}
					\renewcommand{\arraystretch}{1.5}
					\begin{tabular}{cc}
						$0 \leq \nu_2\leq \nu_3\leq\nu_4 \leq 1$ & $\hs<0 , \, \hatt<0 , \quad m_1^2 <0, \, m_2^2>0, \, m_3^2<0, \, m_4^2>0$ \\
						$0 \leq \nu_2\leq \nu_4\leq\nu_3 \leq 1$ & $\hs>0 , \, \hatt<0 , \quad m_1^2 <0, \, m_2^2<0, \, m_3^2>0, \, m_4^2>0$ \\
						$0 \leq \nu_3\leq \nu_2\leq\nu_4 \leq 1$ & $\hs<0 , \, \hatt>0 , \quad m_1^2 >0, \, m_2^2<0, \, m_3^2<0, \, m_4^2>0$  \\
						$0 \leq \nu_3\leq \nu_4\leq\nu_2 \leq 1$ & $\hs<0 , \, \hatt>0 , \quad m_1^2 >0, \, m_2^2<0, \, m_3^2<0, \, m_4^2>0$  \\
						$0 \leq \nu_4\leq \nu_2\leq\nu_3 \leq 1$ & $\hs>0 , \, \hatt<0 , \quad m_1^2 <0, \, m_2^2<0, \, m_3^2>0, \, m_4^2>0$  \\
						$0 \leq \nu_4\leq \nu_3\leq\nu_2 \leq 1$ & $\hs<0 , \, \hatt<0 , \quad m_1^2 <0, \, m_2^2>0, \, m_3^2<0, \, m_4^2>0$  \\
					\end{tabular}
					\caption{Correspondence of the six different box integration regions and field-theory kinematic regions through the mapping of point in eq.~\eqref{eq:string_point}.}
					\label{tab:string_point_regions}
				\end{table}

				\subsubsection{Tensor boxes} \label{sec:tensor_boxes}
				
				Since we are eventually interested in removing the IR regulator (wherever possible) and, thus, reimposing strict momentum conservation in string kinematics\footnote{Let us emphasize that momentum conservation is strictly preserved in field-theory kinematics, where we allow virtualities, see \ref{sec:IR_virtualities}.}, we can, in this intermediate step, look at the field-theory tensor boxes in the first-order expansion in the masses; see section \ref{sec:box_STFT_map}. Let us write here a few representative results of the tensor boxes, where the full set of calculations and results is in the accompanying Mathematica file \href{https://github.com/YonatanZimmerman/Field-theory-limit-of-half-maximal-string-loop-amplitudes/blob/main/Field%20Theory%20Integral%20Calculations.nb}{\faGithub}.
				
				The examples below make explicit the $i\efn$ prescription for the first region, i.e. the first line in table~\ref{tab:string_point_regions}. We use the analytic continuation rule in, e.g. \cite{Weinzierl:2022eaz},
				\begin{equation} \label{eq:analytic_cont}
					\log\left(\frac{s}{t}\right) = \log\left|\frac{s}{t}\right| +i \pi \left[\Theta(-s)-\Theta(-t)\right] \ ,
				\end{equation} 
				where $\Theta$ is the Heaviside step function\footnote{$\Theta(x)=
					\begin{cases}
						1, & x>0,\\
						0, & x<0.
					\end{cases}$}.
				Here, $s$ and $t$ stand for any kinematic variable.
				
				\begin{itemize}
					\item scalar box
					\begin{equation}
						I_4[1] = \frac{\pi^2 + 3\log\left(\frac{m_1^2 \, m_3^2}{\hat{s}\, \hat{t}} \right)\, \log\left(\frac{m_2^2 \, m_4^2}{\hat{s}\, \hat{t}}\right)}{3\,\hat{s}\, \hat{t}} + \mathcal{O}(m_i^2) \ ,
					\end{equation}
					
					\item vector box
					\begin{equation}
						\begin{aligned} \label{eq:vector_box_a1}
							I_4[a_1] &=-\frac{1}{6\,\hat{s}\, \hat{t}\, (\hat{s}+\hat{t})} \Bigg\{\pi^2 \, (\hat{t}-2\hat{s}) - 3\,\hat{t}\, \log\left(\frac{m_1^2 \, m_3^2}{\hat{s}\, \hat{t}} \right)\, \log\left(\frac{m_2^2 \, m_4^2}{\hat{s}\, \hat{t}}\right)
							\\
							&+3\, \hat{t}\bigg[\log\left(\frac{m_1^2 }{\hat{s}} \right)\, \log\left(\frac{m_2^2}{\hat{s}}- i\efn \right)+\log\left(\frac{m_2^2 }{\hat{t}}- i\efn \right)\, \log\left(\frac{m_3^2}{\hat{t}} \right)
							\\
							&+\log\left(\frac{m_3^2 }{\hat{s}} \right)\, \log\left(\frac{m_4^2}{\hat{s}} - i\efn \right)\bigg]
							-3(2\hat{s}+\hat{t}) \log\left(\frac{m_1^2 }{\hat{t}} \right)\, \log\left(\frac{m_4^2}{\hat{t}}- i\efn \right) \Bigg\} + \mathcal{O}(m_i^2)\ ,
					\end{aligned}	\end{equation}
					
					\item two-tensor box
					\begin{equation}
						\begin{aligned}
							I_4[a_1\, a_2] &=-\frac{1}{2\,\hat{s}\, \hat{t}\, (\hat{s}+\hat{t})^2} \Bigg\{\pi^2 \, \hat{s}\, \hat{t} - \hat{s}\,\hat{t}\, \log\left(\frac{m_1^2 \, m_3^2}{\hat{s}\, \hat{t}}\right)\, \log\left(\frac{m_2^2 \, m_4^2}{\hat{s}\, \hat{t}}\right)
							 \\
							&+\log\left(\frac{m_1^2 }{\hat{s}}\right) \left[(\hat{s}+\hat{t})(\hat{s}+2\hat{t})+\hat{s}\,\hat{t} \, \log\left(\frac{m_2^2}{\hat{s}}- i\efn\right)\right]+\log\left(\frac{m_1^2 }{\hat{t}}\right) \bigg[\hat{s}\,(\hat{s}+\hat{t})
							\\
							&+\hat{s}\,\hat{t} \, \log\left(\frac{m_4^2}{\hat{t}}- i\efn\right)\bigg] 
							+\hat{s}\,\hat{t} \log\left(\frac{m_2^2 }{\hat{t}}- i\efn\right)\, \log\left(\frac{m_3^2}{\hat{t}}\right)+\hat{s}(\hat{s}+\hat{t}) \log\left(\frac{m_4^2 }{\hat{t}}- i\efn\right)  
							\\
							&+\log\left(\frac{m_4^2 }{\hat{s}}- i\efn\right) \left[-\hat{s}\, (\hat{s}+\hat{t})+\hat{s}\,\hat{t} \, \log\left(\frac{m_3^2}{\hat{s}}\right)\right] \Bigg\} + \mathcal{O}(m_i^2)\ .
					\end{aligned}	\end{equation}
					
				\end{itemize}
				
				For our integrals, we also calculated the three- and four-tensor boxes which play a role in the boxy box integrals, $\FTM_0^A$. 
				
				\subsubsection{Triangly boxes} \label{sec:triangle_like_boxes}
				
				Similar to the boxy box eq.\eqref{eq:boxybox_in_MW}, we bring the triangly boxes of section \ref{sec:triangly_box_integral} to the form:
				
				\begin{equation} \label{eq:general_trBox}
					\FTM_0^B \sim \sum_{r\in \text{regions}}\int_0^1 \dd^4 a  \, \delta {\textstyle \left(1-\sum_i a_i\right) }\, \frac{P\left(a_i;s_{ij(k)}\right)}{Q_4(a_i; s_{ij(k)})} \bigg|_r\ ,
				\end{equation}
				where the difference to the boxy box lies in the exponent of the denominator, $Q_4^1$.
				In fact, this integral is the $D=6$ equivalent of the four-dimensional \eqref{eq:FT_general_box}, as pointed out in \cite{Bern:1993kr} and explained in appendix \ref{app:BDK}. Specifying eq.~\eqref{eq:D6_box} to our case, we get that a general triangly box in a specific region is obtained by
				\begin{equation}
					I_4^{D=6}[P\left(a_i;s_{ij(k)}\right)]_r = I_4\left[P\left(a_i;s_{ij(k)}\right) \, \hat{Q}_4(a_i; p_i)\right]_r \ ,
				\end{equation}
				which we have calculated in the previous section by subsuming $Q_4$ in the boxy box numerator polynomial and applying the string- and field-theory kinematics mapping.
				
				\subsubsection{Bubbly box} \label{sec:bubble_like_boxes}
				
				In a similar way, where we obtained the triangly boxes by looking at boxes in six dimensions, we can do the same for our bubbly boxes in section \ref{sec:bubbly_box_integral}, of the form
				\begin{equation}\label{eq:general_bubBox}
					\FTM_0^C \sim \sum_{r\in \text{regions}}\int_0^1 \dd^4 a  \,  \delta {\textstyle \left(1-\sum_i a_i\right) }\,\frac{C}{Q_4^{\euv}(a_i; s_{ij(k)}) \big|_r} \ ,
				\end{equation}
				which in turn are equivalent to $D=8-2\euv$ boxes. In this case, it is important to notice that the integrals are UV divergent, and thus, require reintroducing a regulator. The numerator of the bubbly boxes, $C$ in eq.\eqref{cs_integrand_lines}, is position independent. Therefore, we only need the scalar $D=8$ dimension box, given by specifying eq.~\eqref{eq:D8_box}:
				\begin{equation} \label{eq:bub_box_BDK}
					I_4^{D=8-2\euv} [1]= \Gamma(\euv) \, I_4\left[\hat{Q}^2_4(a_i; p_i)\right] \ .
				\end{equation}
				
				\subsection{Triangles - SCET and simplified differentiation method}
				
				Similarly to the box integrals in the previous sections, we translate our triangle integrals \eqref{eq:triangly_triangles}, in the general form
				\begin{equation} \label{eq:triangly_triangle_general_integral}
					\FTM_{2|ij}^A \sim  \int_0^1 \dd \nu_i \dd\nu_k \  \frac{P(\nu ; s_{ij}, s_{ijk})]}{Q_3(\nu ; s_{ij}, s_{ijk})} \Bigg |_{\substack{\nu_i=\nu_j \\ \nu_1\ra 0}} 
				\end{equation}
				into field-theory triangle integrals:
				\begin{equation} \label{eq:FT_triangle}
					I_3[\hat{P}\left(\left\{a_i\right\}\right)] = \int_0^1 \dd^3 a \, {\textstyle \delta\left(1-\sum_{j=1}^3 a_j\right) }\, \frac{\hat{P}\left(\left\{a_i\right\}\right)}{-\pT_1^2 \, a_1a_2-\pT_2^2 \, a_2a_3-\pT_3^2 \, a_1a_3} \ ,
				\end{equation}
				here specifying eq.~\eqref{eq:D4_nGon_BDK} by $n\ra 3$ and neglecting the UV dimensional regulator, $\euv$, in these UV finite integrals.
				
				The triangly triangles, obtained by the collision of two vertices, keep the minahaning IR regularisation prescription inherited from the box, see \ref{sec:minahaning_4point}. As in section \ref{sec:ST_FT_correspondence}, after performing the hypercube transformation on the integratioin variables, here $\{\nu_i,\nu_k\} \ra \{a_1,a_2\}$, in a specific collision and region, e.g. $2\leftrightarrow3$ and $0\leq \nu_2\leq\nu_4\leq 1$, we find an embedding map between the string-theory kinematics and field-theory kinematics,
					\begin{equation}
					\pT_i^2 = \pT_i^2(s_{jk}, s_{jkl}) \ ,
				\end{equation}
				by 
				\begin{equation}
					Q_3(a_i ; s_{ij}, s_{ijk})  \ra \hat{Q}_3=-\pT_1^2 \, a_1a_2-\pT_2^2 \, a_2a_3-\pT_3^2 \, a_1a_3 \Big|_{a_3\ra 1-(a_1+a_2)} \ .
				\end{equation}
				For example, in the region mentioned above, the mapping is
				\begin{equation}
					\pT_1^2=-2s_{23}+s_{123} +s_{234}\, ,  \qquad \pT_2^2 = -s_{123}\, , \qquad \pT_3^2 = -s_{234} \ .
				\end{equation}
				
				We calculated the field-theory triangle integrals by the simplified differentiation method, see section~\ref{sec:simplified_BDK}, where the generating scalar triangle is already given in eq.~\eqref{eq:scal_triangle}. Notice that in \eqref{claim}, $u$ and $v$ are defined differently than in eq.~\eqref{eq:triangle_uv_rule1} -- the former has $\pT_3^2$ in the denominator whereas the latter has $\pT_1^2$ in the denominator, corresponding to the factor pulled out in eqs.~\eqref{mellin} and \eqref{eq:scal_triangle}, respectively. Since we eventually reinstate momentum conservation, i.e. $s_{ijk} \ra 0$, it is prudent to pick the \emph{finite} $1 / \pT_i^2$ factor to pull out, according to the mapping in each region. For example, in the region above, $\pT_1^2$ remains finite, and, thus, we calculate with the definitions in section \ref{sec:BDK}. Other collisions and regions necessitate different $1 / \pT_i^2$ choices. In \href{https://github.com/YonatanZimmerman/Field-theory-limit-of-half-maximal-string-loop-amplitudes/blob/main/Field%20Theory%20Integral%20Calculations.nb}{\faGithub}. we calculate the scalar-, vector- and two tensor-triangles in all three options.
				Evidently, all these triangle building blocks diverge -- \emph{at most} -- as
				 \begin{equation}
				 	\log\left(\frac{\pT_i^2}{\pT^2_{\text{finite}}}\right) \sim \log\left(s_{ijk}\right) \ .
				 \end{equation}
				 Therefore, the triangly triangle, $\FTM_2^A$, does not contribute to our amplitude, as the numerators in eqs.\eqref{eq:A2|23}--\eqref{eq:A2|34} are proportional to $s_{ijk}^2$ or are identically $0$.
			
				\subsection{Bubbles - direct computation} \label{sec:pos_int_bubbles_direct}
				Some integrals were straightforward enough to calculate directly, see \href{https://github.com/YonatanZimmerman/Field-theory-limit-of-half-maximal-string-loop-amplitudes/blob/main/Assembly%20and%20Results.nb}{\faGithub}.
				
				\subsubsection{Bubbly bubble}
				The denominator in the bubbly-bubble integral in eq.\eqref{eq:bubble_integral}, $Q_2^{\euv}$, is obtained by plugging in the kinematics into the Koba-Nielsen factor in the field theory limit, eq.\eqref{Qm_FT_lim}:
				\begin{equation}
					Q_2(k_1,k_{234}) = s_{234} \,  \nu_2 \left(1-\nu_2\right) \ .
				\end{equation}
				Together with the numerator $A_3$, given in eq.\eqref{eq:bubble_numerator}, the integral yields:
				\begin{equation} \label{eq:proper_bubble_calc}
					\begin{aligned}
						\FTM_3 &\sim s_{234}^{1-\euv} \, \Gamma(\euv)\, \frac{\Gamma^2(1-\euv)}{\Gamma(2-2\euv)} \\
						& \xrightarrow{\euv \ll 1} \frac{s_{234}}{\euv} +s_{234}\, \left[2-\gamma-\log(s_{234})+\mathcal{O}(\euv)\right]\ ,
					\end{aligned}
				\end{equation}
				where $\gamma$ is the Euler-Mascheroni constant. Evidently, the contribution of the bubbly bubble term is a matter of hierarchy of limits. In this work, we use dimensional regularization for UV divergences and new mass scales (due to new kinematic variables $s_{ijk}$) for the IR divergences, 
				in contrast to dimensionally regulating both UV and IR, as is often done (see e.g.\ \cite{Cohen:2019wxr} for a review). Here, we  take $s_{234} \rightarrow 0$ and keep the UV regulator $\euv$ finite, so we receive no contribution from the bubbly bubble term above. 
				
				\subsubsection{Bubbly triangles}
				Similarly to the previous section, we plug in the appropriate Koba-Nielsen factor for the integration regions of the bubbly triangles, stemming from the three two-vertex collision regions, in eq.\eqref{eq:bubbly_triangle_integral}. For example, in the collision and region $2\leftrightarrow3$ and $0\leq \nu_2\leq\nu_4\leq 1$, respectively, we get:
				\begin{equation}
					Q_3^{-\euv}(k_1,k_{23},k_4) = \left(2s_{23} \, \nu_2 (\nu_4-\nu_2)\right)^{-\euv} + \mathcal{O}(m_i^2) \ .
				\end{equation}
				By reinstating prefactors and numerator in \eqref{eq:bubbly_triangle_integral}, the contribution of this particular region to the UV divergent terms is
				\begin{equation}
					\FTM_{2|23}^B \Big|_{\nu_2 \leq \nu_4} = 256\, \pi^2 \,  \Gamma(\euv) \sum_{k,k'=1}^{N-1} \hat{\chi}_{k,k'} \, \half \left[s_{23} \, C^m_{1|23|4}\tilde{C}^m_{1|23|4} \right]\ .
				\end{equation}
				Summing over all collisions and regions, we get:
				\begin{equation} \label{eq:bub_triangle_calc}
					\FTM_2^B = 256\, \pi^2 \,   \Gamma(\euv)\, \sum_{k,k'=1}^{N-1} \hat{\chi}_{k,k'} \,  \left[s_{23} \, C^m_{1|23|4}\tilde{C}^m_{1|23|4} + \text{cyc}(2,3,4)\right]  \ ,
				\end{equation}
				when momentum conservation is restored.
				
				\section{Assembly and results} \label{sec:results}
				
				Having calculated the field theory like integrals, we can prepare to assemble the original amplitude's field-theory limit. Firstly, we treat IR divergent and finite terms separately. As previously discussed, all $m_i^2$ vanish when restoring momentum conservation whereas $\hat{s}$ and $\hat{t}$ remain finite. Therefore, terms such as $\log\left(\tfrac{m_2^2}{\hat{s}} -i\efn \right)$ denote our divergent contributions but $\log\left(\tfrac{\hat{s}}{\hat{t}}  \right)$ terms are finite. The latter terms come from standardizing our building blocks to contain only $\log$s of $\left(\frac{m_1^2}{\hs}\right)$, $\left(\frac{m_2^2}{\hs}\right)$,$\left(\frac{m_3^2}{\hatt}\right)$, and $\left(\frac{m_4^2}{\hatt}\right)$.  
				Subsequently, the other terms are modified, e.g.
				\begin{equation}
					\log\left(\frac{m_1^2}{\hatt} \pm i\efn\right) = \log\left(\frac{m_1^2}{\hs} \pm i\efn\right) + \log\left(\frac{\hs}{\hatt} \pm i\efn\right) \ .
				\end{equation} 
				For example, the $a_1$ vector box integral in eq.\eqref{eq:vector_box_a1} contributes
				\begin{equation} \label{eq:vac_box_ai_div}
					I_4[a_1] \Big |_{\text{div., reg.1}} = \frac{1}{\hat{s} \, \hat{t}} \ \left\{ \log\left(\tfrac{m_4^2}{\hat{t}} - i \efn \right)  \left[ \log\left(\tfrac{m_1^2}{\hat{s}} \right)  +  \log\left(\tfrac{\hat{s}}{\hat{t}}  \right)\right] \right\} \ ,
				\end{equation}
				 to the divergent terms and
				\begin{equation} \label{eq:vac_box_ai_finite}
					I_4[a_1] \Big |_{\text{finite, reg.1}} =- \frac{\pi^2\left(\hat{t}-2\hat{s}\right) + 3\hat{t} \,\log^2 \left(\tfrac{\hat{s}}{\hat{t}} \right) }{6\hat{s} \, \hat{t} \left(\hat{s}+\hat{t}\right)} \ ,
				\end{equation}
			to the finite terms. The full assembly calculations and results can be found in \href{https://github.com/YonatanZimmerman/Field-theory-limit-of-half-maximal-string-loop-amplitudes/blob/main/Assembly%20and%20Results.nb}{\faGithub}.
				
				\subsection{Infra-red divergences}
				Arising from the boxy box, we get the following IR divergent contribution in the field-theory first region\footnote{First region, as in the first line in table~\ref{tab:string_point_regions}.}:
			\begin{equation}
				\begin{aligned}
					\FTM^A_0 \Big |_{\text{IR, reg.1}} &= \frac{ 256}{9 \pi^2 \hat{s}\, \hat{t}} \,t_8 \tilde{t}_8 \ \Bigg\{ 9 \left[\log\left(\frac{m_1^2}{\hat{s}}  \right)+\log\left(\frac{m_3^2}{\hat{t}}  \right)\right] \left[\log\left(\frac{m_2^2}{\hat{s}} - i\efn \right)+\log\left(\frac{m_4^2}{\hat{t}} - i\efn \right)\right]
					\\
					&+\sum_{k,k'=1}^{N-1} \hat{\chi}_{k,k'}\, \bigg\{ \left(\pi^2-3F_{1/2}^{(2)}\right)\left(\pi^2-3\tilde{F}_{1/2}^{(2)}\right) \left[\log\left(\frac{m_1^2}{\hat{s}}  \right)+\log\left(\frac{m_3^2}{\hat{t}}  \right)\right] 
					\\
					& \times \left[\log\left(\frac{m_2^2}{\hat{s}} - i\efn \right)+\log\left(\frac{m_4^2}{\hat{t}} - i\efn \right)\right] -  18 \left(\pi^2- \left(F_{1/2}^{(2)}+\tilde{F}_{1/2}^{(2)}\right)\right) \log\left(\frac{m_4^2}{\hat{t}} - i\efn \right) 
					\\
					&+  6 \left(\pi^2-3 \left(F_{1/2}^{(2)}+\tilde{F}_{1/2}^{(2)}\right)\right) \left[ \log\left(\frac{m_1^2}{\hat{s}}  \right)+\log\left(\frac{m_2^2}{\hat{s}} - i\efn \right)+\log\left(\frac{m_3^2}{\hat{t}}  \right)\right] \bigg\} \Bigg\} \ ,
				\end{aligned}
			\end{equation}	
				when plugging in building-blocks, e.g. \eqref{eq:vac_box_ai_div}, in the boxy box eq.\eqref{eq:boxy_box_integral}:
				\begin{equation}\label{properboxintegral}
					\int\limits_{0 \leq \nu_2 \leq \nu_3 \leq \nu_4 \leq 1} \dd^3\nu \  \frac{A_0\left(\nu_i;s_{ij(k)}\right)}{Q_4^2\left(\nu_i;s_{ij(k)}\right)} \ra I_4[A_0\left(\nu_i;s_{ij(k)}\right)]  \Big |_{\text{div., reg.1}} \ . 
				\end{equation}
				
				Notice that the above IR contribution is proportional to $t_8 \tilde{t}_8$ exclusively, as are all IR contributions in all regions. Let us also note that there are no IR divergent contributions from triangly box terms, eq.\eqref{eq:triangly_box_integral}. Once momentum conservation is restored to these terms after assembly, they vanish, i.e.
				\begin{equation}
					I_4\left[B_0(a_i;s_{ij(k)}) \, \hat{Q}_4(a_i,p_i)\right]_r = 0 \ ,
				\end{equation}
				for all regions $r$.
				
				\subsection{Ultra-violet divergences}
				The UV divergent terms come from the bubbly triangles in eq.~\eqref{eq:bub_triangle_calc} and from applying our method to calculating box integrals for the bubbly box in eq.\eqref{eq:bub_box_integral_full}. The latter is assembled, following our prescription, as follows:
					\begin{equation}
					\FTM_0^C = \sum_{r\in \text{regions}} \Gamma(\euv) \, C \, I_4\left[\hat{Q}^2_4(a_i; p_i)\right]_r=4 \pi^2  \, \OrbSum \,  \Gamma(\euv) C \ ,
				\end{equation}
				where $C$ in eq.\eqref{cs_integrand_lines}, is position independent and, therefore, could be taken out of the $I_4$ polynomial calculation, and
				we denote the orbifold sector summation over fixed terms (no twist vector dependency), i.e. the number of fixed points of the orbifold counted with multiplicity, by
				\begin{equation}
					\OrbSum \equiv \sum_{k,k'=1}^{N-1} \hat{\chi}_{k,k'} \ .
				\end{equation}
				Combining all the bubbles yields:
				\begin{equation}
					\FTM^C = 256\,  \pi^2\, \OrbSum \, \Gamma(\euv) \left\{  \frac{1}{2} C_{1|2,3,4}^{mn} \tC_{1|2,3,4}^{mn}+ \left[s_{23} \, C^m_{1|23|4}\tilde{C}^m_{1|23|4} - P_{1|2|3,4}\tP_{1|2|3,4}  + \text{cyc}(2,3,4)\right] \right\} \ ,
				\end{equation}
				which is proportional to eq.(6.34) in \cite{paper1}. Note that the term in curly brackets vanishes in strictly $D=4$, cf.\ \cite{paper1}.
				
				\subsection{Finite terms in field-theory kinematics }
				
				In this section, we present the finite results of our amplitude in the field-theory kinematics region 1, i.e. the first line in table~\ref{tab:string_point_regions}. The building blocks calculated in section~\ref{sec:position_int} are plugged into the appropriate numerator polynomials, as explained for the example in \eqref{eq:example_assembly},
				\begin{equation}
					\FTM_0^{A} \Big |_\text{reg.1} = I_4\left[A_0(a_i;s_{ij(k)})\right]_{\text{reg.1}} \ ,
				\end{equation} 
				for the boxy box terms \eqref{eq:boxy_box_integral}, and
				\begin{equation}
					\FTM_0^{B} \Big |_\text{reg.1} = I_4\left[B_0(a_i;s_{ij(k)}) \, \hat{Q}_4(a_i;p_i)\right]_{\text{reg.1}} \ ,
				\end{equation} 
				for the triangly box terms \eqref{eq:triangly_box_integral}. As mentioned previously, there are no other contributing terms.
				
				The results are summarized in the following representations:
				
				\begin{equation}
					\begin{aligned}
						\FTM_0^{A} \Big |_\text{reg.1}
						&=\frac{256}{3\,\hat{s} \, \hat{t}} \, t_8 \tilde{t}_8
						+\frac{256 \pi^2}{3} \sum_{k,k'=1}^{N-1} \hat{\chi}_{k,k'}\,
						\\ &
						\begin{pmatrix} t_8 & P_{1|2|3,4} & P_{1|3|2,4} \end{pmatrix}
						\begin{pmatrix}
							\mathcal{A}_{11}
							&
							s_{34}\,\mathcal{A}_{12}
							&
							s_{24}\,\mathcal{A}_{13}
							\\[6pt]
							
							s_{34}\,\tilde{\mathcal{A}}_{12}
							&
							s_{34}^{2}\,\mathcal{A}_{22}
							&
							s_{24}s_{34}\,\mathcal{A}_{23}
							\\[6pt]
							
							s_{24}\,\tilde{\mathcal{A}}_{13}
							&
							s_{24}s_{34}\,\tilde{\mathcal{A}}_{23}
							&
							s_{24}^{2}\,\mathcal{A}_{33}
						\end{pmatrix}
						\begin{pmatrix} \tilde{t}_8 \\ \tilde{P}_{1|2|3,4} \\ \tilde{P}_{1|3|2,4} \end{pmatrix} ,
					\end{aligned}
				\end{equation}
				with matrix entries
				\begin{equation}
					\begin{aligned}
						\mathcal{A}_{11}(\hs,\hatt,k,k')
						&=
						\frac{1}{9\,\hs\hatt}
						\left\{
						\pi^2 -72 + 36 \Lgsh - 3 \pi^{-2} \left(\FO+\FOT \right) \left[36 \Lgsh + \pi^2 \right] 
						+ 9 \FO \FOT
						\right\}, 
						\\[10pt]
						\mathcal{A}_{12}(\hs,\hatt,k)
						&=
						-\frac{2}{\hs\hatt(\hs+\hatt)^2}
						\bigg\{
						\hs \hatt\left[\log^2\!\left(\frac{\hs}{\hatt}\right)+\pi^2\right]
						- \left(\hs+\hatt\right)
						\left[ 
						\left(\hatt - \hs \right) \Lgsh
						+ \hs+\hatt
						\right] \\
						& \qquad - 3 \pi^{-2} \FO \, (\hs+\hatt)^2 \Lgsh
						\bigg\},
						\\[6pt]
						\mathcal{A}_{13}(\hs,\hatt,k)
						&=
						\frac{1}{\hs(\hs+\hatt)^3}
						\bigg\{
						\hs (\hs - 3\hatt)\left[\log^2\!\left(\frac{\hs}{\hatt}\right)+\pi^2\right]
						+2 (\hs + \hatt) \left[ (\hatt -3\hs)\Lgsh + 2(\hs+\hatt) \right] \\
						& \qquad - 3 \pi^{-2} \FO  \, (\hs+\hatt) \left[ \hs \, \log^2\!\left(\frac{\hs}{\hatt}\right) - 2(\hs+\hatt) \Lgsh  + \pi^2 \hs \right]
						\bigg\},
						\\[6pt]
						\mathcal{A}_{22}(\hs,\hatt)
						&=
						\frac{2}{\hs\hatt(\hs+\hatt)^2}
						\left\{
						\hs \hatt \left[\log^2\!\left(\frac{\hs}{\hatt}\right)+\pi^2\right]
						+ \left( \hs+\hatt \right)
						\left[\left(\hs-\hatt \right)\Lgsh- 2  \left( \hs+\hatt \right) \right]
						\right\},
						\\[10pt]
						\mathcal{A}_{23}(\hs,\hatt)
						&=
						-\frac{1}{\hs(\hs+\hatt)^3}
						\left\{
						\hs \left(\hs-3 \hatt \right)\left[\log^2\!\left(\frac{\hs}{\hatt}\right)+\pi^2\right]
						-2 \left(\hs+\hatt\right)
						\left[ \left(3\hs-\hatt\right)\Lgsh -2 \left(\hs+\hatt\right)\right]
						\right\},
						\\[10pt]
						\mathcal{A}_{33}(\hs,\hatt)
						&=
						-\frac{2}{\hs(\hs+\hatt)^4}
						\bigg\{
						\hs \hatt 	\left(\hs-2 \hatt \right)\left[\log^2\!\left(\frac{\hs}{\hatt}\right)+\pi^2\right]
						\\
						& \qquad+ \left(\hs+\hatt\right) \left[\hatt \left(\hatt - 5\hs\right)\Lgsh - 2\left(\hs+\hatt\right)\left(\hs-2\hatt\right)\right]
						\bigg\},
					\end{aligned}
				\end{equation}
				where 
				\begin{equation}
					\tilde{\mathcal{A}}_{ij} = \mathcal{A}_{ij} \bigg|_{\FO \rightarrow \FOT} \ ,
				\end{equation}
				and where $\FO$ and $\FOT$ are placeholders for the field-theory limit of these worldsheet functions, given in eq.~\eqref{WS_FTlimit}, and reiterated here:
				\begin{equation*}
					F_{1/2}^{(2)} (kv,\tau) \FTa \pi^2 \left[ \frac{1}{3} - \ \frac{1}{\sin^2(\pi k v)} \right] \ . 
				\end{equation*}
				The boxy box result is given above in terms of $(t_8, P_{1|2|3,4}, P_{1|3|2,4} )$ in order to connect with the well-known maximally-supersymmetric $t_8$. In the Mathematica file, we calculate the above, additionally, in a more symmetric basis of $(P_{1|2|3,4}, P_{1|3|2,4}, P_{1|4|2,3} )$.
				
				The triangly box finite terms in field-theory kinematics are given in terms of coefficients of \\ 
				$(C^m_{1|23,4}, C^m_{1|24,3} , C^m_{1|34,2})$:
				\begin{equation}
					\FTM_0^B \Bigg |_\text{reg.1}
					=
					\frac{64\pi^2 \, \OrbSum}{3(\hat{s}+\hat{t})^3}
					\begin{pmatrix} C^m_{1|23,4} & C^m_{1|24,3}  & C^m_{1|34,2}  \end{pmatrix}
					\begin{pmatrix}
						
						s_{23}^2\,  \mathcal{B}_{11}
						&
						s_{23}s_{24}\,  \mathcal{B}_{12}
						&
						s_{23}s_{34}\,  \mathcal{B}_{13}
						\\[6pt]
						
						s_{23}s_{24}\,  \mathcal{B}_{12}
						&
						s_{24}^2 \,  \mathcal{B}_{22}
						&
						s_{24}s_{34}\,  \mathcal{B}_{23}
						\\[6pt]
						
						s_{23}s_{34}\,  \mathcal{B}_{13}
						&
						s_{24}s_{34}\,  \mathcal{B}_{23}
						&
						s_{34}^2\,  \mathcal{B}_{33}
						
					\end{pmatrix}
					\begin{pmatrix} \tilde{C}^m_{1|23,4}\\ \tilde{C}^m_{1|24,3}  \\ \tilde{C}^m_{1|34,2}  \end{pmatrix}
				\end{equation}
				with matrix entries
				\begin{equation}
					\begin{aligned}
						\mathcal{B}_{11}(\hat{s},\hat{t}) &= 
						-\left(\hat{s}^2+3\hat{t}^2\right)
						\left[\log^2\!\left(\frac{\hat{s}}{\hat{t}}\right) + \pi^2 \right]
						-4 \left(\hat{s} + \hat{t}\right)
						\left[ 2\hat{t} \, \log\!\left(\frac{\hat{s}}{\hat{t}}\right) + \left(\hat{s} + \hat{t}\right) \right] ,
						\\[6pt]	
						\mathcal{B}_{12}(\hat{s},\hat{t}) &= 
						-(\hat{s}+\hat{t})
						\left\{
						\hat{s} \left[\log^2\!\left(\frac{\hat{s}}{\hat{t}}\right) + \pi^2 \right]
						-2(\hat{s}+\hat{t})\log\!\left(\frac{\hat{s}}{\hat{t}}\right)
						\right\},
						\\[6pt]
						\mathcal{B}_{13}(\hat{s},\hat{t}) &= 
						4 \left\{ -\hat{s}\hat{t} \left[\log^2\!\left(\frac{\hat{s}}{\hat{t}}\right) + \pi^2 \right]
						-\left(\hat{s} + \hat{t}\right) \left[\left(\hat{s} - \hat{t}\right)\log\!\left(\frac{\hat{s}}{\hat{t}}\right)
						-\left(\hs+\hatt\right) \right]\right\},
						\\[6pt]
						\mathcal{B}_{22}(\hat{s},\hat{t}) &= 
						-(\hat{s}+\hat{t})^2
						\left[
						\log^2\!\left(\frac{\hat{s}}{\hat{t}}\right)+\pi^2
						\right],
						\\[6pt]
						\mathcal{B}_{23}(\hat{s},\hat{t}) &= 
						-(\hat{s}+\hat{t})
						\left\{
						\hat{t} \left[\log^2\!\left(\frac{\hat{s}}{\hat{t}}\right) + \pi^2 \right]
						+2(\hat{s}+\hat{t})\log\!\left(\frac{\hat{s}}{\hat{t}}\right)
						\right\},
						\\[6pt]
						\mathcal{B}_{33}(\hat{s},\hat{t}) &= 
						-\left(3\hat{s}^2+\hat{t}^2\right)
						\left[\log^2\!\left(\frac{\hat{s}}{\hat{t}}\right) + \pi^2 \right]
						-4 \left(\hat{s} + \hat{t}\right)
						\left[ 2\hat{s} \, \log\!\left(\frac{\hat{s}}{\hat{t}}\right) + \left(\hat{s} + \hat{t}\right) \right] .
					\end{aligned}
				\end{equation}

				\subsection{Finite terms in string kinematics}
				
				Having the results in field-theory kinematics, we calculated the full result in string kinematics. For each box integration region, we first applied the corresponding string kinematics through the embedding map, e.g \eqref{eq:map_reg1} in table~\ref{tab:string_point_regions}. Then we analytically continued the logs of string kinematic ratios through eq.~\eqref{eq:analytic_cont}, having the signs of \eqref{eq:string_point} in mind. Finally, we combined these results from all regions.
				
				The full results are given in the following representations:
				
			\begin{equation}
				\mathcal{M}_0^{A} \Big |_\text{All reg.}
				=\frac{256 \pi^2}{3} \sum_{k,k'=1}^{N-1} \hat{\chi}_{k,k'}\,
				\begin{pmatrix} t_8 & P_{1|2|3,4} & P_{1|3|2,4} \end{pmatrix}
				\begin{pmatrix}
					\mathring{\mathcal{A}}_{11}
					&
					s_{34}\,\mathring{\mathcal{A}}_{12}
					&
					s_{24}\,\mathring{\mathcal{A}}_{13}
					\\[6pt]
					
					s_{34}\,\mathring{\tilde{\mathcal{A}}}_{12}
					&
					s_{34}^{2}\,\mathring{\mathcal{A}}_{22}
					&
					s_{24}s_{34}\,\mathring{\mathcal{A}}_{23}
					\\[6pt]
					
					s_{24}\,\mathring{\tilde{\mathcal{A}}}_{13}
					&
					s_{24}s_{34}\, \mathring{\tilde{\mathcal{A}}}_{23}
					&
					s_{24}^{2}\,\mathring{\mathcal{A}}_{33}
				\end{pmatrix}
				\begin{pmatrix} \tilde{t}_8 \\ \tilde{P}_{1|2|3,4} \\ \tilde{P}_{1|3|2,4} \end{pmatrix} ,
			\end{equation}
			with matrix entries
			\begin{equation}
				\mathring{\mathcal{A}}_{ij} (s_{rl},k,k') = \mathring{\mathcal{A}}_{ij}^{(0)} +\mathring{\mathcal{A}}_{ij}^{(\alpha)}+\mathring{\mathcal{A}}_{ij}^{(\beta)} +\mathring{\mathcal{A}}_{ij}^{(\gamma)}  ,
			\end{equation}
			and we give here a sample of the $\log$ dependencies:
			\begin{equation}
				\begin{aligned}
					\mathring{\mathcal{A}}_{11}^{(\alpha)} &= -\,\frac{3\pi^{-2}\,\left(\FO+\FOT \right)}{s_{23}\,s_{24}}\,\log\left|\frac{s_{23}}{s_{24}}\right| , 
					\\[6pt]	
					\mathring{\mathcal{A}}_{11}^{(\beta)} &= -\,\frac{3\pi^{-2}\,\left(\FO+\FOT \right)}{s_{23}\,s_{34}}\,\log\left|\frac{s_{23}}{s_{34}}\right| , 
					\\[6pt]	
					\mathring{\mathcal{A}}_{11}^{(\gamma)} &= \frac{2}{s_{23}^4} \left\{2\left[3\left(s_{23}(s_{24}-s_{34})+i\pi\left(s_{23}^2-6s_{24}s_{34}\right)\right)\right]	\log\left|\frac{s_{24}}{s_{34}}\right|  +\left(s_{23}^2-6s_{24}s_{34}\right)\log^2\left|\frac{s_{24}}{s_{34}}\right| \right\}
					\\
					& \qquad- \frac{3 \pi^{-2}{\textstyle \left(\FO+\FOT\right)}}{s_{23}^2} \left\{\frac{s_{23}(s_{34}-s_{24})+2\pi i s_{24}s_{34}}{s_{24}s_{34}}\log\left|\frac{s_{24}}{s_{34}}\right|+\log^2\left|\frac{s_{24}}{s_{34}}\right|\right\}
					\\[6pt]	
					\mathring{\mathcal{A}}_{11}^{(0)} &=\frac{12}{s_{23}^3}\left(s_{23}-i\pi \left(s_{34}-s_{24}\right)\right)
					+ \frac{3 i\pi^{-1}{\textstyle \left(\FO+\FOT\right)}}{s_{23}s_{34}}	\ .
				\end{aligned}
			\end{equation}
			The rest can be found in \href{https://github.com/YonatanZimmerman/Field-theory-limit-of-half-maximal-string-loop-amplitudes/blob/main/Assembly%20and%20Results.nb}{\faGithub}. And the full results of the triangly box are:
				
				\begin{equation}
					\FTM_0^B \Bigg |_\text{All reg.}
					=
					\frac{64\pi^2 \, \OrbSum}{3}
					\begin{pmatrix} C^m_{1|23,4} & C^m_{1|24,3}  & C^m_{1|34,2}  \end{pmatrix}
					\begin{pmatrix}
						
						s_{23}^2\,  \mathring{\mathcal{B}}_{11}
						&
						s_{23}s_{24}\,   \mathring{\mathcal{B}}_{12}
						&
						s_{23}s_{34}\,   \mathring{\mathcal{B}}_{13}
						\\[6pt]
						
						s_{23}s_{24}\,   \mathring{\mathcal{B}}_{12}
						&
						s_{24}^2 \,  \mathring{\mathcal{B}}_{22}
						&
						s_{24}s_{34}\,  \mathring{\mathcal{B}}_{23}
						\\[6pt]
						
						s_{23}s_{34}\,   \mathring{\mathcal{B}}_{13}
						&
						s_{24}s_{34}\,   \mathring{\mathcal{B}}_{23}
						&
						s_{34}^2\,  \mathring{\mathcal{B}}_{33}
						
					\end{pmatrix}
					\begin{pmatrix} \tilde{C}^m_{1|23,4}\\ \tilde{C}^m_{1|24,3}  \\ \tilde{C}^m_{1|34,2}  \end{pmatrix} \ ,
				\end{equation}
				with matrix entries
				\begin{equation}
					\mathring{\mathcal{B}}_{ij} (s_{kl}) = \frac{\mathring{\mathcal{B}}_{ij}^{(\alpha)}}{s_{34}^{3}} +\frac{\mathring{\mathcal{B}}_{ij}^{(\beta)}}{s_{24}^{3}}+\frac{\mathring{\mathcal{B}}_{ij}^{(\gamma)}}{s_{23}^{3}} \ ,
				\end{equation}
				and log dependencies
				\begin{align}
						\mathring{\mathcal{B}}_{11}^{(\alpha)} &=
						-4s_{34}\left(s_{34}+2i\pi s_{24}\right)
						+2\left(4s_{24}s_{34}+i \pi \left(s_{23}^2+3s_{24}^2\right)\right)
						\log\left|\frac{s_{23}}{s_{24}}\right|
						-\left(s_{23}^{2}+3s_{24}^{2}\right)
						\log^{2}\left|\frac{s_{23}}{s_{24}}\right|,
						\nonumber \\
						\mathring{\mathcal{B}}_{11}^{(\beta)} &=
						-\left(4s_{24}^2+\pi^2\left(s_{23}^2+3s_{34}^2\right)\right)
						+8s_{24}s_{34} \, 
						\log\left|\frac{s_{23}}{s_{34}}\right|
						-\left(s_{23}^2+3s_{34}^2\right)
						\log^{2}\left|\frac{s_{23}}{s_{34}}\right|,
						\nonumber\\
						\mathring{\mathcal{B}}_{11}^{(\gamma)} &=
						-2i \pi s_{23}^2
						\log\left|\frac{s_{24}}{s_{34}}\right| 
						- s_{23}^2
						\log^2\left|\frac{s_{24}}{s_{34}}\right|,
						\nonumber\\
						\mathring{\mathcal{B}}_{12}^{(\alpha)} &=
						-2\left(s_{34}^2+2 i \pi \left(s_{23}^2-s_{24}^2\right)\right) 
						+4\left(s_{23}^{2}-s_{24}^{2}-2 i\pi\,s_{23}s_{24}\right)
						\log\left|\frac{s_{23}}{s_{24}}\right|
						+4s_{23}s_{24}
						\log^{2}\left|\frac{s_{23}}{s_{24}}\right|,
					\nonumber	\\
						\mathring{\mathcal{B}}_{12}^{(\beta)} &=
						\pi^2 s_{23}s_{24}
						+2s_{24}^2
						\log\left|\frac{s_{23}}{s_{34}}\right|
						+s_{23}s_{24}
						\log^{2}\left|\frac{s_{23}}{s_{34}}\right|,
						\\
						\mathring{\mathcal{B}}_{12}^{(\gamma)} &=
						2 i \pi s_{23}^2
						+2s_{23}	\left(s_{23}+ i\pi\,s_{24}\right)
						\log\left|\frac{s_{24}}{s_{34}}\right|
						+s_{23}s_{24}
						\log^{2}\left|\frac{s_{24}}{s_{34}}\right|,
					\nonumber	\\
						\mathring{\mathcal{B}}_{13}^{(\alpha)} &=
						2i\pi s_{34}^2
						-2s_{34}\left(s_{34}+i \pi s_{23}\right)
						\log\left|\frac{s_{23}}{s_{24}}\right|
						-s_{23}s_{34}
						\log^{2}\left|\frac{s_{23}}{s_{24}}\right|,
					\nonumber	\\
						\mathring{\mathcal{B}}_{13}^{(\beta)} &=
						4\left(s_{24}^2-\pi^2 s_{23}s_{34}\right)
						+4s_{24}\left(s_{23}-s_{34}\right)
						\log\left|\frac{s_{23}}{s_{34}}\right|
						-4s_{23}s_{34}
						\log^{2}\left|\frac{s_{23}}{s_{34}}\right|,
					\nonumber	\\
						\mathring{\mathcal{B}}_{13}^{(\gamma)} &=
						2i\pi s_{23}^2
						+2s_{23}\left(s_{23}-i\pi s_{34}\right)
						\log\left|\frac{s_{24}}{s_{34}}\right|
						-s_{23}s_{34}
						\log^{2}\left|\frac{s_{24}}{s_{34}}\right|,
					\nonumber	\\
						\mathring{\mathcal{B}}_{22}^{(\alpha)} &=
						2 i \pi s_{34}^2
						-2s_{34}\left(s_{34}-i\pi s_{23}\right)
						\log\left|\frac{s_{23}}{s_{24}}\right|
						-s_{23}s_{34}
						\log^{2}\left|\frac{s_{23}}{s_{24}}\right|,
					\nonumber	\\
						\mathring{\mathcal{B}}_{22}^{(\beta)} &=
						4\left(s_{24}^2-\pi^2s_{23}s_{34}\right)
						+4s_{24}\left(s_{23}-s_{34}\right)
						\log\left|\frac{s_{23}}{s_{34}}\right|
						-4s_{23}s_{34}
						\log^{2}\left|\frac{s_{23}}{s_{34}}\right|,
					\nonumber	\\
						\mathring{\mathcal{B}}_{22}^{(\gamma)} &=
						2 i \pi s_{23}^2
						+ 2s_{23}\left(s_{23}-i\pi s_{34}\right)
						\log\left|\frac{s_{24}}{s_{34}}\right|
						-s_{23}s_{34}
						\log^{2}\left|\frac{s_{24}}{s_{34}}\right|,
					\nonumber	\\
						\mathring{\mathcal{B}}_{23}^{(\alpha)} &=
						2 i \pi s_{34}^2
						+2s_{34}\left(-s_{34} - i \pi s_{24}\right)
						\log\left|\frac{s_{23}}{s_{24}}\right|
						+s_{24}s_{34}
						\log^{2}\left|\frac{s_{23}}{s_{24}}\right|,
					\nonumber	\\
						\mathring{\mathcal{B}}_{23}^{(\beta)} &=
						\pi^2 s_{24}s_{34}
						-2s_{24}^2
						\log\left|\frac{s_{23}}{s_{34}}\right|
						+s_{24}s_{34}
						\log^{2}\left|\frac{s_{23}}{s_{34}}\right|,
					\nonumber	\\
						\mathring{\mathcal{B}}_{23}^{(\gamma)} &=
						-4s_{23} \left(s_{23} - i \pi \left(s_{34}-s_{24}\right)\right)
						+4\left(s_{23}(s_{34}-s_{24})+2 i\pi\,s_{24}s_{34}\right)
						\log\left|\frac{s_{24}}{s_{34}}\right|
						+4s_{24}s_{34}
						\log^{2}\left|\frac{s_{24}}{s_{34}}\right|,
					\nonumber	\\
						\mathring{\mathcal{B}}_{33}^{(\alpha)} &=
						2i \pi s_{34}^2
						\log\left|\frac{s_{23}}{s_{24}}\right|
						-s_{34}^2
						\log^{2}\left|\frac{s_{23}}{s_{24}}\right|,
					\nonumber	\\
						\mathring{\mathcal{B}}_{33}^{(\beta)} &=
						-2\left(2s_{24}^2+\pi^2(3s_{23}^2+s_{34}^2)\right)
						-8s_{23}s_{24}
						\log\left|\frac{s_{23}}{s_{34}}\right|
						-(3s_{23}^2+s_{34}^2)
						\log^{2}\left|\frac{s_{23}}{s_{34}}\right|,
					\nonumber	\\
						\mathring{\mathcal{B}}_{33}^{(\gamma)} &=
						-4 s_{23}\left(s_{23}+ 2 i \pi s_{24}\right)
						-2\left(4s_{23}s_{24}+ i\pi \left(3s_{24}^2+s_{34}^{2}\right)\right)
						\log\left|\frac{s_{24}}{s_{34}}\right|
						-\left(3s_{24}^{2}+s_{34}^{2}\right)
						\log^{2}\left|\frac{s_{24}}{s_{34}}\right|. \nonumber
					\end{align}
				
				Ideally we would like to express our result involving bilinears in the $C^m$'s in terms of the six combinations
				\bea
				t_8 (R_1 R_2 R_3 R_4) & = & t_8 F^4\, (F_1 F_2 F_3 F_4)\ , \label{t8R4} \\
				t_8 (R_1 R_2) (R_3 R_4) & = & t_8 F^4\, (F_1 F_2) (F_3 F_4)\quad {\rm and} \ {\rm cyclic}(2,3,4)\ , \label{t8R22}
				\eea
				as was done for the one-loop graviton amplitude in ${\cal N} = 4$ supergravity in \cite{Bern:2017tuc} (obtained via an asymmetric double copy therein). In \eqref{t8R4} and \eqref{t8R22} we used the notation 
				\bea
				(F_i F_j F_k F_l) & = & F_i^{\mu \nu} F_{j\, \nu \rho} F_k^{\rho \sigma} F_{l\, \sigma \mu}\ , \\
				(F_i F_j) (F_k F_l) & = & F_i^{\mu \nu} F_{j\, \mu \nu} F_k^{\rho \sigma} F_{l\, \rho \sigma}\ .
				\eea
				Moreover, the authors of \cite{Bern:2017tuc} use the definition
				\be
				t_8 F^4 = 2 (F_1 F_2 F_3 F_4) - 2 (F_1 F_2) (F_3 F_4) + {\rm cyclic}(2,3,4)\ .
				\ee
				Following \cite{Bern:2017tuc}, we also use the notation
				\bea
				&& F^4_{tu} = (F_1 F_4 F_2 F_3)\ , \quad F^4_{us} = (F_1 F_3 F_4 F_2)\ , \quad F^4_{st} = (F_1 F_2 F_3 F_4)\ , \\[4pt]
				&& (F^2_t)^2 = (F_1 F_4) (F_2 F_3)\ , \quad (F^2_u)^2 = (F_1 F_3) (F_4 F_2)\ , \quad (F^2_s)^2 = (F_1 F_2) (F_3 F_4)\ .
				\eea
				Linear combinations of \eqref{t8R4} and \eqref{t8R22} correspond to the generators $G_{{\bf 3}, 1}$ and $G_{{\bf 3}, 5}$ of the parity even part of the local graviton S-matrix module in $D=4$ analysed in section 6.5.1 of \cite{Chowdhury:2019kaq}. As discussed there, these generators are not all independent. Moreover, there is another generator (called $G_{{\bf S}, 2}$) which, however, is of higher order in momenta and does not appear in the one-loop graviton 4-point amplitude in ${\cal N} = 4$ supergravity discussed in \cite{Bern:2017tuc}. We did not manage to rewrite our result using \eqref{t8R4} and \eqref{t8R22} yet but in a first step we found two relations involving the $C^m$s that we give in appendix \ref{app:Cmrelations}.

			\section{Outlook}
			In essence, this paper is about connecting a few existing amplitude techniques used in field theory and string theory, and implementing
			them in such a way that they can at least in principle be automated. 
			We have merely scratched the surface of this topic, so there are many directions left unexplored.
			
			Perhaps the most glaring omission is that of helicity amplitudes. It would be straightforward to plug in the
			helicity expressions for the Berends-Giele currents from \cite{paper2} here. But doing only this would be incomplete:
			to take our method seriously, one should recalculate amplitudes in a massive helicity formalism, like \cite{Arkani-Hamed:2017jhn},
			and then take the soft/collinear limit. 
			
			On a related note, we intentionally structured this work so it will apply straightforwardly to six dimensions,
			where the hexagon amplitude is a priori the generating function. (We say ``a priori'', because here it
			was the box, but we got away with using the triangle, since for all-mass, it is related to the box.)
			The massless helicity formalism in $D=6$ \cite{Cheung:2009dc,Bern:2010qa} 
			is related but not identical to some massive helicity formalisms in $D=4$ , so this could usefully be connected
			to the previous question.
			
			Clearly there is much left to do about details of KLT relations at one-loop. The  amplitudes we study here
			do not obviously show double-copy structure after integration, but see e.g.\ \cite{Stieberger:2022lss,Stieberger:2023nol}
			for an argument that this should be possible to make manifest.
			
			One aspect of amplitudes that was important in this work is the sub-hierarchy of graph topologies
			for each topology (``bubbly box'', etc.). It would be interesting to learn whether techniques like in \cite{Hidding:2022ycg}
			to represent this graphically could make this more manifest.
			
			As mentioned in section \ref{SCET}, there is by now an extensive literature on single-valued polylogarithms and generalizations thereof,
			the work most closely related to that here being	 \cite{Schlotterer:2025qjv}. We have nothing new to say about it here,
			but we believe that our work could help connect these powerful mathematical methods
			to more conventional Feynman integral techniques used in particle physics. 
			
			Then there is the purely mathematical side of amplitudes. In \cite{brown2024generalised}, Brown discusses Feynman integrals from the point of view of cohomology. He effectively also uses all-mass objects, like here. It would be interesting to understand this connection better.
			
			\section*{Acknowledgments}
			
			We thank Paolo Di Vecchia, Hofie Hannesdottir, Henrik Johansson, Julio Parra-Martinez, Oliver Schlotterer, Matt Schwartz 
			and Kostja Zarembo for useful discussions. M.B.\ thanks the ASC in Munich and we all
			thank the GGI in Florence and Nordita in Stockholm for hospitality during part of this work. 
			The work of M.H.\ and Y.Z.\ is supported by the Origins Excellence Cluster in Munich. Y.Z.\ was supported by a Minerva Fellowship of the Minerva Stiftung Gesellschaft f\"ur die Forschung mbH.

				\appendix
				
				\section{Berends-Giele currents at one-loop} \label{app:bg}
				For the purposes of this paper, a sufficient motivation for introducing the kinematic factors $C_{i|jkl}$, $C^m_{i|jkl}$ and $C^{mn}_{i|jkl}$
				is that they compactly represent our amplitude of interest. However, here we provide a little more detail
				on the background, for readers that are not familiar with Berends-Giele currents. We hope this can entice such readers to study the original papers,
				where motivation for the structure of these objects is provided from resummation of Feynman diagrams in field theory,
				and from operator product expansions (OPEs) and BRST invariance in string theory.

				Linearized gauge transformations take polarizations $e_i^m$ into momenta $k_i^m$:
				\begin{equation}
					\delta e_i^m = k_i^m \ .
					\label{BG11a}
				\end{equation}
				The gauge-invariant linearized field strength in terms of the polarization vector $e^m$ is:
				\begin{equation}
					f_i^{mn} \equiv k_{i}^m e_{i}^n - k_{i}^n e_{i}^m \co \delta f_i^{mn} =0 \ .
					\label{BG11b}
				\end{equation}
				The purpose of this appendix is to illustrate how to construct multiparticle generalizations of the gauge-invariant linearized field strength \eqref{BG11b}.
				The strategy of Berends-Giele  \cite{Berends:1987me} is to allow one off-shell leg and weaken the demand of invariance under gauge transformations
				to only covariance. Also, following BCJ, we want to weaken the requirement of 
				no negative powers of Mandelstams, i.e.\ allow some nonlocal-looking expressions
				in intermediate steps, but only in judiciously chosen places. Then in the final step, we demand that gauge variations
				and negative powers of certain Mandelstams
				cancel, and assemble physical, gauge-invariant kinematic factors $C_{i|jkl}$, $C^m_{i|jkl}$ and $C^{mn}_{i|jkl}$ for one-loop amplitudes.\footnote{As discussed in \cite{paper1}, $C^{mn}_{i|jkl}$ is only  gauge-pseudo-invariant in $D=6$. Here we focus our discussion on $D=4$. Moreover, with ``physical'' we mean that the only allowed poles in Mandelstam variables are those which appear in a local theory, i.e.\ a theory based on a local Lagrangian. For instance, $C_{1|234}$ in \eqref{polC} below is equal to the color ordered tree level 4-point partial gluon amplitude and consists of the sum of contact terms and exchange terms with simple poles. Hence, all terms in the bracket in \eqref{polC} contain at least one power of a Mandelstam variable, cf.\ e.g.\ (4.27) in \cite{paper1}. \label{pseudo}}
				
				\subsection{Local multiparticle polarizations}
				\label{sect21}
				
\begin{figure}[htbp]
					\begin{center}
						\includegraphics[scale=0.4]{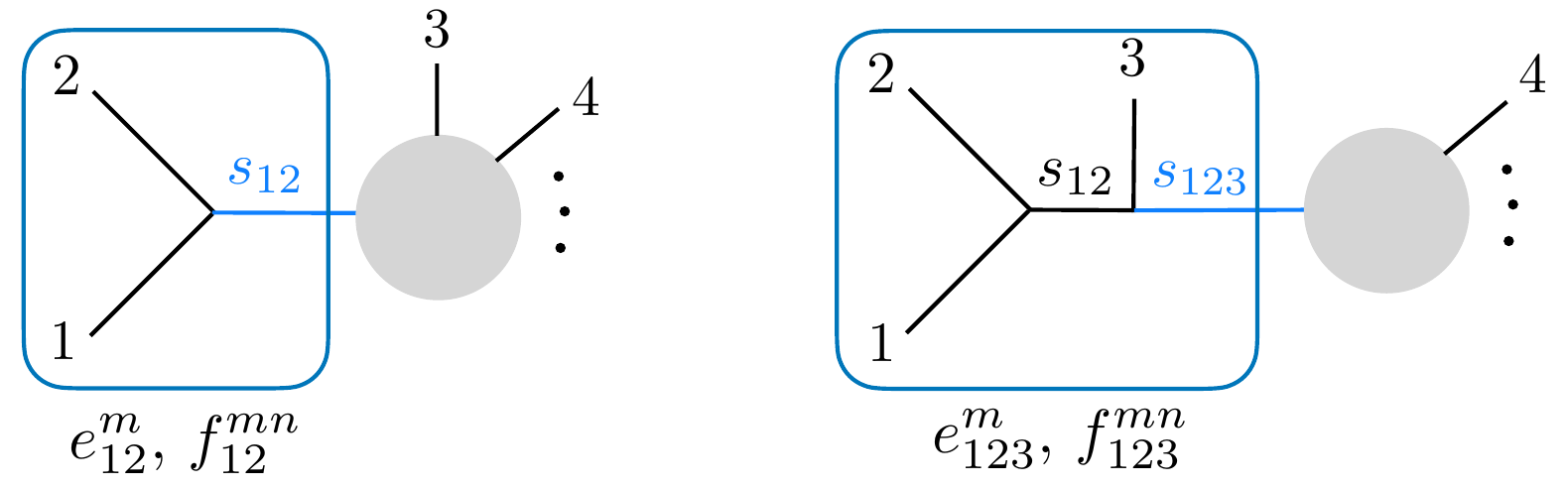}
						\caption{The cubic-vertex subdiagrams in the rounded boxes are represented by local multiparticle polarizations $e^{m}_{12},f^{mn}_{12}$ and $e^{m}_{123},f^{mn}_{123}$, respectively. They each have an off-shell leg (in blue).}
						\label{f:loc}
					\end{center}
				\end{figure}

				To construct kinematic expressions that factorize correctly onto tree-level diagrams, 
				we start from the opposite direction: attach tree-level diagrams to a given one-loop diagram, as in figure \ref{f:loc}.
				
				To do so, we define local 2- and 3-particle generalizations of polarizations and field-strengths. With the convention 
				\begin{equation}
					s_{12}= k_1\cdot k_2 \co
					s_{12\ldots p} = \frac{1}{2} (k_{12\ldots p})^2 \co
					k_{12\ldots p}^m \equiv k_1^m + k_2^m +\ldots+k_p^m \ ,
					\label{mand}
				\end{equation}
				we have the 2-particle polarization and field-strength
				\begin{align}
					e^{m}_{12}  &\equiv  e_2^m (k_2\cdot e_1) -  e_1^m (k_1\cdot e_2) +\frac{1}{2}(k_1^m - k_2^m) (e_1 \cdot e_2)\; ,
					\label{BG31} \\
					f^{mn}_{12}&\equiv k_{12}^m e_{12}^n - k_{12}^n e_{12}^m
					-s_{12}\big( e_1^m e_2^n - e_1^n e_2^m \big)  \; . 
					\label{BG32}
				\end{align}
				We will use them to relate the factorization limit of a $4$-point amplitude on a 2-particle channel $\sim (k_i{+}k_j)^{-2}$ 
				to a $3$-point amplitude with one gluon polarization replaced by $e^m_{ij}$. 
				
				The next step is to factorize to a $2$-point amplitude on a
				3-particle  channel $\sim (k_i{+}k_j{+}k_l)^{-2}$, which should produce the 3-particle polarization and field-strength
				\begin{align}
					e^{m}_{123}  &\equiv  e_3^m (k_3\cdot e_{12}) -  e_{12}^m (k_{12}\cdot e_3) +\frac{k_{12}^m - k_{3}^m}{2} (e_{12} \cdot e_3) + \frac{s_{12}}{2} \big( e_2^m (e_1\cdot e_3) -  e_1^m (e_2\cdot e_3) \big)\ ,
					\label{BG33} \\
					f^{mn}_{123}&\equiv k_{123}^m e_{123}^n - k_{123}^n e_{123}^m
					-(s_{13}\!+\!s_{23})\big( e_{12}^m e_3^n - e_{12}^n e_3^m \big)  -
					s_{12} \big( e_1^m e_{23}^n- e_1^n e_{23}^m
					-(1\!\leftrightarrow\! 2) \big) \ .
					\label{BG34}
				\end{align}
				
				Gauge covariance in this context means
				that the variations of an $n$-particle polarization closes into $(n-1)$-particle polarizations and lower, but not higher,
				as well as $n$-particle momenta and lower. 
				For example, $\delta e_{12}$ generates $k_{12}$ and $s_{12}$, but not  $e_{12}$. Explicit expressions are given in \cite{paper2}, section 2. 
				
				Note that so far, all objects are strictly local:
				no negative powers of Mandelstam variables.

				\subsection{Berends--Giele currents}
				\label{sect22}
				So far, we temporarily gave up gauge invariance and only required gauge covariance. Now, we also give up manifest locality,
				by allowing negative powers of Mandelstam variables $s_{ij}$ and $s_{ijk}$. A key step is that we even allow negative powers of those Mandelstam variables that are zero on-shell, but only those required by minahaning, here $s_{ijk}\neq 0$, but $s_{ijkl}=0$. 
				
\begin{figure}[htbp]
					\begin{center}
						\includegraphics[scale=0.4]{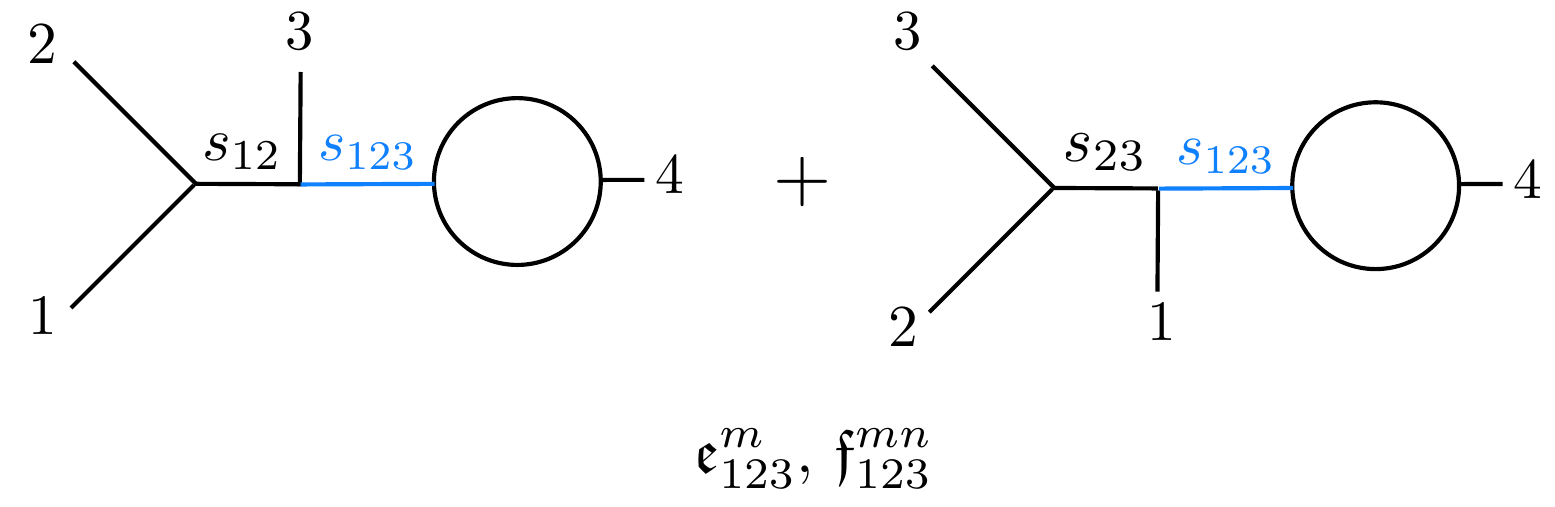}
						\caption{Berends--Giele currents $\efrak^m_{123}$ and $\ffrak^{mn}_{123}$ combine multiparticle 
							polarizations and propagators  to reproduce the cubic-vertex subdiagrams in a 
							$4$-point one-loop amplitude with an off-shell leg $4$.}
						\label{f:bg}
					\end{center}
				\end{figure}

				These gauge-covariant objects with negative powers of Mandelstam variables  are the  Berends--Giele currents \cite{Berends:1987me}. We use the $\mathfrak{Fraktur}$ typeface to distinguish them from the local multiparticle polarizations $e^m_{12}$, $f^{mn}_{12}, \ldots$ \ in eqs.\ (\ref{BG31}) to (\ref{BG34}):
				\begin{align}
					\efrak_{12}^m &\equiv \frac{e_{12}^{m}}{s_{12}} \co \ffrak_{12}^{mn} \equiv \frac{f_{12}^{mn}}{s_{12}} \ , \notag \\
					\efrak_{123}^m &\equiv \frac{e_{123}^{m}}{s_{12} s_{123}}+\frac{e_{321}^{m}}{s_{23} s_{123}} \ ,
					\label{BG81}  \\
					\ffrak_{123}^{mn} &\equiv \frac{f_{123}^{mn}}{s_{12} s_{123}}+\frac{f_{321}^{mn}}{s_{23} s_{123}} \ . \notag
				\end{align}
				As one can see from the 3-particle instances $\efrak_{123}^m$ and $\ffrak_{123}^{mn}$, the cubic graphs in figure \ref{f:loc} are combined according to a color-ordered 4-point amplitude with one off-shell leg, see figure \ref{f:bg}.
				
				Here is a compact recursion formula from which the 2- and 3-particle examples (\ref{BG81}) follow \cite{Berends:1987me, Mafra:2015vca}
				\begin{align}
					\efrak^{m}_{P}  &=  \frac{1}{2\,s_P} \sum_{XY=P}  \bigl[ \efrak_{Y}^m (k^Y\cdot  \efrak^{X})
					+ \efrak^{Y}_n  \ffrak_X^{mn}
					- (X \leftrightarrow Y)\bigr] 
					\label{BG01} \\
					\ffrak^{mn}_P &= k_P^m \efrak_P^n - k_P^n \efrak_P^m
					- \sum_{XY=P}\!\!\big( \efrak_X^m \efrak_Y^n - \efrak_X^n \efrak_Y^m \big)  \ ,
					\label{BG02}
				\end{align}
				with the multiparticle label $P=12\ldots p$ and initial conditions $\efrak^{m}_{1}\equiv e^m_1$, $\ffrak^{mn}_{1}\equiv f^{mn}_1$
				for the recursion.  The instruction $XY=P$ on the sum means to deconcatenate (i.e.\ split up) the word $P = 12\ldots p$ 
				into two non-empty words $X = 12\ldots j$ and $Y=j+1\ldots p$ with $j=1,2,\ldots,p-1$. As an example:
				\be
				\sum_{XY=1234} \efrak_X^m \efrak_Y^n = \efrak_1^m \efrak_{234}^n+\efrak_{12}^m \efrak_{34}^n+\efrak_{123}^m \efrak_4^n \; . 
				\ee

				\subsection{Parity-even one-loop building blocks}
				\label{sectappBG}
				
\begin{figure}[htbp]
					\begin{center}
						\includegraphics[scale=0.4]{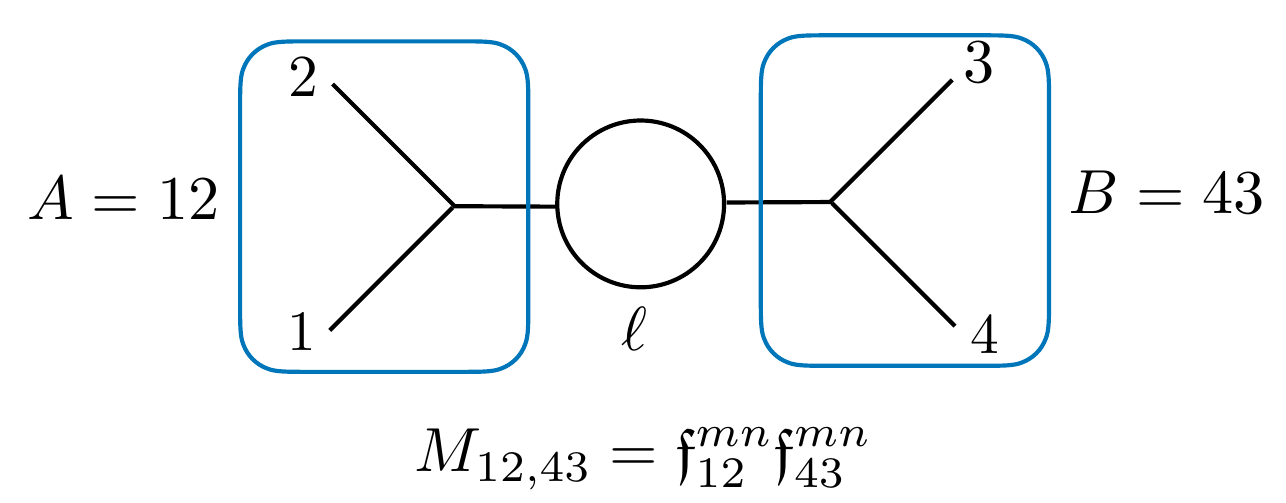}
						\caption{4-point diagram with kinematic numerator $M_{12,43}$, and $\ell$ the loop momentum.}
						\label{f:bub}
					\end{center}
				\end{figure}

				We now use the Berends-Giele currents  $\efrak^m_A$  and $\ffrak^{mn}_A$ from the previous subsection to
				construct local and gauge-invariant building blocks for one-loop amplitudes.
				These building blocks can be thought of as pieces of numerators in Feynman integrals. As motivated in string theory \cite{paper1}, we
				begin with bubble diagrams $\ffrak^{mn} \ffrak^{mn}$ as in figure \ref{f:bub}, and continue as follows:
				\begin{align}
					M_{A,B} &=  -\frac{1}{2} \ffrak^{mn}_A\ffrak^{mn}_B\ , \notag\\
					M^m_{A|B,C} &=  \efrak^m_A M_{B,C} +  (A\leftrightarrow B,C)\ , \\
					M^{mn}_{A|B,C,D} & = \big[  \efrak_B^{n} M^{m}_{A|C,D} + (B\leftrightarrow C,D) \big]+ \efrak_A^{n} M^{m}_{B,C,D}   \ ,
					\notag
				\end{align}
				where the relative coefficient is fixed by the string amplitude \cite{paper1}. The reason we need vector and 2-tensor 
				objects, but not for example 3-tensor objects, is that in Feynman-integral numerators, we have contractions with at most two powers of the loop momentum $\ell^m$,
				since we are computing a four-point amplitude.
				
				Finally, we define sums of permutations of these  combinations of Berends-Giele currents that are actually gauge invariant. 
				
				Scalar invariants:
				\begin{align}
					C_{1|2} &\equiv M_{1,2}\ ,
					\label{BG2pt}
					\\
					C_{1|23} &\equiv M_{1,23} + M_{12,3}  - M_{13,2} \ ,
					\label{BG17}
					\\
					C_{1|234} &\equiv M_{1,234} + M_{123,4}  + M_{412,3} + M_{341,2} + M_{12,34} + M_{41,23}   \; . 
					\label{BG19} 
				\end{align}
				Note that \eqref{BG19} contains three types of building blocks: $M_{A,B}$ with $A$ having length 1,2 and 3. 
				
				Vector invariants:\footnote{for maximal supersymmetry, cf.\  the BRST invariants defined in section 5 of \cite{Mafra:2014oia}.}
				\begin{align}
					C^m_{1|2,3} &\equiv M^m_{1|2,3} + k_2^m M_{12,3} + k_3^m M_{13,2}\ ,
					\label{BG18}
					\\
					C^m_{1|23,4} &\equiv M^m_{1|23,4} + M^m_{12|3,4}  - M^m_{13|2,4}  - k_2^m M_{132,4} + k_3^m M_{123,4} - k_4^m (M_{41,23} + M_{412,3} - M_{413,2}) \; . 
					\label{BG20}
				\end{align}
				A tensor invariant:\footnote{cf.\  (3.14) of \cite{Mafra:2014gsa}}
				\begin{align}
					C^{mn}_{1|2,3,4} &\equiv M^{mn}_{1|2,3,4} +  2\big[ k_2^{(m} M^{n)}_{12|3,4}+(2\leftrightarrow 3,4) \big] - 2 \big[ k_2^{(m} k_3^{n)} M_{213,4}+ (23\leftrightarrow 24,34)\big] \; .
					\label{BGtens}
				\end{align}
				We show in the auxiliary repository that these Berends-Giele currents (or more specifically, 
				combinations of Berends-Giele currents) are gauge-invariant (in $D=4$, cf.\ footnote \ref{pseudo}):
				\begin{equation}
					\delta C_{1|A}=0 \co \delta C^m_{1|A,B}=0
					\co
					\delta C^{mn}_{1|2,3,4} =0
					\label{BG21var}
				\end{equation}
				using momentum conservation for the vectors and the tensor. 
				
				One way to 
				build intuition for the above sums of permutations is in terms of how the amplitudes arise from permutations of vertex operators. 
				(In this appendix, we are not making direct use of string theory, but plenty of indirect use.)
				In particular, worldsheet functions  like $X_{ij,k}$
				that appear as coefficients of the above kinematic factors $C_{1|jkl}$
				have permutation-type symmetries for vertex positions, such as Fay identities. Each index of $C_{1|jkl}$ is obviously correlated with 
				the same index on $X_{ij,k}$, since they arise from the same vertex operator. 
				When the vertex position dependence in $X_{ij,k}$ is all integrated
				over, there must still remain a ``memory'' of the symmetries of $X_{ij,k}$, now transferred to functions of the Mandelstam variables that are contained in $X_{ij,k}$. Let us discuss the original symmetries of $X_{ij,k}$ and $C_{1|jkl}$  in a little more detail in the following three paragraphs.
				
				The basic properties of $X_{ij,k}$ and $C_{1|ijk}$ are that $X_{ij,k}$ is antisymmetric in $ij$ and $C_{1|ijk}$ enjoys reflection symmetry due to $A_{\rm tree}(1,3,2,4)=A_{\rm tree}(4,2,3,1)$. The  $X_{ij,k}$ and $C_{1|jkl}$ also enjoy  ``cycle-to-zero'' three-term 
				symmetries $X_{23,4} + X_{34,2} + X_{42,3} = 0$ and $C_{1|234} + C_{1|342} + C_{1|423}$ = 0
				that can be viewed as ``dual'' to each other\footnote{Note that the ``cycle to zero'' property for $A_{\rm tree}$ is the {\it photon decoupling identity} (see e.g.\ \cite{ElvangHuang}).}.
				When combined, these symmetries allow us to implement full
				permutation invariance in 2,3,4 on just {\it two} terms $X_{23,4}C_{1|234} + X_{24,3}C_{1|243}$, cf.\ eq.\ \eqref{cs_integrand_lines}.				
				
				Here is a quick reminder of two elementary ways to show why only two terms are needed. The obvious way is to implement the symmetries as rules
				acting on the numbers $2$, $3$, $4$ (recall we always have leg 1 fixed), i.e.\ a three-dimensional representation of the symmetric group $S_3$. One then implements
				the above symmetries to show by explicit computation that we only need the quoted two terms to
				have $S_3$ symmetry (with all six generators).

				The inconvenience of the three-dimensional representation of $S_3$ is that it is reducible: the sum $1+2+3$ is automatically invariant.
				Another way is to go to the irreducible standard representation, which is 2-dimensional. In this representation, it becomes clear that $X$ is represented as
				a dual  ${\bf 2}^*$, and ${\bf 2}^* \otimes {\bf 2} = {\bf 1}\mbox{ (trivial)} \oplus {\bf 1}'\mbox{ (alternating)}\oplus {\bf 2}\mbox{ (standard)}$,
				i.e.\ there is a singlet, which corresponds to the given invariant combination of two terms.

				Finally, in addition to all the above, one can construct another scalar invariant as follows:
				\begin{equation}
					P_{1|2|3,4} \equiv e_{2m} e^m_1 M_{3,4}  + \frac{1}{2} \big[ (e_2 \cdot e_3) M_{1,4} + (3\leftrightarrow 4) \big]+ k_2^m M^m_{12|3,4} + s_{23} M_{123,4} + s_{24} M_{124,3} \ .
					\label{BG24}
				\end{equation}
				It is invariant as well:
				\begin{equation}
					\delta P_{1|2|3,4} =0 \ .
					\label{BG24var}
				\end{equation}
				
				To connect this to perhaps more familiar Lorentz traces over linearized field strengths $f_{i}^{mn} \equiv k_i^m e_i^n - k_i^n e_i^m$,
				\begin{align}
					t(1,2) & \equiv
					(e_1\cdot k_2) (e_2 \cdot k_1) - (e_1\cdot e_2) (k_1\cdot k_2)\ ,
					\label{red28} \\
					t(1,2,3,4) &\equiv (e_1 \cdot k_2) (e_2\cdot k_3)(e_3 \cdot k_4)(e_4\cdot k_1)  - \te{antisymmetrization in all} \ (k_j\leftrightarrow e_j) \ , \notag
				\end{align}
				we obtain \cite{paper1}
				\begin{align}
					C_{1|234} &=    \frac{1}{s_{12} s_{23}} \, \Big[   t(1,2) t(3,4)  - t(1,2,3,4) + {\rm cyc}(2,3,4) \Big]\ ,
					\label{polC}
					\\
					P_{1|2|34} &=  \frac{1}{s_{12}} \, \Big[ t(1,3)t(2,4) + t(1,4)t(2,3) - t(1,2)t(3,4) - t(1,4,2,3) \Big] \ .
					\label{polP}
				\end{align}
				This shows that the scalar invariant $C_{1|234}$
				is in fact proportional to the $t_8$ tensor, familiar from maximal supersymmetry:
				\begin{equation}
					t_8(1,2,3,4) \equiv \frac{1}{2} \, \Big[
					t(1,2) t(3,4) -  t(1,2,3,4) + {\rm cyc}(2,3,4) \Big] 
					= \frac{1}{2} \, s_{12} s_{23} C_{1|234}  \ ,
					\label{t8}
				\end{equation}
				but note that as emphasized above, a proportionality factor of Mandelstam variables affects symmetry properties. 
				
				The additional scalar invariant   $P_{1|2|34}$  of (\ref{polP}) is independent of the $t_8$ tensor, and not present in the corresponding amplitude in maximal supersymmetry,
				as it only preserves 8 supercharges, see e.g.\  (4.8) and (4.9) of \cite{Johansson:2014zca}. Above,the
				$P_{1|2|34}$ in \eqref{BG24} was less systematically constructed than $C_{1|234}$, $C^m_{1|234}$ and $C^{mn}_{1|234}$ , so this helps explain where it comes from.

				This concludes our brief review of how to construct
				the local, gauge-invariant kinematic factors $C_{1|234}$, $C^m_{1|234}$ and $C^{mn}_{1|234}$ for one-loop amplitudes from Berends-Giele currents.
				A few comments:\\[2mm]
				\noindent $\bullet$ We used momentum conservation to show 
				that the vector $\delta C^m_{1|A,B}=0$ and tensor $\delta C^{mn}_{1|2,3,4} =0$. In the context of minahaning in quarter-maximal supersymmetry (which we do not consider in this paper) it could be useful to relax momentum conservation in intermediate steps, like we did for their scalar counterparts. \\
				\noindent $\bullet$ In $D=4$, as we study here, there are no parity-odd building blocks. In $D=6$, for example, they contribute,
				and then $P_{1|2|3,4}$ is only invariant up to anomaly terms (see \cite{paper2}, section 2).\\
				\noindent $\bullet$ Several of the construction steps can be improved. For example, 
				the OPE and BRST properties for lower supersymmetry have not been explored in detail. \\
				\noindent $\bullet$  As mentioned in the main text,
				it is useful to connect this discussion to that of the most general basis for the $S$-matrix.
				
			\subsection{Identities}\label{sec:BCJ_ids}
				Scalar identities:
				\begin{equation}
					\begin{aligned}
						s_{23}s_{34}C_{1|234} &= 2\, t_8 \, ,
						\\[4pt]
						s_{23}s_{34}C_{1|234} &= s_{12}P_{1|2|3,4}+s_{13}P_{1|3|2,4}+s_{14}P_{1|4|3,2} \, ,
						\\[4pt]
						s_{23}C_{1|234} &= s_{24}C_{1|243} \, ,
						\\[4pt]
						s_{34}C_{1|234} &= s_{24}C_{1|324} \, ,
						\\[4pt]
						P_{1|2|3,4} &= P_{1|2|4,3} \, .
					\end{aligned}
				\end{equation}
				The photon decoupling identity:
				\begin{equation}
					C_{1|234} + C_{1|342}+C_{1|423}  = 0 \ .
				\end{equation}
				Vector currents identity:
				\be \label{Cmpermute}
				C^{m}_{1\mid 23,4} = - C^{m}_{1\mid 32,4}\ .
				\ee
				One can find the validation of these identities in \href{https://github.com/YonatanZimmerman/Field-theory-limit-of-half-maximal-string-loop-amplitudes/blob/main/Berends%20Giele%20BCJ%20Identities.nb}{\faGithub}.
				\section{Details of field theory limit}
				
				\subsection{Skinny limit} \label{sec:skinny_limit}
				
				As $\tho(z,\tau)$ has a simple zero if and only if $z \in \Lambda \equiv \mathbb Z \oplus \tau \mathbb Z$ (see, e.g., \cite{Polchinski1, paper1, OSthpole}), it follows that
				\begin{equation} 
					\frac{\tho'(0)}{\tho(z)} = \frac{\partial \,[c (z - \Lambda)]|_{z \in \Lambda}}{c (z - \Lambda)} + \mathcal O [(z - \Lambda)]= \frac{1}{z - \Lambda} + \mathcal O [(z - \Lambda)] 
				\end{equation}
				for $z\ra \Lambda$, by which we mean $z\ra w$ for $w\in \Lambda$. Accordingly, and due to the quasi-periodicity of $\tho$, the Kronecker-Eisenstein series in \eqref{red11} reduces to
				\begin{equation}
					\Omega(z,\alpha, \tau) \SKa \lambda(z,\tau) \, \frac{1}{z - \Lambda} + \mathcal O [(z - \Lambda)] \ ,
				\end{equation}
				where 
				\begin{equation} 
					\lambda(z,\tau) = \exp [\pi i (n-m-m^2\tau)] \ ,
				\end{equation} 
				for $z=n+m\tau \in \Lambda$, and thus bounded. Since there is no dependence on $\alpha$ in the skinny limit, we can conclude that only $f^{(1)}$ admits a pole, and it is a simple pole on the lattice $\Lambda$. Notice that for the often-used representative $z=0 \in \Lambda$, $\lambda(0,\tau) = 1$.
				
				Next, we consider the skinny-limit implications for the Green's function. Based on the discussion regarding the zeros of $\tho$, we can immediately obtain the ratio that appears in \eqref{G_closed}
				\begin{equation}
					\frac{\tho(z)}{\tho'(0)} = z-\Lambda + \mathcal O [(z-\Lambda)^{3}] \ .
				\end{equation}
				In turn, the Green's functions \eqref{G_closed} reduce to
				\begin{equation} \begin{aligned} \label{G_SK}
						G(z) & \SKa  \frac12 \alpha' \Big \{ \ln |z-\Lambda|^2 - 2\pi \, \frac{(m \, \IM \, \tau)^2}{\IM \, \tau} \Big \} \\
						& = \ln |z-\Lambda|^{\alpha'} - \pi \alpha' \, m^2 \, \IM \, \tau \ ,
				\end{aligned} \end{equation} 
				where, again $z \ra n+m\tau \in \Lambda$. Subsequently, the relevant exponentials appearing in the Koba-Nielsen factor \eqref{KNdef} reduce to
				\begin{equation}
					\exp [s\, G(z)] \SKa |z-\Lambda|^{\alpha' s} \, e^{-\pi \alpha'  m^2 s\, \IM \tau} \ .
				\end{equation}

				Adhering to the permutation rules noted immediately after the classification of regions (i)-(iii), cf.\ the beginning of sec.\ \ref{sec_3col}, the possible quartet combinations for cases (ii) and (iii) are obtained by permuting \eqref{3col_contribution1}, resulting in
				\begin{equation} \begin{aligned} \label{3col_contribution2}
						|X_{24}|^2 \ \text{along with} \ \ |X_{23}|^2 \, , \, |X_{43}|^2 \, , \, X_{23}\bX_{43} \, \text{or} \, X_{43}\bX_{23} \ 
				\end{aligned} \end{equation} 
				and
				\begin{equation} \label{3col_contribution3}
					|X_{34}|^2 \ \text{along with} \ \ |X_{32}|^2 \, , \, |X_{42}|^2 \, , \, X_{32}\bX_{42} \, \text{or} \, X_{42}\bX_{32} \ ,
				\end{equation}
				respectively.
				
				The first calculation reproduces the first part of eq. (6.21) in \cite{paper1}:
				\begin{equation} \begin{aligned} 
						X_{23,4}\bX_{23,4} & = X_{23}[X_{24}+X_{34}] \, \bX_{23}[\bX_{24}+\bX_{34}] \\
						& \SKa \frac{1}{s_{234}} \times  \Big \{ \\
						& \qquad + s_{23} \, \{ (s_{24}+ s_{34})^2 \} \, , \quad \text{for} \ \text{(i)} \\
						& \qquad + s_{24} \, \{ s_{23}^2 \} \, , \quad \quad \qquad \text{for} \ \text{(ii)} \\
						& \qquad + s_{34} \, \{ s_{23}^2 \} \Big \} \, , \quad \qquad \text{for}  \ \text{(iii)} \\
						& = \frac{1}{s_{234}}  \{ s_{23}(s_{24}+s_{34}) s_{234} \} \\
						& = s_{23}(s_{24}+s_{34}) \ ,
				\end{aligned} \end{equation} 
				where we used $s_{234} = s_{23} + s_{24} + s_{34}$, a factor of $\frac{16 \pi^2}{\alpha'^2}$ was omitted but not forgotten and $\SKa$ is understood to be under integration. Similarly, the other squared combination,
				\begin{equation}
					X_{24,3}\bX_{24,3} \SKa s_{24}(s_{23}+s_{34}) \ ,
				\end{equation}
				reproduces the other part of eq. (6.21) in \cite{paper1}, remembering that $X_{ij}=-X_{ji}$. The remaining combinations are also straightforwardly calculated:
				\begin{equation} \begin{aligned}
						X_{23,4}\bX_{24,3} & = X_{23}[X_{24}+X_{34}] \, \bX_{24}[\bX_{23}+\bX_{43}] \\
						& \SKa \frac{1}{s_{234}} \times  \Big \{ \\
						& \qquad + s_{23} \, \{ s_{24}(s_{24}+s_{34}) \} \, , \quad \text{for} \ \text{(i)} \\
						& \qquad + s_{24} \, \{ s_{23}(s_{23}+s_{34}) \} \, , \quad  \text{for} \ \text{(ii)} \\
						& \qquad - s_{34} \, \{ s_{23}s_{24} \} \Big \} \, , \ \ \qquad \quad \text{for}  \ \text{(iii)} \\
						& = s_{23}s_{24} \ ,
				\end{aligned} \end{equation} 
				and it is the same result for the mirrored combination $\bX_{23,4}X_{24,3}$.
				\subsection{Field theory limit of worldline functions}\label{sec:WL_functions_limit}
				
				Most of the components affected by the field theory limit depend on $\tho$, which, assuming the conventions of \cite{Polchinski1}, eqs. (7.2.37d) and (7.2.38d) therein, can be expressed by
				\begin{equation} \begin{aligned}
						\vartheta_1 (z,\tau) & = i \sum_{n= - \infty}^{\infty} (-1)^n\,  q^{(n-\half)^2/2} \,  \zeta^{n- \half}\\
						&= 2 q^{\frac{1}{8}}\, \sin \, \pi z \prod_{m=1}^{\infty} (1-q^m)(1-\zeta q^m)(1-\zeta^{-1}q^m) \ ,
				\end{aligned} \end{equation} 
				where $q = e^{2\pi i \tau}$ and $\zeta = e^{2\pi i z}$. Since $q = e^{2\pi i \RE \tau} e^{-2\pi \IM \tau}$, and since we are examining the collapse of the worldsheet to a worldline, i.e., $\tau_2= \IM \tau \ra \infty$, we are interested in the $q \ra 0$ expansion:
				\begin{equation} \label{tho_ft_limit}
					\vartheta_1 (z,\tau) \FTa 2 q^{\frac{1}{8}}\, \sin \, \pi z + \mathcal O [q] \ ,
				\end{equation}
				where this ``limit" is slightly misleading as it is only partial, suppressing the limit of the argument $z = \RE(z) + i \tau_2 \nu$. Therefore, the calculations of the limits below will be carried out in two steps: first plugging in \eqref{tho_ft_limit} and later performing $\tau_2 \ra \infty$ once more.
				
				The limit of the bosonic Green's functions, given in eq. \eqref{G_closed}, is therefore
				\begin{equation} \begin{aligned}
						G(z,\tau) & \FTa  \frac12 \alpha' \Big \{ \log \Big |\frac{\sin \, \pi z}{\pi} \Big |^2 - \frac{2 \pi}{\IM \, \tau} \IM ^2 (z) \Big \} \\
						& =  \frac12 \alpha' \Big \{ \log \Big [ \frac{\sin ^2(\pi x) \cosh ^2 (\pi \tau_2 \nu)}{\pi^2 } +  \frac{\cos ^2(\pi x) \sinh ^2 (\pi \tau_2 \nu)}{\pi^2 } \Big ] -2 \pi \tau_2 \,\nu^2\Big \}  \\
						& =  \frac12 \alpha' \Big \{ \log [e^{2\pi \tau_2 |\nu|}] - \log[\pi^2] + \log[\sin^2x \pm \cos^2 x] -2 \pi \tau_2 \,\nu^2\Big \}  \\
						&  \FTa - \pi t (\nu^2 - |\nu|) \ ,
				\end{aligned} \end{equation} 
				where $x \equiv \RE(z)$. 
				
				Next we move to the worldsheet functions. The limit of $f^{(1)}$, defined in eq. \eqref{red12}, is
				\begin{equation} \begin{aligned}
						f^{(1)}(z,\tau) &\FTa \pi \, \text{cot} \, \pi z  + 2\pi i \frac{\IM z}{\tau_2} \\
						&= -\frac{\pi \sin(2\pi x)}{\cos (2\pi x) - \cosh (2\pi \tau_2 \nu)} + i\pi \frac{\sinh(2\pi \tau_2 \nu)}{\cos (2\pi x) - \cosh (2\pi \tau_2 \nu)} + 2\pi i \nu \\
						& = -i\pi \, \text{sgn}(\nu)  + \frac {\pi \, \sin(2\pi \, x) - i \pi \text{sgn} (\nu) \cos(2\pi \, x) }{e^{2 \pi \tau_2 \nu}} + 2\pi i \nu + \mathcal O[(q^{\nu})^2] \\
						&  =  2\pi i \Big [ \nu - \frac{1}{2} \, \text{sgn}(\nu) \Big ]+ \mathcal O[q^{\nu}] \ ,
				\end{aligned} \end{equation} 
				noting that the ratio $\frac{\sinh}{\cosh}$ is the signum function of the argument in the $\tau_2 \ra \infty$ limit. Similarly, $f^{(2)}$ and $F_{1/2}^{(2)}$ in \eqref{red13} and \eqref{F2_def} reduce to
				\begin{equation} \begin{aligned}
						f^{(2)}(z,\tau) &\FTa \half \{ (\pi \, \text{cot} \, \pi z  + 2\pi i \nu)^2  - \frac{\pi^2}{\sin^2 (\pi \, z)} + \frac{\pi^2}{3} \} \\
						&  =  (2\pi i)^2 \Big [ \half \nu^2 - \half \nu \, \text{sgn}(\nu)  + \frac{1}{12} \Big ]+ \mathcal O[(q^{\nu})^2] \ ,
				\end{aligned} \end{equation} 
				and
				\begin{equation} \begin{aligned}
						F_{1/2}^{(2)} (kv,\tau)  &\FTa \half \{ (\pi \, \text{cot} ( \pi kv)  + 2\pi i \nu)^2  - \frac{\pi^2}{\sin^2 (\pi \, kv)} + \frac{\pi^2}{3} \} \\
						&  =  \pi^2 \Big [ \frac{1}{3} - \ \frac{1}{\sin^2(\pi k v)} \Big ]+ \mathcal O[(q^{\nu})^2] \ .
				\end{aligned} \end{equation} 

				\subsection{Identities} \label{app:FT_limit_id}
				
				In this appendix we shall prove the following integral identities that are used in section \ref{sec_2col}:
				\begin{equation} \label{col_id2}
					\int \dd^2 z \, |z|^{\esm -1} \, g(z) = 0 + \mathcal O[\esm] \ ,
				\end{equation}
				and
				\begin{equation} \label{col_id3}
					\int \dd^2 z \, |z|^{\esm -2} \, g(z) = \frac{4\pi}{\esm} \, g(0)+ \mathcal O[\esm^0] \ ,
				\end{equation}
				for functions $g(z)$ that are regular at the origin.\footnote{\label{double_int_conventions_2} We remind the reader of our conventions $\int \dd^2z = i\int \dd z \dd \bar{z} =2 \int \dd(\IM z) \, \dd(\RE z)$.}
				
				Consider a $z=0$ singularity $|z|^{-2}$ that comes from some vertex collision(s)  and that $g(z)$ is smooth at $z=0$.  The first observation is that \eqref{col_id3} cannot be true if $\tfrac{1}{|z|^{2-\esm}}$ is viewed as a smooth function in the usual sense: integrals give antiderivatives evaluated on integral endpoints, not the unintegrated function in the middle at a single point, like $z=0$. One way to proceed is to multiply by $\esm/\pi$ and write out explicitly that we are considering $\esm\rightarrow 0$:
				\begin{equation} \label{M_id2}
					\lim_{\esm\rightarrow 0}\int \dd^2 z  { \frac{\esm }{\pi}}{\frac{1}{|z|^{2-\esm}}}
					g(z) =4g(0) \; . 
				\end{equation}
				This is supposed to look familiar:\footnote{The $\bar{z}$ in, e.g. $\delta^{(2)}(z,\bar{z})$, is omitted. We do not follow the convention that writing $f(z)$ implies holomorphy of $f$.} 
				\begin{equation}
					\int \dd^2 z\,  \delta^{(2)}(z)g(z)=g(0) \; .
				\end{equation}
				Going back to \eqref{M_id2}, one is tempted to move the limit into the integral, but then the integrand would vanish. So the only hope of being nonzero is to claim an ``order of limits'' situation: the integration is singular, as opposed to smooth. It is an order of limits problem for ordinary Riemann integration, since it involves a limit of rectangles with widths that become smaller than any variation of the function being integrated. We have to give up that assumption here, i.e. in principle allow variation that even a very narrow rectangle can't resolve, but still generalize to these ``singular integrals'' in a way that preserves features of integration that we like.
				
				One way to express now what needs to be done is to separate the domain of integration into a small disk $D_{\delta}$ of radius $\delta$ and ``the rest'', where the rest had better be suitably negligible, if the answer is to depend only on the value of $g$ at $z=0$. We can think of it as a way to make the order of limits in \eqref{M_id2} explicit, in that the operation of performing the integral depends on $\delta$. Set $z=re^{i\theta}$, then $\dd^2 z=2r \dd r \dd \theta$, and use that $g$ is smooth at $z=0$,
                \begin{eqnarray}\label{M_id3}
						\lim_{\esm\rightarrow 0}\int_{D_{\delta}} \dd^2 z  { \frac{\esm }{\pi}}{\frac{1}{|z|^{2-\esm}}}
						g(z) &=&\lim_{\esm\rightarrow 0}\int_0^{\delta}\!dr \int_0^{2\pi}\! d\theta\,  { 2\esm \over \pi}{1 \over r^{1-\esm}}\Big(g(0)+r e^{i\theta} \partial g(0) + r e^{-i\theta} \bar \partial g(0) + \ldots \Big) \nonumber \\
						&=&\lim_{\esm\rightarrow 0} {2\cdot 2\pi \over \pi}\left[r^{\esm}g(0)+ \mathcal{O}(r^{\esm + 2}) + \ldots\right]_0^{\delta}=4g(0) \;, 
				\end{eqnarray}
				if we set $\delta^{\esm}=1$, i.e\ take $\esm\rightarrow 0$ first. A similar example is discussed in Appendix H in \cite{Berg:2014ama}, where ``proportional to'' is cleverly written, so we can't read off the overall factor.
				
				In ch.12, equation (12.6.11) of \cite{Polchinski2} we have a similar expression to \eqref{col_id3}, but with the holomorphic $1/z^2$ instead of $1/|z|^2$ (because Polchinski's example is in the heterotic string):
				\begin{equation}
					I_{\rm Polchinski} = \int_{|{\rm Im}\, z|<\delta} \dd^2 z {1 \over z^2}|z|^{2\esm}
				\end{equation}
				where $\esm=k\cdot k'$. Note that this is {\it not} a cut-out disk $D_{\delta}$,
				but a cut-out horizontal strip $S_{\delta}$,  in the torus variable. Note also
				that at this point, he explicitly distinguishes between $\esm$ in the integrand and $\delta$
				in the domain. 
				
				Instead of the cut-out strip in Polchinski, let us do the same calculation but with a small disk around the origin. Take the radius of this disk to be $\delta$, independent of $\esm$, as Polchinski does in (12.6.11). Use the divergence theorem, eq.\ (2.1.9) in Polchinski.

				As above, only the leading constant term in a Taylor-expansion of $g$ contributes, i.e.
				\be \label{M_taylor}
				\int_{D_\delta} \dd^2 z \frac{1}{|z|^{2-\esm}} g(z) = \int_{D_\delta} \dd^2 z \frac{1}{|z|^{2-\esm}} (g(0) + z \partial g(0) + \bar z \bar \partial g(0)+ \ldots ) = \int_{D_\delta} \dd^2 z \frac{1}{|z|^{2-\esm}} g(0)\ .
				\ee
				All the higher terms in the expansion either vanish via angular integration or in the limit $\delta \rightarrow 0$, as the integrand for the higher terms involving only $|z|$ is regular for $\esm > 0$. 
				
				Proceed as \cite{Polchinski2} (12.6.12):
				\be \label{M_step2}
				\frac{1}{|z|^{2-\esm}} = \frac{2}{\esm} \partial_z \Big( z^{\esm / 2} \bar z^{-1 + \esm / 2} \Big)\ .
				\ee
				With this \eqref{M_taylor} becomes
				\be
				\int_{D_\delta} \dd^2 z \frac{1}{|z|^{2-\esm}} g(0) = \int_{D_\delta} \dd^2 z \frac{2}{\esm} \partial_z \Big( z^{\esm / 2} \bar z^{-1 + \esm / 2} \Big) g(0) \ .
				\ee
				This can be rewritten using the divergence theorem, eq.(2.1.9) in \cite{Polchinski2}, leading to
				\be
				\int_{D_\delta} \dd^2 z \frac{1}{|z|^{2-\esm}} g(0) = \frac{2}{\esm} g(0) i \oint_{\partial D_\delta} \dd \bar z z^{\esm / 2} \bar z^{-1 + \esm / 2} = \frac{2}{\esm} g(0) i \oint_{\partial D_\delta} \frac{\dd \bar z}{\bar z} |z|^{\esm} \ .
				\ee
				In polar coordinates 
				\be \label{M_polar1}
				z = r e^{i \theta}\ , \quad \bar z = r e^{-i \theta}\ , 
				\ee
				the measure becomes
				\be \label{M_polar2}
				\frac{\dd \bar z}{\bar z} = \dd \ln \bar z \stackrel{r = \delta}{=} - i \dd \theta\ ,
				\ee
				and we obtain
				\be
				\int_{D_\delta} \dd^2 z \frac{1}{|z|^{2-\esm}} g(0) = \frac{2}{\esm} g(0) i (-i) \delta^{\esm} \int_0^{2 \pi} \dd \theta = \frac{4 \pi}{\esm} g(0) \delta^{\esm} = \frac{4 \pi}{\esm} g(0) + {\cal O}[\esm^0]\ ,
				\ee
				which confirms \eqref{col_id3}.

				Finally, we move to prove eq.\eqref{col_id2}. Consider a $z=0$ singularity $|z|^{\esm -1} $ for a small positive $\esm$ and a smooth $g(z)$ at the origin. We follow the steps of the previous proof with the exchange $2-\esm \ra 1-\esm$. Similarly to \eqref{M_step2},
				\be 
				\frac{1}{|z|^{1-\esm}} = \frac{2}{\esm +1} \partial_z \Big( z^{\esm / 2 + 1/2} \bar z^{ \esm / 2 -1/2} \Big)\ .
				\ee
				Hence,
				\begin{equation}
					\begin{aligned}
						\int_{D_\delta} \dd^2 z \frac{1}{|z|^{1-\esm}} g(0) &= \int_{D_\delta} \dd^2 z \frac{2}{\esm +1} \partial_z \Big( z^{\esm / 2 + 1/2} \bar z^{ \esm / 2 -1/2} \Big) g(0) \\ & ~\\
						&= \frac{2}{\esm+1} g(0) i \oint_{\partial D_\delta} \dd \bar z z^{\esm / 2 + 1/2} \bar z^{ \esm / 2 -1/2} \\ &~ \\
						& = \frac{2}{\esm+1} g(0) i \oint_{\partial D_\delta} \frac{\dd \bar z}{\bar z} |z|^{\esm+1} \ ,
					\end{aligned}
				\end{equation}
				by the divergence theorem. Applying the polar coordinates transformation of eqs.\eqref{M_polar1} and \eqref{M_polar2}, we get
				\be
				\int_{D_\delta} \dd^2 z \frac{1}{|z|^{1-\esm}} g(0) = \frac{4 \pi}{\esm+1} g(0) \delta^{\esm+1} \stackrel{\esm \rightarrow 0}{\longrightarrow} 4 \pi g(0) \delta +  {\cal O}[\esm] \stackrel{\delta \rightarrow 0}{\longrightarrow} 0 + {\cal O}[\esm]\ ,
				\ee
				which confirms \eqref{col_id2}.
				
				\section{Brief review of all-mass three-point one-loop scalar amplitudes}
				
				``All-mass'' means three external masses (that we will interpret as virtualities, but here we call them ``mass'' as is conventional in this context) for the triangle. Here we only consider massless internal lines. In the main text we use the result given in (4.54) and (6.45) of \cite{Bourjaily:2020wvq}. In this appendix we comment on other representations in the literature for the 3-mass scalar triangle for massless scalars, in particular formulas (38) of \cite{Duplancic:2002dh}, (11)-(14) of \cite{Usyukina:1992jd} and (2.8) \& (2.11) of \cite{Davydychev:1992xr}. (B.9) \& (B.10) of \cite{Weinzierl:2022eaz} have two typos and we give the corrected result below. 
				
				Let us begin with the result of Duplancic and Nizic \cite{Duplancic:2002dh}. The result for
				\begin{equation} \label{I3DNint}
					I_3 = \int_0^1 \dd a_1 \dd a_2 \dd a_3 \frac{\delta (1-a_1-a_2-a_3)}{(a_1 a_2 p_1^2 + a_2 a_3 p_2^2 + a_1 a_3 p_3^2 + {\rm i} \efn)} 
				\end{equation}
				is given by 
				\begin{equation}
					I_3 = \frac{1}{p_2^2 (x_1 - x_2)} \left\{ 2 {\rm Li}_2 \left( \frac{1}{x_2} \right) - 2 {\rm Li}_2 \left( \frac{1}{x_1} \right) + \ln[x_1 x_2 + {\rm i} \efn {\rm sign}(p_2^2)] \left[ \ln \left( \frac{1-x_1}{-x_1} \right) - \ln \left( \frac{1-x_2}{-x_2} \right) \right] \right\} .  \label{I3DN}
				\end{equation}
				Here we have distinguished $p_2^2$, following the discussion in the appendix of \cite{Duplancic:2002dh}. In view of the permutation symmetry of \eqref{I3DNint}, this is an arbitrary choice. $x_1$ and $x_2$ are given by\footnote{Note that the terms under the square-root are given by $\Delta_3 / (p_2^2)^2$ in the notation of Weinzierl \cite{Weinzierl:2022eaz}, cf.\ \eqref{Delta3} below.}
				\begin{equation}
					x_{1,2} = \frac12 \left[ 1 - \frac{p_1^2}{p_2^2} +  \frac{p_3^2}{p_2^2} \pm \sqrt{\left( 1- \frac{p_1^2}{p_2^2} - \frac{p_3^2}{p_2^2} \right)^2 - 4 \frac{p_1^2}{p_2^2} \frac{p_3^2}{p_2^2}} \right] ,
				\end{equation}
				i.e.\ $x_1$ is defined with the plus and $x_2$ with the minus. 
				
				Next, the result of Davydychev \cite{Davydychev:1992xr} is given by
				\begin{equation}\label{I3D}
					\begin{aligned}
						I_3 & = \frac{1}{p_3^2 \lambda} \left\{ 2 \ln \left( \frac{1+x-y-\lambda}{2} \right) \ln \left( \frac{1-x+y-\lambda}{2} \right) -\ln x \ln y \right. 
						\\ 
						&  \left. \hspace{1.3cm} - 2 {\rm Li}_2 \left( \frac{1+x-y-\lambda}{2} \right) - 2 {\rm Li}_2 \left( \frac{1-x+y-\lambda}{2} \right) + \frac{\pi^2}{3} \right\} ,  
					\end{aligned} 
				\end{equation} 
				where 
				\begin{equation}
					x = \frac{p_1^2}{p_3^2}\ , \quad y = \frac{p_2^2}{p_3^2}\ , \quad \lambda = \sqrt{(1-x-y)^2 - 4 x y}\ .
				\end{equation}
				This is related by dilogarithm transformations to the result of \cite{Usyukina:1992jd}:
				\begin{equation} \label{I3UD}
					I_3 = \frac{1}{p_3^2 \lambda} \left( 2 {\rm Li}_2 ( - \rho x) + 2 {\rm Li}_2 ( - \rho y) + \ln(\rho x) \ln(\rho y) + \ln \frac{y}{x} \ln \frac{1 + \rho y}{1 + \rho x} + \frac{\pi^2}{3} \right) 
				\end{equation}
				with 
				\begin{equation}
					\rho = \frac{2}{1-x-y+\lambda}\ .
				\end{equation}
				In \eqref{I3D} and \eqref{I3UD} the ${\rm i} \efn$-prescription is left implicit. According to (14) of \cite{Usyukina:1992jd} a term $\ln \alpha$ (with $\alpha = \rho x$ or $\alpha = \rho y$) should be understood as $\ln(\alpha + {\rm i} \efn\, {\rm sign}(p_3^2))$, i.e.\ $\ln(\alpha - {\rm i} \efn)$ for negative $p_3^2$. This distinguishes $p_3^2$ and leads to agreement with \cite{Duplancic:2002dh}.
				
				Formula \eqref{I3UD} can be split into real and imaginary part according to \cite{Weinzierl:2022eaz}:\footnote{Note the typo in (B.9) of \cite{Weinzierl:2022eaz} in the penultimate term of the real part, cf.\ also the errata page \cite{Weinzierl_errata}.}
				\begin{equation} \begin{aligned}
						I_3 & = & {\rm Re}\left[ \frac{1}{p_3^2 \lambda} \left( 2 {\rm Li}_2 ( - \rho x) + 2 {\rm Li}_2 ( - \rho y) + \ln(\rho x) \ln(\rho y) + \ln \frac{y}{x} \ln \frac{1 + \rho y}{1 + \rho x} + \frac{\pi^2}{3} \right)  \right] \\
						&& - \frac{i \pi \theta(p_2^2)}{p_3^2 \lambda} \ln \left( \frac{(\delta_1 + \sqrt{\Delta_3}) (\delta_3 + \sqrt{\Delta_3})}{(\delta_1 - \sqrt{\Delta_3}) (\delta_3 - \sqrt{\Delta_3})} \right) \label{I3W}
				\end{aligned} \end{equation} 
				with
				\begin{equation} \begin{aligned}
						&& \delta_1 = p_1^2 - p_2^2 -p_3^2\ , \quad \delta_2 = p_2^2 - p_1^2 -p_3^2\ , \quad \delta_3 = p_3^2 - p_1^2 -p_2^2\ , \\
						&& \Delta_3 = (p_1^2)^2 + (p_2^2)^2 + (p_3^2)^2 - 2 p_1^2 p_2^2 - 2 p_1^2 p_3^2 - 2 p_2^2 p_3^2\ . \label{Delta3} 
				\end{aligned} \end{equation} 
				Here we adjusted the sign of the imaginary part such that it agrees with \cite{Duplancic:2002dh}. Note that the imaginary part can also be expressed via $\delta_2$, leading to 
				\begin{equation} \begin{aligned}
						I_3 & = & {\rm Re}\left[ \frac{1}{p_3^2 \lambda} \left( 2 {\rm Li}_2 ( - \rho x) + 2 {\rm Li}_2 ( - \rho y) + \ln(\rho x) \ln(\rho y) + \ln \frac{y}{x} \ln \frac{1 + \rho y}{1 + \rho x} + \frac{\pi^2}{3} \right)  \right] \\
						&& +\frac{i \pi \theta(p_2^2)}{p_3^2 \lambda} \ln \left( \frac{\delta_2 + \sqrt{\Delta_3}}{\delta_2 - \sqrt{\Delta_3}} \right) ,
				\end{aligned} \end{equation} 
				cf.\ (10) in \cite{Lu:1992ny}. This can also be verified numerically with mathematica. Weinzierl writes that \eqref{I3W} is valid for $\Delta_3 > 0$ if either $p_1^2, p_2^2, p_3^2<0$ or if $p_1^2, p_3^2<0$ and $p_2^2>0$.
				
				We make the following observations  plugging in explicit values for $p_i^2$ in mathematica: 
				\begin{itemize}
					\item Duplancic and Nizic claim that their result \eqref{I3DN} is only valid if $p_2^2$ either has the opposite sign of $p_1^2$ and $p_3^2$ (in which case $\Delta_3>0$  and one of $x_1$ and $x_2$ is smaller than 0 and the other is larger than 1 (both of them are real)) or if all three $p_i^2$ have the same sign and $p_2^2$ has the smallest absolute value (in which case $x_1$ and $x_2$ could be the complex conjugate of each other, they could both be smaller than 0 or both larger than 1). However, it turns out that in the second case, $p_2^2$ does not have to have the smallest absolute value. It is sufficient that neither $x_1$ nor $x_2$ is in the interval $[0,1]$. If that is the case, \eqref{I3DN} agrees with the other formulas like \eqref{I3UD}.
					\item If neither $x_1$ nor $x_2$ is in the interval $[0,1]$, it turns out that \eqref{I3W} gives the correct result (i.e.\ agrees with \eqref{I3DN}) even if $\Delta_3 < 0$ (and then $p_1^2, p_2^2, p_3^2$ necessarily have to have the same sign). In that case, $I_3$ is real and one could use either \eqref{I3W} or (B.7) of \cite{Weinzierl:2022eaz}, i.e.
					\begin{equation}
						I_3 = -\frac{2}{\sqrt{-\Delta_3}} \left[ 
						{\textstyle
							{\rm Cl}_2 \left( 2 \arctan \left( \frac{\sqrt{-\Delta_3}}{\delta_1} \right) \right) + {\rm Cl}_2 \left( 2 \arctan \left( \frac{\sqrt{-\Delta_3}}{\delta_2} \right) \right) + {\rm Cl}_2 \left( 2 \arctan \left( \frac{\sqrt{-\Delta_3}}{\delta_3} \right) \right)
						}
						\right] .
					\end{equation}
				\end{itemize}

				\section{BDK differentiation method} \label{app:BDK}
				
				In this section, we very briefly review the parts of \cite{Bern:1993kr} which are relevant for our calculations. As stated in section \ref{SCET}, after specifying to a particular position integration region, and transforming said position  to Hypercube variables, $\nu \leftrightarrow a$, we obtain in eq.~\eqref{eq:FT_general_box} a Feynman integral which is the specification of eq.~(2.5) in the reference
				\begin{equation} \label{eq:D4_nGon_BDK}
					I_n[P\left(\left\{a_i\right\}\right)] = \Gamma(n-2+\euv) \, \int_0^1 \dd^n a_i \, \delta {\textstyle \left(1-\sum_i a_i\right) }\, \frac{P\left(\left\{a_i\right\}\right)}{\left[\sum_{i,j}^{n} S_{ij}a_ia_j -i \efn\right]^{n-2+\euv}} \ ,
				\end{equation}
				to the box ($n\rightarrow4$) with numerator polynomial $P$. In this appendix we differ from the notation in section \ref{sec:BDK}, and have the general polynomial in field-theory kinematics $P$ and not $\hat{P}$. Applying a change of integration variables to projective ones
				\begin{equation}
					\begin{aligned}
						a_i &= \frac{\alpha_i \, u_i}{\sum_{j=1}^n \alpha_j \, u_j} \ ,
						\\[6pt]
						a_n & =   \frac{\alpha_i \left(1-\sum_{j=1}^{n-1} u_j\right)}{\sum_{j=1}^n \alpha_j \, u_j}  \ ,
					\end{aligned}
				\end{equation}
				see \cite{tHooft:1978jhc}, and introducing the \emph{reduced} integrals
				\begin{equation}
					\hat{I}_n[P\left(\left\{a_i\right\}\right)] = {\textstyle \left(\prod_{j=1}^{n} \alpha_j\right)^{-1} \, I_n\left[P\left(\left\{a_i \over \alpha_i \right\}\right)\right] } \ ,
				\end{equation}
				the above integral becomes
				\begin{equation} \label{eq:BDK_general_red_int}
					\hat{I}_n[P_m\left(\left\{a_i\right\}\right)] = \Gamma(n-2+\euv) \, \int_0^1 \dd^n u_i \, \delta {\textstyle \left(1-\sum_i u_i\right) }\, \frac{P_m\left(\left\{u_i\right\}\right)\, \left(\sum_{j=1}^{n}\alpha_j \, u_j\right)^{n-4-m+2\euv}}{\left[\sum_{i,j}^{n} \rho_{ij} u_i u_j -i \efn\right]^{n-2+\euv}} \ ,
				\end{equation}
				when the numerator polynomial is of homogeneous degree $m$. Additionally, in this step the $\alpha_i$ are defined such that the elements
				\begin{equation}
					\rho_{ij} = S_{ij}\,\alpha_i \alpha_j \ ,
				\end{equation}
				are independent of $\alpha_i$ (see eq.~\eqref{eq:alpha_choice} for our particular choice). It is then straightforward to see that one can obtain tensor integrals by differentiating the scalar generator:
				\begin{equation} \label{eq:BDK_diff}
					\hat{I}_n[P_m\left(\left\{a_i\right\}\right)] = \frac{\Gamma(n-3-m+2\euv)}{\Gamma(n-3+2\euv)}\, P_m\left(\left\{\frac{\partial}{\partial \alpha_i}\right\}\right)\, \hat{I}_n[1] \ .
				\end{equation}
				
				Furthermore, our amplitude requires calculating triangly and bubbly boxes, see eqs.~\eqref{eq:general_trBox} and \eqref{eq:general_bubBox}, respectively. Having calculated sufficiently high-rank tensor $n$-point integrals as above, this becomes algorithmically straightforward by the trick pointed out in section 3 of \cite{Bern:1993kr}. Simply redefining the dimension of the original $n$-point integral in eq.~\eqref{eq:D4_nGon_BDK} by $\euv \rightarrow \euv -1$, we get
				\begin{equation} \label{eq:D6_box}
					I_n^{D=6-2\euv} [P\left(\left\{a_i\right\}\right)]= \frac{\Gamma(n-3+\euv)}{\Gamma(n-2+\euv)} \, I_n {\textstyle \left[\left(\sum_{i,j}^{n} S_{ij}a_ia_j \right) \, P\left(\left\{a_i\right\}\right)\right] }\ ,
				\end{equation}
				and
				\begin{equation} \label{eq:D8_box}
					I_n^{D=8-2\euv}[P\left(\left\{a_i\right\}\right)] = \frac{\Gamma(n-4+\euv)}{\Gamma(n-2+\euv)} \, I_n {\textstyle \left[\left(\sum_{i,j}^{n} S_{ij}a_ia_j \right)^2 \, P\left(\left\{a_i\right\}\right)\right] }\ ,
				\end{equation}
				respectively.
				
				\subsection{Vector boxes as generators} \label{app:BDK_vec}
				
\begin{figure}
					\begin{center}
						\includegraphics[scale=0.2]{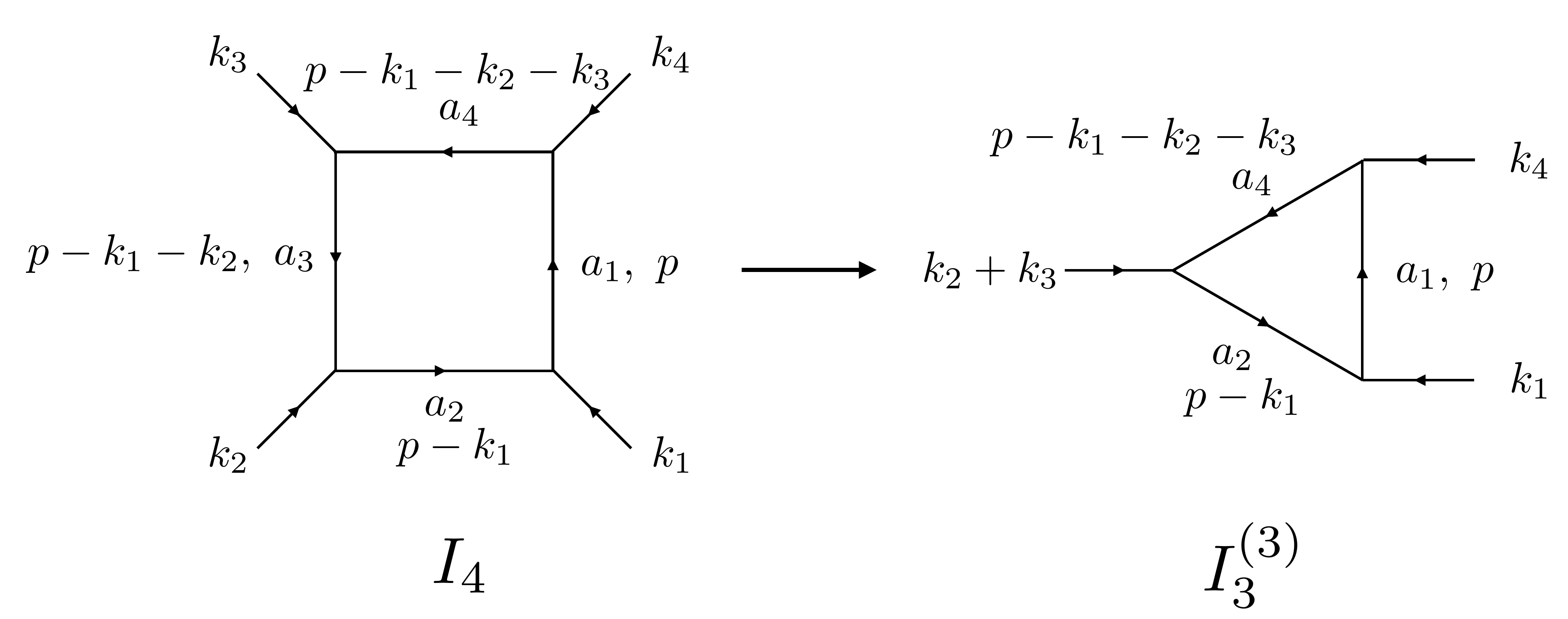}
						\caption{$I_3^{(3)}$ is obtained from $I_4$ via removing the propagator parametrized by $a_3$ (we chose $I_3^{(3)}$ exemplarily for concreteness). Similarly one can obtain $I_3^{(m)}$ for $m \neq 3$. $I_3^{(m)}$ should then be compared with $I_3$ of figure \ref{fig_I3} in order to read off which invariants of the four momenta appear in the $u$ and $v$ variables of the triangle.\label{fig:daughter_int}}
					\end{center}
				\end{figure}

\begin{figure}
					\begin{center}
						\includegraphics[scale=0.2]{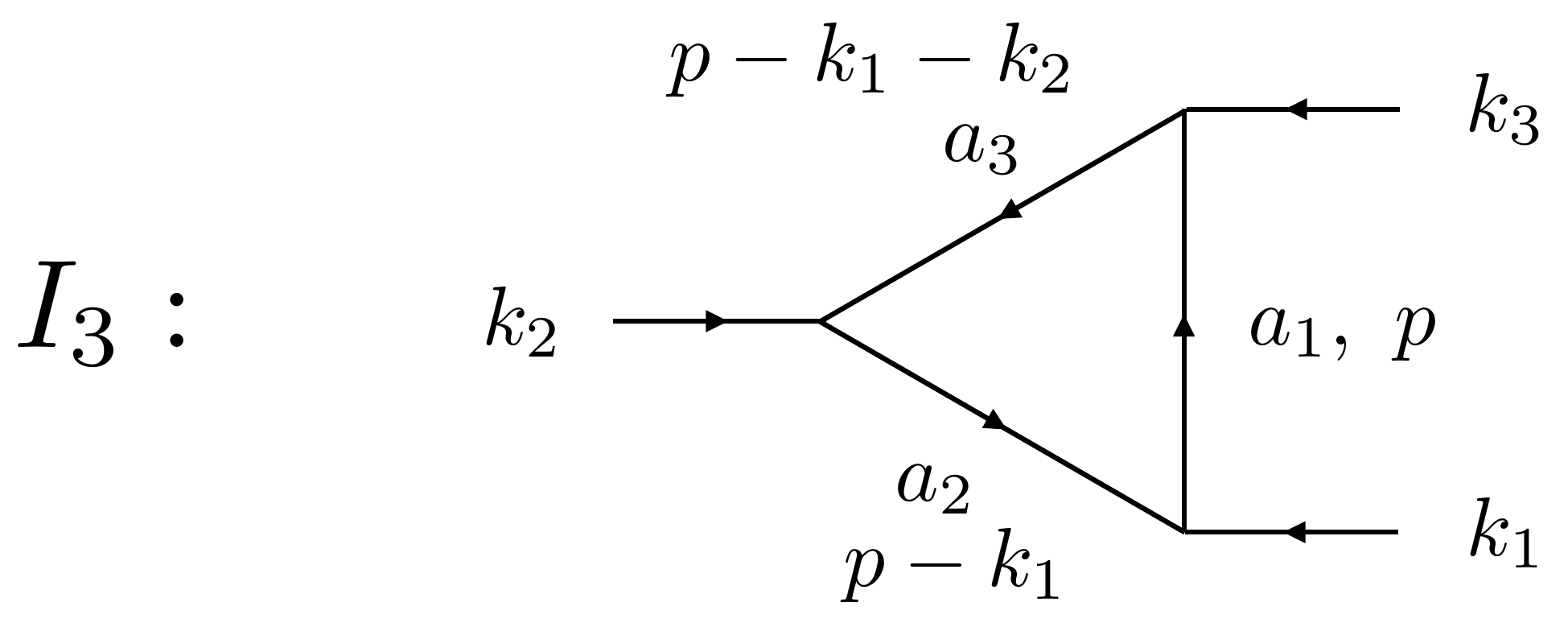}
						\caption{\label{fig_I3}}
					\end{center}
				\end{figure}

				As mentioned in section \ref{sec:BDK}, our specific case, where the box is not dimensionally regulated, i.e. $\euv \rightarrow 0$, excludes the use of the scalar box, $I_4[1]$, as our generator in eq.~\eqref{eq:BDK_diff}. However, we notice that one can alter this differentiating equation thus,
				\begin{equation} \label{eq:BDK_diff_vec}
					\hat{I}_n[P_m\left(\left\{a_j\right\}\right)\, a_i] = \frac{\Gamma(n-4-m+2\euv)}{\Gamma(n-4+2\euv)}\, P_m\left(\left\{\frac{\partial}{\partial \alpha_j}\right\}\right)\, \hat{I}_n[a_i] \ ,
				\end{equation}
				where the vector $n$-point integrals play the role of the generating functions. Conveniently, BDK give an algebraic prescription how to obtain just that in section 3 of \cite{Bern:1993kr}. Eq.~(3.16) there spells for the box:
				\begin{equation}
					\hat{I}_4[a_i] = \frac{1}{2N_4}\, \sum_{j=1}^{4}\left(\eta_{ij} -\frac{\gamma_i \gamma_j}{\hat{\Delta}_4}\right)\hat{I}_{3}^{(j)} 
					+ \frac{\gamma_i}{\hat{\Delta}_4} \hat{I}_4 [1] \ ,
				\end{equation}
				where
				\begin{equation}
					N_4 = 8\, {\rm Det} [\rho] = \frac12 \Big(u^2 + (v-1)^2 - 2 u (v+1)\Big) \ ,
				\end{equation}
				for
				\begin{equation}
					\rho = \frac12 \left( \begin{array}{cccc} 
						0 & u & 1 & 1 \\
						u & 0 & v & 1 \\
						1 & v & 0 & 1 \\
						1 & 1 & 1 & 0 
					\end{array} \right) \ ,
				\end{equation}
				that we have chosen in eq.~\eqref{eq:alpha_choice},
				\begin{equation}
					\eta = N_4\, \rho^{-1} \ ,
				\end{equation}
				\begin{equation}
					\hat \Delta_4 = \sum_{i,j=1}^4 \eta_{ij} \alpha_i \alpha_j \ ,
				\end{equation}
				\begin{equation}
					\gamma_i = \sum_{j=1}^4 \eta_{ij} \alpha_j \ ,
				\end{equation}
				and the ``daughter" integrals $\hat{I}^{(m)}_{n-1}$ correspond to removing the propagator between lines $(m-1)$ and $m$ from the original $n$-point integral, see figs. \ref{fig:daughter_int} and \ref{fig_I3}.
				
				\subsection{Warm-up calculations} 
				
				As a preliminary step to calculating our integrals using the differentiation method explained above, we looked at the massless box integrals in \cite{Tourkine:2012vx}, section IV. In particular, we completely recalculated the integral in eq.~(IV.17)\footnote{For eqs. (IV.15-17) to  be consistent with eqs.(IV.12-14) in \cite{Tourkine:2012vx}, we think that the prefix in eq.(IV.16) should  changed so: $\frac{t_8t_8\, R^4}{64} \ra \frac{t_8t_8\, R^4}{16}$.}.
				
				The procedure of calculating the box integral was similar to the one presented in section \ref{sec:ST_FT_correspondence} and earlier in this appendix: \\
				$(i)$ Splitting the integration region into its simplexes according to the prescription in \cite{Montag:1991hd}, eq.(8). \\
				$(ii)$ Changing of integration variables to "cube" them,  as in \cite{Montag:1991hd}, eq.(12). \\
				$(iii)$ Performing the integral by the BDK prescription \cite{Bern:1993kr}, section 2, namely the one recited in eq.~\eqref{eq:BDK_diff}, here specified to the box:
				\begin{equation}
					\hat{I}_4[P_m(\{a_i \})] = \frac{\Gamma(1-m+2\euv) }{\Gamma(1+2\euv)} \ P_m\left(\left\{ \frac{\partial}{\partial \alpha_i} \right\} \right) \hat{I}_4[1] \ ,
				\end{equation} 
				where here the reduced \emph{scalar} box generating function is given in \cite{Bern:1993kr}, eq~(4.22):
				\begin{equation} \label{I4hat1}
					\hat{I}_4[1] = \frac{2 r_{\Gamma}}{\euv^2} \ \left[(-\alpha_2 \alpha_4)^{\euv} \, {}_2 F_1\left(-\euv,-\euv; 1-\euv;1+\frac{\alpha_1\alpha_3}{\alpha_2\alpha_4} \right) +(-\alpha_1 \alpha_3)^{\euv} \, {}_2 F_1\left(-\euv,-\euv; 1-\euv;1+\frac{\alpha_2\alpha_4}{\alpha_1\alpha_3} \right)  \right] \ ,
				\end{equation}
				with $\alpha_1=\alpha_3= 1/\sqrt{-s}$, $\alpha_2=\alpha_4= 1/\sqrt{-t}$. For example, in the case of the simplex $0\leq \omega_1 \leq \omega_2 \leq \omega_3\leq 1$ of \cite{Tourkine:2012vx} (IV.17), the polynomial is $P_4 = a_3^2(a_1+a_2)^2$, and so
				\begin{equation}
					\hat{I}_4[a_3^2(a_1+a_2)^2] = \frac{\Gamma(-3+2\euv) }{\Gamma(1+2\euv)} \ \{ \partial_{\alpha_3}^2\, (\partial_{\alpha_1}+\partial_{\alpha_2})^2 \} \ \hat{I}_4[1] \ .
				\end{equation}
				\\
				$(iv)$ Taking the $\euv \ra 0$ limit with special attention to the hypergeometric functions, see \ref{app:Lerch}. In order to bring the results of the different integration regions in his limit to a similar for, we needed to use Pfaff transformation carefully, e.g.
				\begin{equation}  \label{pfaff}
					{}_2F_1(a,b;c;z) = (1-z)^{-a}\, {}_2F_1(a,c-b;c;\frac{z}{z-1}) \ .
				\end{equation} 
				Combining the expanded results from all the simplexes, we retrieved the result in \cite{Tourkine:2012vx} eq.~(IV.11).
				
				\subsubsection{Expanding hypergeometric representations} \label{app:Lerch}
				Hypergeometric representations are not necessarily the most convenient. But it is a standard language that symbolic manipulation software like Mathematica likes to use.
				\paragraph*{Polylogarithms and the Lerch function}
				
				Here is the Lerch function:
				\begin{equation}
					\Phi(z,s,\alpha)=\sum_{n=0}^{\infty}{z^n \over (n+\alpha)^s}
				\end{equation}
				Note that it starts at zero,
				so if we set $\alpha=1$ ($\euv=0$) it reduces to $1/z$ times a polylogarithm:
				\begin{equation}
					\Phi(z,s,1)=\sum_{n=0}^{\infty}{z^n \over (n+1)^s}=\sum_{n=1}^{\infty}{z^{n-1} \over n^s}={1 \over z} \, {\rm Li}_s(z) 
					\; . 
				\end{equation}
				To connect with e.g.\ \eqref{pfaff} above, note that one of the specific hypergeometric
				function of interest in the basic amplitude calculations has arguments $a=1$, $b=\euv$, $1+\euv$, so it is  a Lerch function:
				\begin{equation} \label{LerchF}
					{}_2F_1(1,\euv;1+\euv;z)
					\equiv\sum_{n=0}^{\infty} {(1)_n(\euv)_n \over (1+\euv)_n}{z^n \over n!}
					=\sum_{n=0}^{\infty}  {n!\, \euv  \over n+\euv} {z^n \over n!}
					=\sum_{n=0}^{\infty}  {\euv \over n+\euv} {z^n}=\euv \, \Phi(z,1,\euv)
				\end{equation}
				since these particular Pochhammer symbols combine:
				\begin{equation} \begin{aligned}
						(1)_n &= {\Gamma(1+n) \over \Gamma(1)} = n! \\
						(\euv)_n &= {\Gamma(\euv+n) \over \Gamma(\euv)} \\
						(1+\euv)_n &= {\Gamma(1+\euv+n) \over \Gamma(1+\euv)} 
						={(\euv+n)\Gamma(\euv+n) \over \euv \Gamma(\euv)}={n+\euv \over \euv}(\euv)_n \; . 
				\end{aligned} \end{equation} 
				There is an overall $1/\euv$ in the usual representation of the scalar box, so in fact
				$\Phi(z,1,\euv)$ does seem like a more natural way to express it
				than $(1/\euv)\cdot {}_2F_1(1,\euv,1+\euv;z)$. In particular,
				the middle argument of 
				$\Phi$ is redundantly equal to 1, but it can be deformed to $s\neq 1$ for zeta function regularization, that the hypergeometric function doesn't provide,
				since by definition it has only integer powers. 
				
				What 
				is the leading term of the hypergeometric function in the above form
				as  $\euv\rightarrow 0$? It is trivial to see that $\Phi(z,1,\euv)$ has a pole at $\euv=0$ of residue one:
				\begin{equation}
					\Phi(z,1,\euv) =\sum_{n=0}^{\infty}  {1 \over n+\euv} {z^n}=
					{1 \over \euv} + \sum_{n=1}^{\infty}  {1 \over n+\euv} {z^n} = {1 \over \euv} +  \mathcal{O}(\euv^0) \; ,
				\end{equation}
				so in fact ${}_2F_1(1,\euv,1+\euv;z)\rightarrow 1$ as $\euv \rightarrow 0$, which we will be able to use below.
				
				Here is a minor generalization: for any ${}_2F_1(a,b,c;z)$ with $a=1$ and $c=b+1$, there will be a representation
				in terms of $\Phi$.
				
				\paragraph*{Lerch as generating function of polylogs}
				It is easy to expand $\Phi$ in the third argument,
				since this follows directly from the sum representation:
				\begin{equation}
					{d \over d\alpha}\Phi(z,s,\alpha)=-s\Phi(z,s+1,\alpha)
				\end{equation}
				so the $4-2\euv$ expansion is
				\begin{equation}
					\Phi(z,s,1+\euv)=\Phi(z,s,1)-s\euv\Phi(z,s+1,1) + \ldots
					={1 \over z}\left({\rm Li}_{s}(z)-s\euv {\rm Li}_{s+1}(z)+\ldots\right)
				\end{equation}
				and in particular we want $s=1$, and we bring up the $z$ in front
				and find the simple result
				\begin{equation}  \label{Liexp}
					z\Phi(z,1,1+\euv)
					={\rm Li}_{1}(z)-\euv {\rm Li}_{2}(z)+\ldots
				\end{equation}
				so $z\Phi(z,1,1+\euv)$ generates polylogarithms. 
				
				Now note the general shift relation in the last argument:
				\begin{equation} \label{Phishift}
					\Phi(z,s,1+\alpha) = \sum_{n=0}^{\infty}{z^n \over (n+1+\alpha)^s}
					=\sum_{n=1}^{\infty}{z^{n-1} \over (n+\alpha)^s}={1 \over z}\left(\Phi(z,s,\alpha)-{1 \over \alpha^s}\right) \; ,
				\end{equation}
				where we explicitly subtracted the $n=0$ term in the definition of $\Phi$. Inverting, we see that
				\begin{equation}
					\Phi(z,s,\euv)=z\Phi(z,s,1+\euv)+{1 \over \euv^s}
				\end{equation}
				so the form we had in \eqref{LerchF} for
				the hypergeometric function ${}_2F_1(1,\euv,1+\euv;z)$ has the expansion
				\begin{equation} \begin{aligned}  \label{regexp}
						{}_2F_1(1,\euv;1+\euv;z)&=\euv \, \Phi(z,1,\euv)\\
						&=1+\euv z\Phi(z,1,1+\euv)\\
						&=1+\euv{\rm Li}_{1}(z)-\euv^2 {\rm Li}_{2}(z)+\ldots \ ,
				\end{aligned} \end{equation} 
				where we inserted the polylogarithm-generating expansion \eqref{Liexp} for $z\Phi(z,s,1+\euv)$.
				
				\subsection{\texorpdfstring{Triangles in $D=4$ - simplified differentiation method}{Triangles in 4D - simplified differentiation method}} \label{sec:simplified_BDK}
				In the triangle there are three Feynman parameters, call them $a_1, a_2$ and $a_3$. The tensor triangle 
				$I_3[a_1^m a_2^n]$ depends on the three virtualities $\pT_i^2$:
				\begin{equation} \label{tensor}
					I_3[a_1^m a_2^n](\pT_1^2,\pT_2^2,\pT_3^2) = \int_0^1 \int_0^1 \int_0^1 da_1 da_2 da_3 \; {a_1^m a_2^n \; \delta(a_1+a_2+a_3-1) \over {\mathcal F}}
				\end{equation}
				through the second Symanzik polynomial (in $D=4$, for now):
				\begin{equation}
					{\mathcal F} = -a_1 a_2 \pT_1^3 - a_2 a_3 \pT_2^2 - a_1 a_3 \pT_3^2 \; . 
				\end{equation}
				Here we assume that although the virtualities $p_i^2$ can be arbitrarily small, they are all nonzero. 
				Then all integrals considered here are finite, so we can freely move derivatives through integral signs.
				\begin{center}
					\fbox{\parbox{\textwidth}{
							{\bf Claim}: With $u = \pT_1^2/\pT_3^2$, $v =\pT_2^2/\pT_3^2$, the tensor triangles 
							are obtained from the scalar triangle as
							\be \label{claim}
							I_3[a_1^m a_2^n](u,v) ={(-1)^{m+n} \over (m+n)!} \left( u^m {\partial^m \over \partial u^m}\right)  \! \left( v^n {\partial^n \over \partial v^n}\right) 
							I_3[1](u,v) \; . 
							\ee}}
				\end{center}
				Here's the proof. 
				In  Duplancic-Nizic \cite{Duplancic:2002dh} eq.\ (25), we find the
				double Mellin representation of the all-mass {\bf  scalar}  triangle in $D=4$  (we suppress the Feynman  $i\efn$ for now):
				\begin{equation} \label{mellin}
					I_3[1](u,v) = {1 \over \pT_3^2}
					\int_{C_s} \! ds \int_{C_t} \! dt \; M^2(s,t) \, u^{-s} v^{-t}
				\end{equation}
				for Mellin contours $C_s$ and $C_t$, that are vertical contours in the complex $s$ and $t$ planes, respectively,
				with real parts between 0 and 1. 
				The Mellin kernel $M^2$ couples the Mellin variables $s$ and $t$:
				\begin{equation} \label{kernel}
					M^2(s,t) = \Big(\Gamma(s)\Gamma(t)\Gamma(1-(s+t))\Big)^2
				\end{equation}
				This Mellin kernel is a product of two equal factors, hence the name $M^2$.
				In  \cite{Duplancic:2002dh} eq.\ (25), one factor $M$ arises from a something that looks suspiciously like a string tree integral, but over the two remaining Feynman parameters (i.e.\ after using the delta function):
				\begin{equation} \label{tree}
					M(s,t) = \int_0^1 da_1 \; a_1^{s-1} (1-a_1)^{-s}\int_0^1 d\tilde{a}_2 \; \tilde{a}_2^{t-1} (1-\tilde{a}_2)^{-(s+t)}
					=\Gamma(s)\Gamma(t)\Gamma(1-(s+t)) \; . 
				\end{equation}
				The original Feynman parameter $a_2$ is related to $\tilde{a}_2$ as $a_2= (1-a_1)\tilde{a}_2$. 
				
				Now consider the  all-mass {\bf tensor} triangle \eqref{tensor}. We need to generalize the
				integral \eqref{tree} to insertion of arbitrary positive integer powers $a_1^m a_2^n$ of the original Feynman parameters. This is easy:
				\begin{equation} \label{tensortree}
					\int_0^1 da_1 \; a_1^m (1-a_1)^n  a_1^{s-1} (1-a_1)^{-s}\int_0^1 d\tilde{a}_2 \;  \tilde{a}_2^n\,  \tilde{a}_2^{t-1} (1-\tilde{a}_2)^{-(s+t)}
					={\Gamma(s+m)\Gamma(t+n)\Gamma(1-(s+t)) \over (m+n)!} \; . 
				\end{equation}
				
				Now we make the trivial observation that differentiating $u^{-s} v^{-t}$ in \eqref{mellin}, we have
				\begin{eqnarray}
						\left( u^m {\partial^m \over \partial u^m}\right)  \! \left( v^n {\partial^n \over \partial v^n}\right) 
						u^{-s} v^{-t} &=&(-1)^{m}s(s+1)\cdots (s+m-1) u^{-s} \cdot
						(-1)^{n} t(t+1)\cdots  (t+n-1) v^{-t} \nonumber \\
						&=&(-1)^{m+n} {\Gamma(s+m)\Gamma(t+n) \over \Gamma(s)\Gamma(t)}  u^{-s} v^{-t}  \; . \label{diff}
				\end{eqnarray}
				Now combine the forefactor in \eqref{diff} with the scalar integral \eqref{tree}. This converts the scalar integral \eqref{tree} to the tensor integral \eqref{tensortree}, up to a constant overall factor that is written explicitly in \eqref{claim}. This concludes the proof of the claim \eqref{claim}.

				\section{Comparison with Mellin representation} \label{app:Mellin}
				Here we briefly review the Mellin representation of the all-mass box, e.g.\ from \cite{Duplancic:2002dh},
				and in particular a residue expansion that we found useful for numerical checks.
				
				Begin from the all-mass triangle in eq.\eqref{mellin}. We can perform one of the contour integrals, say $s$, to get a hypergeometric (or Meijer G) representation:
				\begin{equation} \label{mellindouble}
					I_3(u,v) = {1 \over \pT_3^2}
					\int_{C_t} \dd t \,N_u(t) v^{-t}
				\end{equation}
				with the single Mellin kernel
				\begin{equation} \label{mellinhypF}
					N_u(t) = \csc^2(\pi t) \Gamma(1-t)^2 {}_2\widetilde{F}_1(1-t,1-t,2-2t,1-u)\ ,
				\end{equation}
				where $\tilde F$ is the regularized hypergeometric function. 
				Contrast this \eqref{mellindouble} with the double Mellin representation in \eqref{mellin}. 
				
				If we restrict $u$ to a region where the hypergeometric sum representation \eqref{mellinhypF} converges, the integral
				\eqref{mellindouble} can be computed as a sum over residues:
				\begin{equation} \label{sumrepmellin}
					I_3(u,v) = { 2\pi i  \over \pT_3^2}\sum_{n=0}^{\infty} \mathop{\mathrm{Res}}_{t=-n} N_u(t) v^{-t}\ .
				\end{equation}
				
				Now, a very similar representation can  written for the all-mass box, where the Mandelstam variables and masses-squared appear only in certain combinations (products). 
				From here one can derive a relationship between the 
				all-mass triangle and the all-mass box \cite{Duplancic:2002dh}:
				\begin{equation}
					I_4(s,t;m_1^2,m_2^2,m_3^2,m_4^2) = I_3(st,m_1^2m_3^2,m_2^2m_4^2) \qquad \mbox{($u,v$ not both negative)}\ .
				\end{equation}
				If $u,v$ are both negative, there is a correction term \cite{Duplancic:2002dh}.
				
				As mentioned at the beginning of this appendix, the sum representation \eqref{sumrepmellin}
				can be useful for numerics: by suitably restricting $u$ and $v$, the sum can be truncated and used as alternative  to
			          calculating integrals numerically.

			\section{Two relations for quadratic expressions in \texorpdfstring{$C^m$}{Cm}} \label{app:Cmrelations}
			
				In this appendix we look for examples of relations of the type\footnote{Note that we used $C^{m}_{1\mid 42,3}$ instead of $C^{m}_{1\mid 24,3}$ as in \cite{paper2} (cf.\ (6.32) therein, for instance) so that the indices $2,3,4$ always appear in cyclic order in each term. Both expressions, however, only differ by an overall sign, cf.\ \eqref{Cmpermute}.}
				\begin{equation}
					\begin{aligned}
						\beta_1 \left(C^{m}_{1\mid 23,4}\right)^{2} \!
						+ \beta_2 \left(C^{m}_{1\mid 42,3}\right)^{2}\!
						+ \beta_3 \left(C^{m}_{1\mid 34,2}\right)^{2} \!
						+ \beta_4\, C^{m}_{1\mid 23,4}\,C^{m}_{1\mid 42,3}
						+ \beta_5\,& C^{m}_{1\mid 23,4}\, C^{m}_{1\mid 34,2}
						+ \beta_6\, C^{m}_{1\mid 42,3}\,C^{m}_{1\mid 34,2}
						\\
						&= \sum_{i=1}^7 \sum_{j=i}^7 \gamma_{ij} f[i] f[j]
					\end{aligned}
				\end{equation}
				where we use the notation
				\be
				f = \{ \tft, F^4_{tu}, F^4_{us}, F^4_{st}, (F^2_t)^2, (F^2_u)^2, (F^2_s)^2 \}  
				\ee
				with\footnote{$\tft$ is proportional to $T_{F^3}$ of \cite{Bern:2017tuc}. Moreover, the definitions \eqref{Af3} and \eqref{tg} follow \cite{paper2}.}
				\bea
				&& \tft = - s_{23} s_{34} A^{F^3} (1,2,3,4)\ ,  \\[4pt]
				&& A^{F^3} (1,2,3,4) = s_{24} \left\{ \frac{t(1,2) t(3,4)}{s^2_{34}} + \frac{t(1,3) t(2,4)}{s^2_{24}} + \frac{t(1,4) t(2,3)}{s^2_{23}} - \frac{g_1 g_2 g_3 g_4}{s_{12}^2 s_{13}^2 s_{23}^2} \right\} \ , \label{Af3} \\[4pt]
				&& t(i,j) = (e_i \cdot k_j) (e_j \cdot k_i) - (e_i \cdot e_j) (k_i \cdot k_j)\ , \quad g_i = (k_{i-1} \cdot e_i) s_{i, i+1} - (k_{i+1} \cdot e_i) s_{i-1, i} \ . \label{tg}  
				\eea
				The elements of $f$ correspond to the 7 generators of the even part of the local photon S-matrix module given in (5.17) of \cite{Chowdhury:2019kaq}. Note that they use $(F_1 F_3 F_2 F_4)$ instead of $F^4_{tu}$. However, these are actually the same:
				\be
				(F_1 F_3 F_2 F_4) = (F_1 F_3 F_2 F_4)^T = (F_4^T F_2^T F_3^T F_1^T) = (F_4 F_2 F_3 F_1) =  (F_1 F_4 F_2 F_3)\ ,
				\ee
				where we used ${\rm Tr} A^T = {\rm Tr} A$ and the cyclicity of the trace. We also note that the generator $E_{\bf S}$ of \cite{Chowdhury:2019kaq} (given more explicitly in their (5.20)) can be expressed as\footnote{We use the relations $s_{23} = t/2, s_{24} = u/2, s_{34} = s/2$, cf.\ \cite{Bern:2017tuc} which, however, differs from the convention used in \cite{Chowdhury:2019kaq}.}
				\begin{equation}
					\begin{aligned}
						E_{\bf S} & =  - 2 \tft + 2 s_{34} F^4_{tu} + 2 s_{23} F^4_{us} + 2 s_{24} F^4_{st}\\
						&= - 2 \tft + s F^4_{tu} + t F^4_{us} + u F^4_{st}\ .
					\end{aligned}
				\end{equation}
				
				We found the two relations
					\begin{equation}
					\begin{aligned}
						& s_{23}^2 (s_{34} - s_{24}) \left(C^{m}_{1\mid 23,4}\right)^{2} + s_{24}^2  (s_{23} - s_{34}) \left(C^{m}_{1\mid 42,3}\right)^{2}
						+ s_{34}^2 (s_{24} - s_{23}) \left(C^{m}_{1\mid 34,2}\right)^{2} \\
						& + s_{23} s_{24} (s_{24} - s_{23}) C^{m}_{1\mid 23,4}\,C^{m}_{1\mid 42,3}
						+ s_{23} s_{34} (s_{23} - s_{34}) C^{m}_{1\mid 23,4}\,C^{m}_{1\mid 34,2}
						+ s_{24} s_{34} (s_{34} - s_{24}) C^{m}_{1\mid 42,3}\,C^{m}_{1\mid 34,2} \\
						=& \ \frac{t-u}{2 s} \Big[ F^4_{tu} - \frac14 \Big( (F^2_t)^2 + (F^2_u)^2 + (F^2_s)^2 \Big) \Big]^2 
						+ \frac{u \left(s-t\right)}{t s} \Big[ F^4_{tu} F^4_{us} - \frac14 F^4_{st} \Big( (F^2_t)^2 + (F^2_u)^2 + (F^2_s)^2 \Big) \Big] \\
						& +\ {\rm cyclic} \ ,
					\end{aligned}
				\end{equation}
				and
				\begin{equation}
					\begin{aligned}
						& \frac{s_{23}\,s_{24}\,s_{34}}{(s_{23}-s_{24})\,(s_{23}-s_{34})} \left(C^{m}_{1\mid 23,4}\right)^{2} + \frac{s_{23}\,s_{24}\,s_{34}}{(s_{24}-s_{34}) (s_{24}-s_{23})} \left(C^{m}_{1\mid 42,3}\right)^{2}\\
						& + \frac{s_{23}\,s_{24}\,s_{34}}{(s_{34}-s_{23})\,(s_{34}-s_{24})} \left(C^{m}_{1\mid 34,2}\right)^{2}
						- \frac{2\,s_{23}\,s_{24}\,s_{34}}{(s_{23}-s_{34})\,(s_{24}-s_{34})} C^{m}_{1\mid 23,4}\,C^{m}_{1\mid 42,3} \\
						&- \frac{2\,s_{23}\,s_{24}\,s_{34}}{(s_{34}-s_{24})\, (s_{23}-s_{24})} C^{m}_{1\mid 23,4}\,C^{m}_{1\mid 34,2}
						- \frac{2\,s_{23}\,s_{24}\,s_{34}}{(s_{24}-s_{23})\,(s_{34}-s_{23})} C^{m}_{1\mid 42,3}\,C^{m}_{1\mid 34,2}  \\
						=&  \ -2 {\cal M}^{R^2} (1,2,3,4) + \frac{4}{(t-s)\, (u-s)} \Big( F^4_{tu} \Big[ F^4_{tu} - \frac14 \Big((F^2_t)^2 + (F^2_u)^2 + (F^2_s)^2\Big) \Big] - F^4_{us} F^4_{st} \Big) + {\rm cyclic}\ , 
					\end{aligned}
				\end{equation}
				where 
				\bea
				{\cal M}^{R^2} (1,2,3,4) & = & \tft \left( -\frac{1}{s_{23}\,s_{24}\,s_{34}}\right) t_{8}(1,2,3,4) \label{MR2} \\ 
				& = &  - s_{23} s_{34} A^{F^3} (1,2,3,4) \left( -\frac{1}{s_{23}\,s_{24}\,s_{34}}\right) \left( \frac12 s_{23} s_{34} C_{1|234} \right)\\
				& = & - s_{23} s_{34} A^{F^3} (1,2,3,4) \left( -\frac{1}{s_{23}\,s_{24}\,s_{34}}\right) \left( \frac12 s_{23} s_{34} \frac{s_{24}}{s_{23}} C_{1|243} \right)\\
				& = &  A^{F^3} (1,2,3,4) s_{12} A^{\rm tree}(1,2,4,3)
				\eea
				was defined in (6.34) of \cite{paper1}, cf.\ also (6.41) therein.
				Moreover, we used the definition\footnote{The $t_8$ tensor used here follows the definition in \cite{paper2} and differs from the one used in \cite{Green:1987mn} by an overall minus sign, cf.\ also (12.4.25) in \cite{Polchinski2}.} 
				\be
				t_{8}(1,2,3,4) = -\frac14 t_8 F^4 = -\frac12 (F_1 F_2 F_3 F_4) + \frac18 (F_1 F_2) (F_3 F_4) + {\rm cyclic}(2,3,4)\ .
				\ee 
				Note that ${\cal M}^{R^2} (1,2,3,4)$ of \eqref{MR2} vanishes in strictly $D=4$.
				
				In the above discussion we used some of the relations in appendix \ref{sec:BCJ_ids}.

				
				~
				
				\bibliographystyle{JHEP}
				\bibliography{bibliography}

			\end{document}